\documentclass{aa}
\usepackage{graphicx}
\newcommand{\oldfootnoterule}{\footnoterule}
\renewcommand{\footnoterule}{ }
\begin{document}
   \title{Searching for signs of triggered star formation toward IC 1848}

   \author{M.A. Thompson\inst{1},
           G.J. White\inst{1},
	   L.K. Morgan\inst{1},
	   J. Miao\inst{1},
	   C.V.M. Fridlund\inst{2}
	  \and
	  M. Huldtgren-White\inst{3}
	 \fnmsep
          }
   \authorrunning{M.A.~Thompson et al.}
   
   \titlerunning{Searching for triggered star formation toward IC 1848}

   \offprints{M.A. Thompson}

   \institute{Centre for Astrophysics \& Planetary Science,
   School of Physical Sciences,
   University of Kent,
   Canterbury,
   Kent CT2 7NR,
   UK
   \and  Astrophysics Division, Space Science Department, ESTEC, P.O.Box 299,
         2200 AG Noordwijk, The Netherlands       
   \and  Stockholm Observatory, Roslagstullsbacken 21, SE-106 91 Stockholm, Sweden.
	  }

   \date{}

   \abstract{We have carried out an in-depth study of three bright-rimmed clouds \object{SFO 11},
SFO 11NE and \object{SFO 11E} associated with the HII region IC 1848, using observations carried
out at  the James Clerk Maxwell Telescope (JCMT) and the  Nordic Optical Telescope (NOT),
plus archival data from IRAS, 2MASS and the NVSS. We show that the overall morphology of
the clouds is reasonably consistent with that of radiative-driven implosion (RDI) models
developed to predict the evolution of cometary globules. There is evidence for a
photoevaporated flow from the surface of each cloud and, based upon the morphology and pressure
balance of the clouds, it is possible that  D-critical ionisation fronts are propagating
 into the molecular gas. The primary O star responsible for ionising the surfaces of the clouds is
the 06V star  \object{HD17505}. Each cloud is associated with either recent or ongoing star formation: 
we have detected
8 sub-mm cores which possess the hallmarks of protostellar cores and 
identify  YSO candidates from 2MASS data.  We infer the past and future
evolution of the clouds and demonstrate via a simple
pressure-based argument that the UV illumination may have induced the collapse of the
dense molecular cores found at the head of \object{SFO 11} and \object{SFO 11E}.   \keywords{Stars: formation -- ISM: HII
regions -- ISM: Individual object: IC 1848 -- ISM: Clouds -- ISM: Dust -- ISM: Molecules} }

   \maketitle
%
%________________________________________________________________

\section{Introduction}

 \object{IC 1848} is a large HII region, forming part of the   \object{W5} HII
region-molecular cloud complex in the Perseus Arm. The HII region component of  W5
(also known as S 199) is made up of two roughly circular thermal shells W5 East and W5
West, which are  separated by a dust lane. IC 1848 is located to the south
of W5 West  (Braunsfurth 1983). W5 West is excited by the open cluster OCl 364, which
comprises four O stars, whereas only one O star is  visible within W5 East (Normandeau,
Taylor \& Dewdney.~\cite{normandeau}).  The whole complex lies at a distance of 1.9 kpc
(Ishida \cite{ishida}) and is a well known and well-studied star-forming region
(e.g.~Val\'ee, Hughes \& Viner \cite{vhv79}; Braunsfurth \cite{braun}; Normandeau,
Taylor \& Dewdney \cite{normandeau}; Heyer \& Terebey \cite{heyer}; Carpenter, Heyer \&
Snell \cite{chs00}). 

Numerous small bright-rimmed  clouds (sometimes also known as bright-rimmed globules) are 
found at the rims of W5 East and West,   which may be  star-forming regions triggered via
the expansion of the HII regions (Sugitani, Fukui \& Ogura \cite{sfo}). The expansion of
the HII regions drives shocks into the surrounding  molecular gas and these
photoionisation-induced shocks are thought to trigger the collapse of sub-critical
molecular cores within the clouds in a process known as radiative-driven implosion or RDI
(Bertoldi \cite{bertoldi}; Bertoldi \& McKee \cite{bk90}; Lefloch \& Lazareff \cite{ll94},
\cite{ll95}).   Radiative-driven implosion of molecular cores at the periphery of HII
regions may thus be responsible for a subsequent generation of star formation, amounting
to a possible cumulative total of several hundred new stars per HII region (Ogura,
Sugitani \& Pickles \cite{osp02}) and perhaps 15\% or more of the low-to-intermediate mass
stellar mass function (Sugitani, Fukui \& Ogura \cite{sfo}). Confirming bright-rimmed
clouds as star-forming can provide important insights about the clustered mode of star
formation and the overall star-formation efficiencies of molecular clouds.  

Sugitani, Fukui \& Ogura(\cite{sfo}), hereafter referred to as SFO91, searched the
Sharpless HII region catalogue (Sharpless \cite{sharpless}) for bright-rimmed clouds
associated with IRAS point sources, in order to identify potential star-forming clouds
via their far-infrared emission.  Later, Sugitani \& Ogura (\cite{so94}) -- SO94 --
extended their search to  include bright-rimmed clouds from the ESO(R) Southern
Hemisphere Atlas. At least 89 bright-rimmed clouds have been found to be associated
with IRAS point sources. For brevity (and consistency with SIMBAD) we will refer to the
combined  SFO91 and SO94 catalogues as the SFO catalogue.  Whilst a few individual
clouds from the SFO catalogue have been studied in detail (e.g.~Lefloch, Lazareff \&
Castets~\cite{llc97}; Megeath \& Wilson \cite{mw97}; Codella et al.~\cite{cbnst01}) and
shown to harbour protostellar cores, the vast majority of the SFO clouds  have not been
associated with star-forming regions and the state of these clouds remains unknown.  We
have carried out a star formation census of SFO bright-rimmed clouds (Thompson et
al.~\cite{thompsona}, \cite{thompsonb}) to investigate the star-forming nature of the SFO bright-rimmed clouds and
determine whether any star formation in the SFO sample was likely to have been
triggered by the RDI process.  In this paper we report the results for three
bright-rimmed clouds from our census.

\begin{figure}
\begin{center}
\includegraphics*[scale=0.7,trim=255 70 270 80]{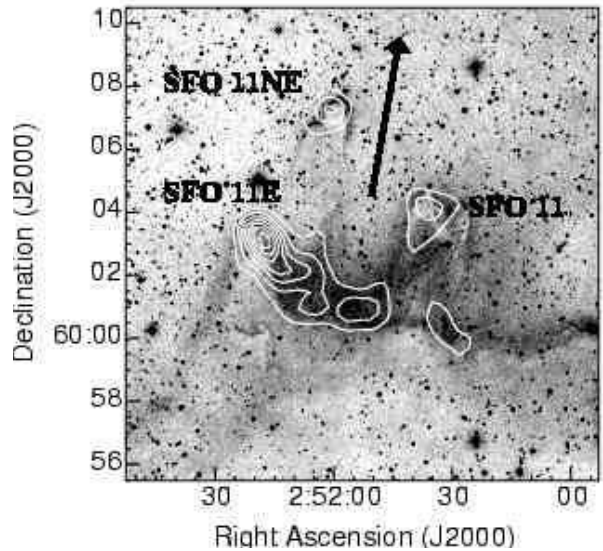}
\label{fig:dssnvss}
\caption{A red Digitised Sky Survey image of the three clouds \object{SFO 11}, \object{SFO 11NE} and SFO
11E. The image is centred on \object{SFO 11} and the other two clouds are the cometary
bright-rimmed clouds found to the NE and E respectively. High pixel
values have been excluded to emphasise the bright optical rims of the clouds. The arrow marks the direction
toward the suspected exciting star of the clouds, HD 17505. The NVSS 20 cm radio
emission is shown by the grey contours which start at 1.5 mJy and are spaced by
1.5 mJy.  The bright
ridge to the south probably marks the southern ionisation boundary of IC 1848
and the broken appearance in the radio map may be partly a consequence of the limited 
sensitivity of NVSS to large angular size scales.}
\end{center}
\end{figure}

The bright-rimmed cloud \object{SFO 11} is found at the southern edge of IC 1848. It is positionally
associated with the IRAS point source 02476+5950 and in optical images is
double-lobed and cometary in appearance (see Fig.~\ref{fig:dssnvss}). 
Two other bright-rimmed clouds that are not in the SFO
catalogue are found within 6\arcmin\ of \object{SFO 11}. Following the terminology of Ogura,
Sugitani \& Pickles (\cite{osp02}) they are described in this paper as \object{SFO 11NE} and SFO
11E. \object{SFO 11NE} is a cometary cloud found 4\arcmin\ NE of \object{SFO 11}, with a ``shoulder'' to the
east side of the cloud. \object{SFO 11E} is found 6\arcmin\  E of SFO11 and is associated with a
bright ridge of nebular emission at the Southern ionisation boundary of IC 1848. A red
optical image of the three clouds obtained from the Digitised Sky Survey is shown in
Fig.~\ref{fig:dssnvss}.

Neither \object{SFO 11NE} or \object{SFO 11E} are associated with any IRAS point sources.  This may be
either due to  confusion caused by the limited IRAS resolution or  simply because SFO
11NE and E do not contain embedded protostars or stars. We included \object{SFO 11} and the two
neighbouring clouds in our study of the SFO catalogue  as they represent an ideal
opportunity to study three possibly star-forming clouds close to each other on the sky
and lying at a similar distance from the UV illumination source. In this paper  we
present JCMT molecular line, SCUBA sub-mm  continuum, Nordic Optical Telescope narrowband
H$\alpha$ and archival observations (IRAS
HIRES, 2MASS and VLA NVSS) to investigate the star-forming activity and general
environment of these three clouds. These data are then modelled to try to understand
their future evolution as potential star-forming regions.

In Sect.~\ref{sect:obs} we describe the observational procedure. The data are analysed in
 Sect.~\ref{sect:anal} where evidence for 
star formation, protostellar cores,
embedded IR sources or molecular outflows is discussed. In Sect.~\ref{sect:discuss} we
explore implications for present and future star formation within the clouds. 
Finally in Sect.~\ref{sect:conc} we
present a summary of our conclusions.

%__________________________________________________________________

\section{Observations}
\label{sect:obs}

\subsection{SCUBA dust continuum maps}

We obtained simultaneous 450 and 850 $\mu$m images of all three clouds using the sub-mm
bolometer camera  SCUBA (Holland et al.~\cite{scuba}) on the James Clerk Maxwell
Telescope (JCMT\footnote{The
JCMT is operated by the Joint Astronomy Centre on behalf of PPARC for the United
Kingdom, the Netherlands Organisation of Scientific Research, and the National Research
Council of Canada.}). SCUBA is comprised of two bolometer arrays, a short-wave array of 91
pixels optimised for operation at 450 $\mu$m and a long-wave array of 37 pixels
optimised for operation at 850 $\mu$m. Both arrays simultaneously sample  a similar
field of view (approx 2\arcmin\ square), although the spacing in between individual
bolometers on the array means that not all of the field of view is sampled 
instantaneously.  To fill in the gaps in spatial coverage the telescope secondary mirror
is moved in a 64-point pattern (``jiggling''), whilst also chopping at a frequency of 1
Hz to remove the sky emission. This procedure is commonly known as a ``jiggle-map''  and
provides  maps with full spatial sampling at both wavelengths.

As each of our three target clouds is less than 2\arcmin~in diameter (from
inspection of the DSS images) we obtained a single jiggle-map of each cloud,
approximately centred on the cometary ``head'' of the clouds. The jiggle-maps were
taken on the nights of the 7th January 2002 (\object{SFO 11} \& \object{SFO 11E}) and the 5th
June 2002 (\object{SFO 11NE}), as part of a wider SCUBA survey of bright-rimmed clouds
(Thompson et al.~\cite{thompsona}). Observational parameters for each cloud are
summarised in Table \ref{tbl:obs}. The maps taken on the 7th January had chop
throws set to 120\arcsec~to avoid chopping onto the array. During the data
reduction of these maps it was noticed that the extended nature of the sources
had led to some chopping onto emission at the edges of the field-of-view.The maps
 taken on the 5th June had chop throws set to the maximum
of 180\arcsec~to avoid this problem. The chop directions for each cloud were
chosen so that the sky positions did not lie on any of the other clouds in the
complex or the extended dark nebular region to the south. SCUBA is installed at
one of the Nasmyth foci on the JCMT and is not equipped with a beam rotator.
The chopping was performed in sky coordinates so  that the chop position stayed
constant over each integration and did not rotate onto cloud emission. In
addition to the maps of the three clouds we also obtained absolute flux
calibration and beam maps of the primary flux calibrator Uranus 
and the secondary calibrator CRL 618. 
Hourly skydips at the azimuth of each observation  were carried out
to estimate the atmospheric zenith optical depth. These values were contrasted
with the fixed-azimuth measurements at 225 GHz made every 10 minutes by the CSO
tipping radiometer and both sets of measurements were found to be consistent.

The data were reduced using a combination of the automated SCUBA reduction
pipeline ORACDR (Economou et al.~\cite{oracdr}),  the SCUBA reduction package SURF
(Jenness \& Lightfoot \cite{surf}) and the Starlink image analysis package KAPPA (Currie
\& Bell \cite{kappa}).
The reduction procedure for 450 and 850 $\mu$m data was the same and
followed the outline given in this paragraph. Initially the chopping and
nodding positions were subtracted from the on-source data to form a
time-ordered series of sky-subtracted bolometer measurements. As the
bright-rimmed clouds are embedded in the larger molecular cloud complex
W5, it is likely that the chopping procedure resulted in a subtraction of
extended cloud emission from the flux levels in each map (particularly in the
case of \object{SFO 11E}, see Sect.~\ref{sect:anal}). The measured fluxes are thus strictly
lower limits to the true flux.  The time-ordered bolometer data were then
corrected for atmospheric extinction using an optical depth value interpolated
from skydips carried out before and after the jiggle-map. At this stage 
bolometers with a mean noise in excess of 100 nV were blanked and transient bolometer 
noise spikes were removed by applying a 
 5$\sigma$ clip to the data.
 Residual sky variations between individual bolometers were
removed using the SURF task \emph{remsky}. The time-ordered data were then regridded
to J2000 sky coordinates with the SURF task \emph{rebin}. In the case of \object{SFO 11},
where two separate jiggle-maps were obtained on the same night, \emph{rebin} was
used to co-add the maps.

Absolute flux calibration was carried out using the calibration maps of Uranus and CRL
618. Predicted fluxes for Uranus and CRL 618 were estimated using the values given by
the Starlink package FLUXES (Privett, Jenness \& Matthews \cite{fluxes}) and on the
JCMT calibrator webpage respectively. Flux correction factors (FCFs) for each
wavelength were then determined by dividing the predicted flux by the measured peak
value of the calibrator. Each jiggle-map was calibrated in units of Jy/beam by multiplying
by the appropriate FCF. The FWHMs and peak values of the  telescope main and error
beams were determined by fitting two Gaussians to azimuthal averages of the maps of the
primary calibrator (Uranus).  These parameters are shown in Table \ref{tbl:fcf}.

\renewcommand{\thefootnote}{\alph{footnote}}
\setcounter{footnote}{0}
\begin{table*}
\caption{Observation parameters for the SCUBA jiggle-maps}
\label{tbl:obs}
\begin{minipage}{\linewidth}
\begin{tabular}{lccccccc} \hline
Map name & Night & Chop throw & Integrations\footnotemark &
\multicolumn{2}{c}{Average
$\tau$} & \multicolumn{2}{c}{R.m.s.~noise (Jy/14\arcsec~beam)\footnotemark} \\
 & & & & 450 $\mu$m & 850 $\mu$m & 450 $\mu$m & 850 $\mu$m \\\hline\hline
SFO 11 & 7th Jan 2002 & 120\arcsec & 20 & 1.82 & 0.33 & 0.2 & 0.013 \\
SFO 11NE & 5th June 2002 & 180\arcsec & 6 & 1.94 & 0.34 & 0.1 & 0.020 \\
SFO 11E & 7th Jan 2002 & 120\arcsec & 10 & 1.82 & 0.33 &  0.3& 0.018 \\ \hline
\end{tabular}
\footnotetext[1]{One SCUBA ``integration'' is equivalent to 64 seconds of
integration}
\footnotetext[2]{450 $\mu$m data smoothed to 14\arcsec~resolution}
\end{minipage}
\end{table*}

\begin{table*}
\caption{Gaussian fit parameters to azimuthal averages of the
 primary calibrator beam maps}
\label{tbl:fcf}
\begin{minipage}{\linewidth}
\begin{tabular}{lcccccc} \hline
Night & Wavelength & \multicolumn{2}{c}{Main Beam} & \multicolumn{2}{c}{Error Beam}\\
 & ($\mu$m) & FWHM (\arcsec) & Relative peak & FWHM (\arcsec) & Relative peak \\ \hline\hline
7th Jan 2002 & 450 & 9.0 & 0.952 & 32.4 & 0.048  \\
 & 850 & 15.3 & 0.986 & 65.7 & 0.014  \\
5th June 2002 & 450 & 8.4 & 0.928 & 25.8 & 0.072  \\
 & 850 & 15.0 & 0.982 & 58.2 & 0.018  \\ \hline
\end{tabular}
\end{minipage}
\end{table*}

\begin{figure*}[p!]
  % \centering
   \includegraphics*[scale=0.4,trim=200 30 200
   50]{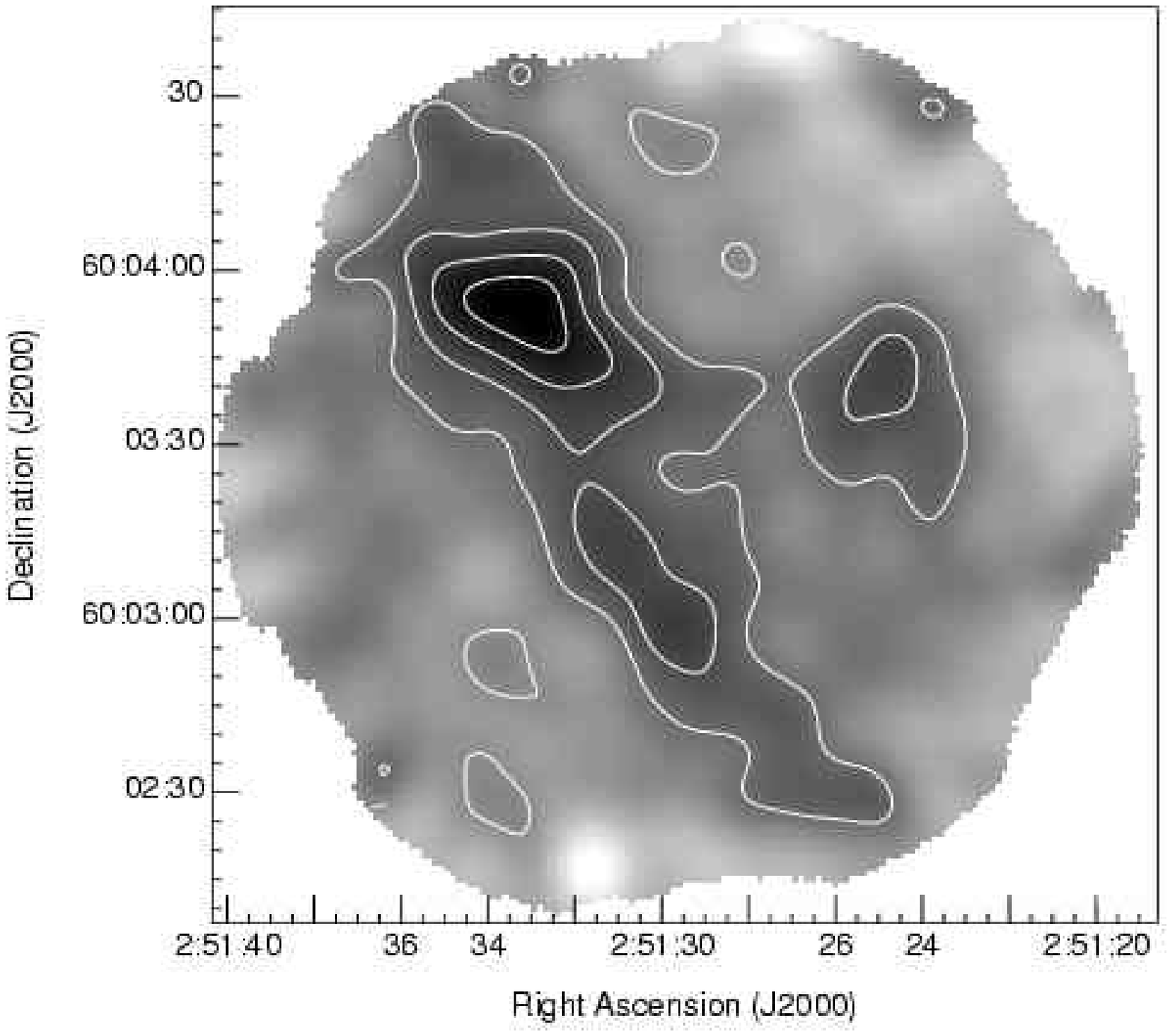}\includegraphics*[scale=0.4,trim=180 0 200 50]{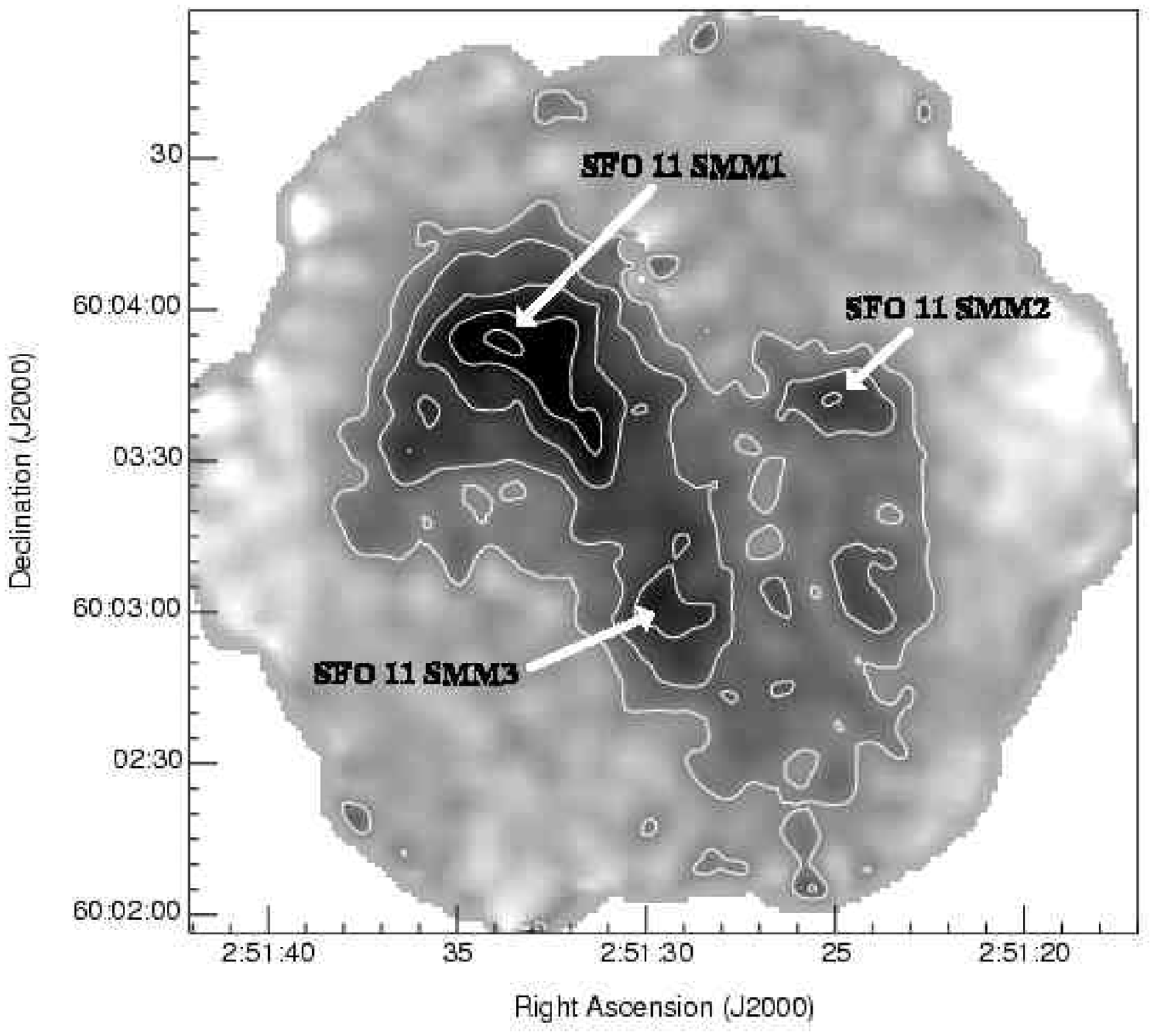}
   \includegraphics*[scale=0.4,trim=200 30 200
   50]{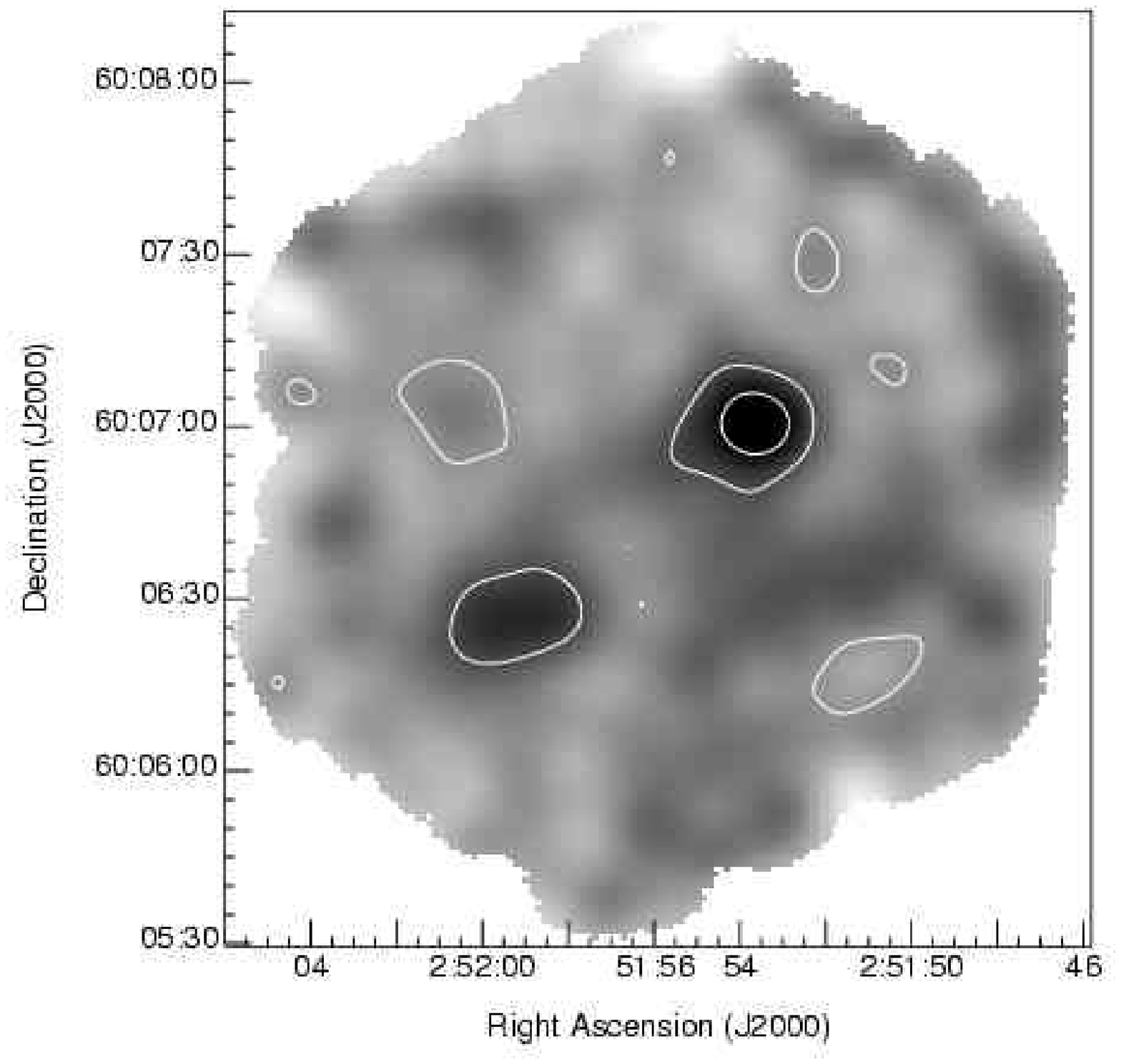}\includegraphics*[scale=0.4,trim=180 0 200 50]{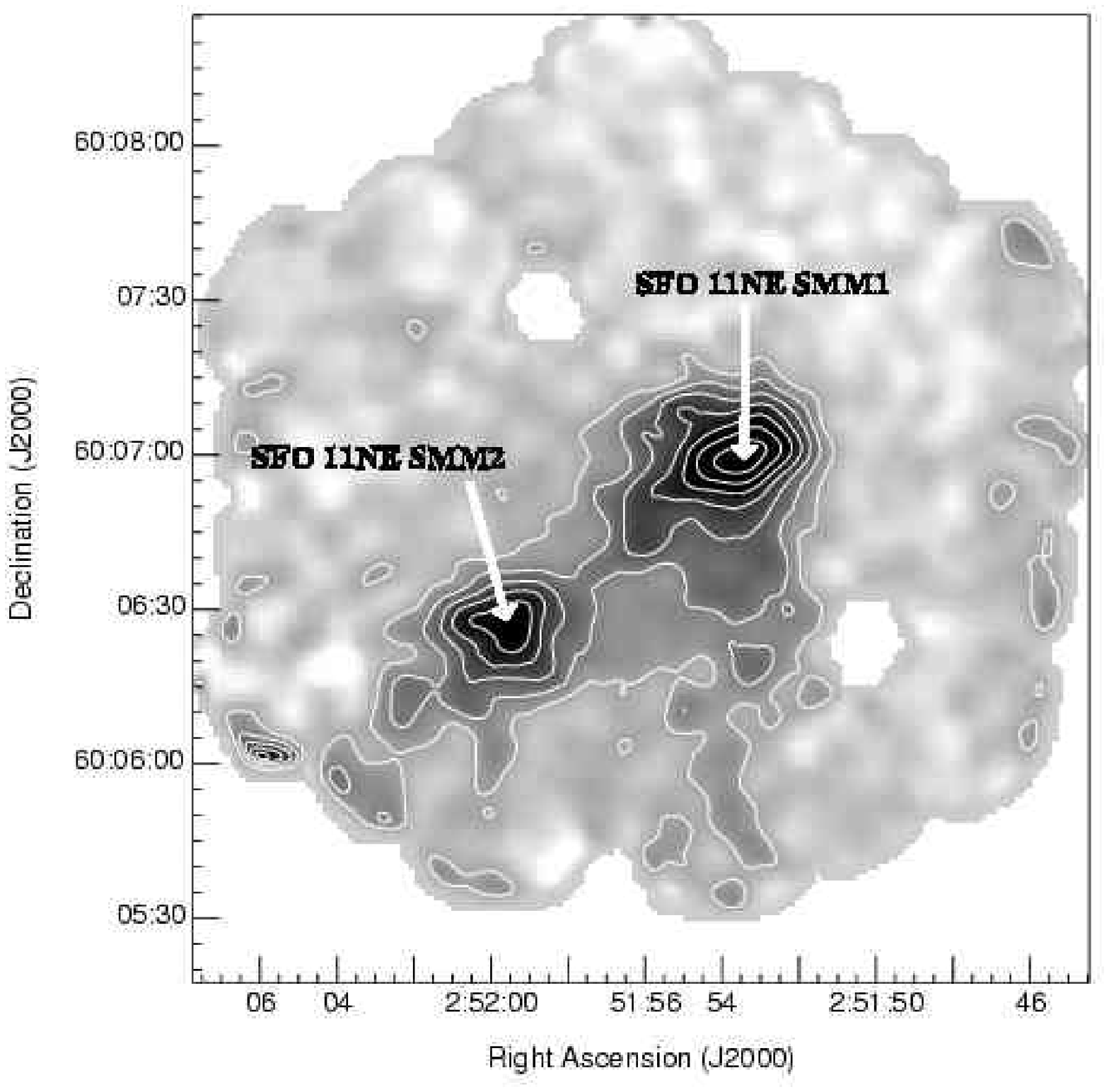}
   \includegraphics*[scale=0.4,trim=200 30 200
   50]{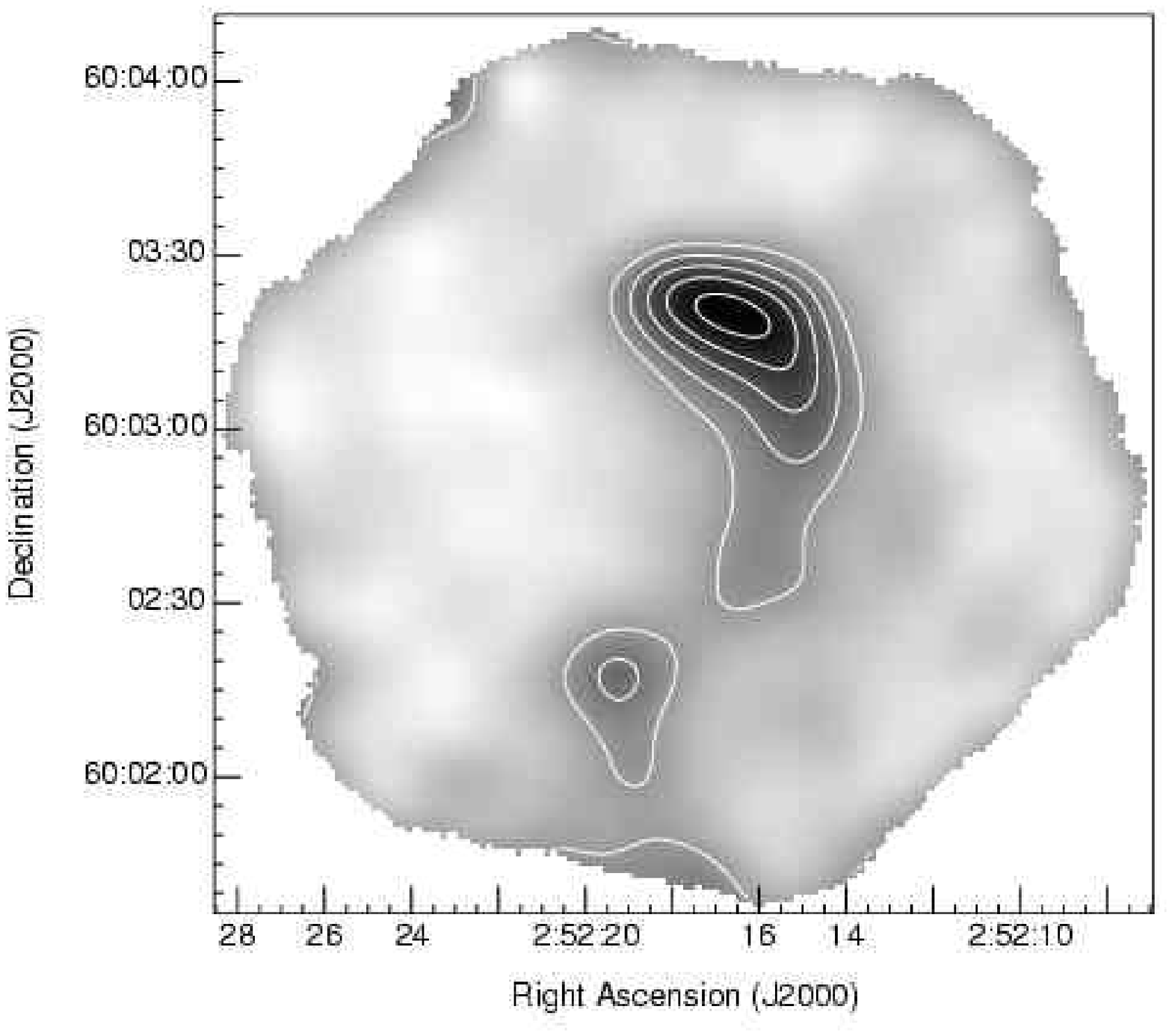}\includegraphics*[scale=0.4,trim=180 0 200 50]{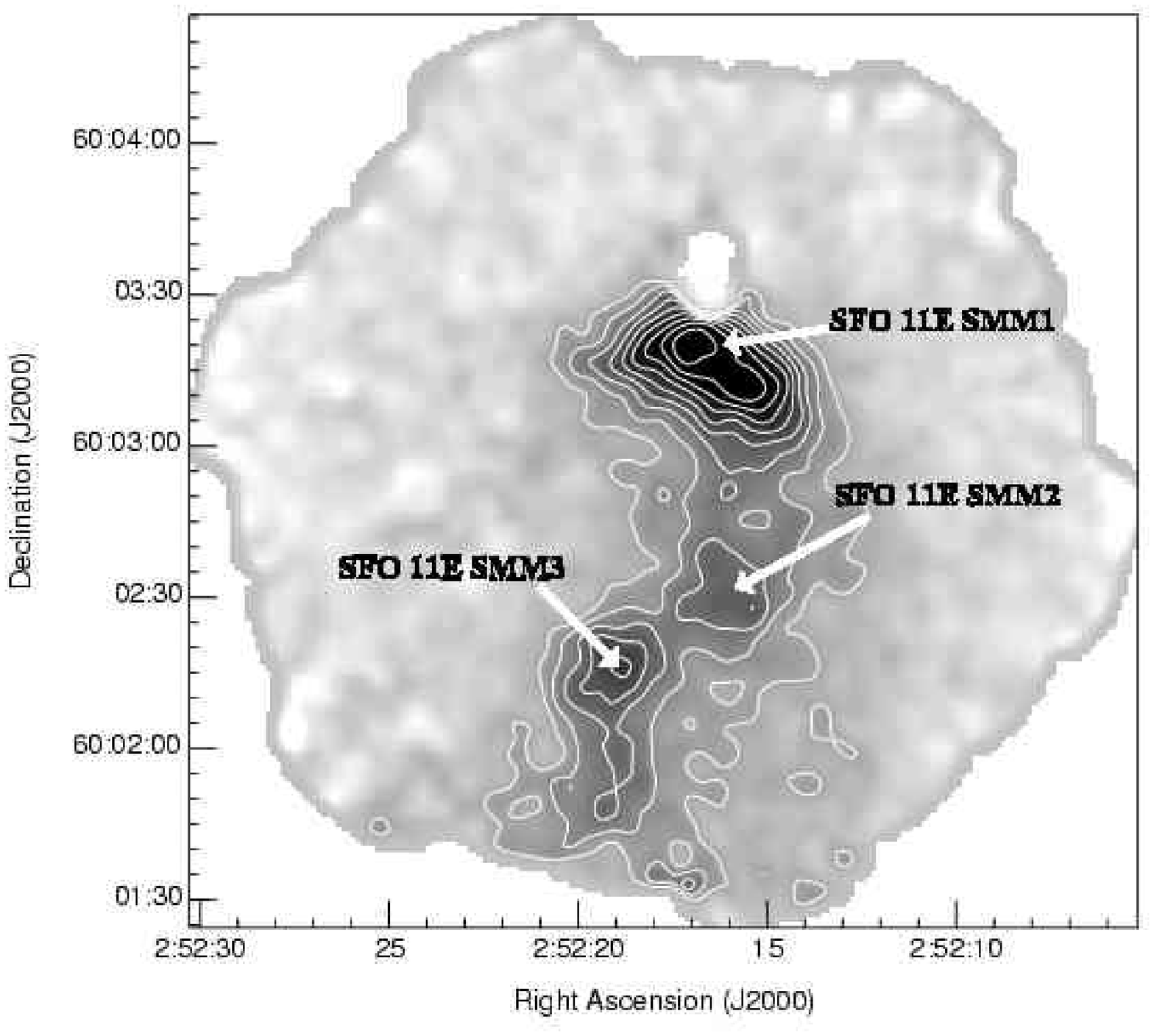}
      \caption{450 $\mu$m (\emph{left}) and 850 $\mu$m (\emph{right}) SCUBA maps of \object{SFO 11}, \object{SFO 11NE} and \object{SFO 11E} (from top to
      bottom). The 450 $\mu$m contours start at 600 mJy/beam and are spaced by 600 mJy/beam,
      except for the map of \object{SFO 11E} (bottom left panel) where the contours start at 900
      mJy/ebam with a spacing of 900 mJy/beam. The 850 $\mu$m contours start at 45 mJy/beam and are spaced by 45 mJy/beam, which is
      approximately 3$\sigma$. The core names are labelled in the 850 $\mu$m maps. The ``holes'' in the maps of \object{SFO 11} E and the 
      ``scallop'' in the contours at the N of \object{SFO 11E} are caused by the removal of
      noisy bolometers from the SCUBA data. }
       \label{fig:scumaps}
   \end{figure*}

The calibrated images were then converted into FITS format and deconvolved to
remove the contribution from the error beam. The deconvolution was performed
using the \emph{clean} task in MIRIAD (Sault, Teuben \& Wright \cite{miriad}) 
with a circularly symmetric two-component
Gaussian beam derived from azimuthal averages of the primary calibrator maps 
(see Table \ref{tbl:fcf} for the Gaussian fit parameters). 
Each image was cleaned down to a cutoff level of twice the 1$\sigma$ r.m.s.~noise and
then restored back to a resolution appropriate for the wavelength (8\arcsec~for
450 $\mu$m and 14\arcsec~for 850 $\mu$m). The advantage of this technique is that
the different error beam contributions from each wavelength are removed,
facilitating comparison of 450 $\mu$m and 850 $\mu$m maps and allowing
the integrated fluxes to be determined more accurately. The clouds were found to be
marginally detected at 450 $\mu$m, with a peak S/N ratio of 4 at most. 
The native resolution 8\arcsec\ 450 $\mu$m maps were smoothed to the
same resolution as the 850 $\mu$m maps (14\arcsec) to increase their signal to noise ratio.
The cleaned calibrated 450 \& 850 $\mu$m maps are shown in Fig.~\ref{fig:scumaps}.

\subsection{JCMT CO mapping}

Maps of all three clouds in the $^{12}$CO and $^{13}$CO J=2--1 lines were
obtained with the JCMT during April 1996 and January 1997. The heterodyne
front-end receiver A2 was used along with a back-end digital autocorrelation
spectrometer (DAS). All maps were obtained in raster-mapping mode, in which
the telescope is scanned along a line on the sky, sampling 
spectra at regular time intervals to provide Nyquist or better spatial sampling.
At the end of each line the telescope position-switches to a clean offset
position so that the spectra may be sky-subtracted. To reduce the likelihood of
rastering artifacts the maps were sampled at better than Nyquist sampling, with 
6\arcsec\ sampling intervals both parallel
and perpendicular to the scan lines. The FWHM of the telescope beam at 230 GHz
is $\sim$ 21\arcsec. 

The integration time per map position was typically 6 seconds 
and in most cases two or three raster maps of each
cloud were co-added to improve the signal to noise ratio. The only
exception was the $^{12}$CO map of \object{SFO 11}, where only one raster map was observed.
 The atmospheric conditions during the observations
were stable and good, with typical system temperatures of 400--500 K. The
resulting 1$\sigma$ sensitivity of each map is between 0.5--1 K per 0.2 km s$^{-1}$ 
channel. The velocities of the lines
were observed relative to the Local Standard of Rest (LSR) and all the velocities
quoted in this paper are on this scale.

   \begin{figure*}
\vspace*{0.5cm}
\hspace*{3.3cm}{\large {\bf\textsf{SFO 11}}}\hspace*{3.8cm}{\large {\bf\textsf{SFO
11NE}}}\hspace*{3.4cm}{\large {\bf\textsf{SFO 11E}}}\hspace*{-3cm}

  \centering
\vspace*{0.5cm}\hspace*{0.45cm}\includegraphics*[scale=0.3,trim=0 0 0 
68]{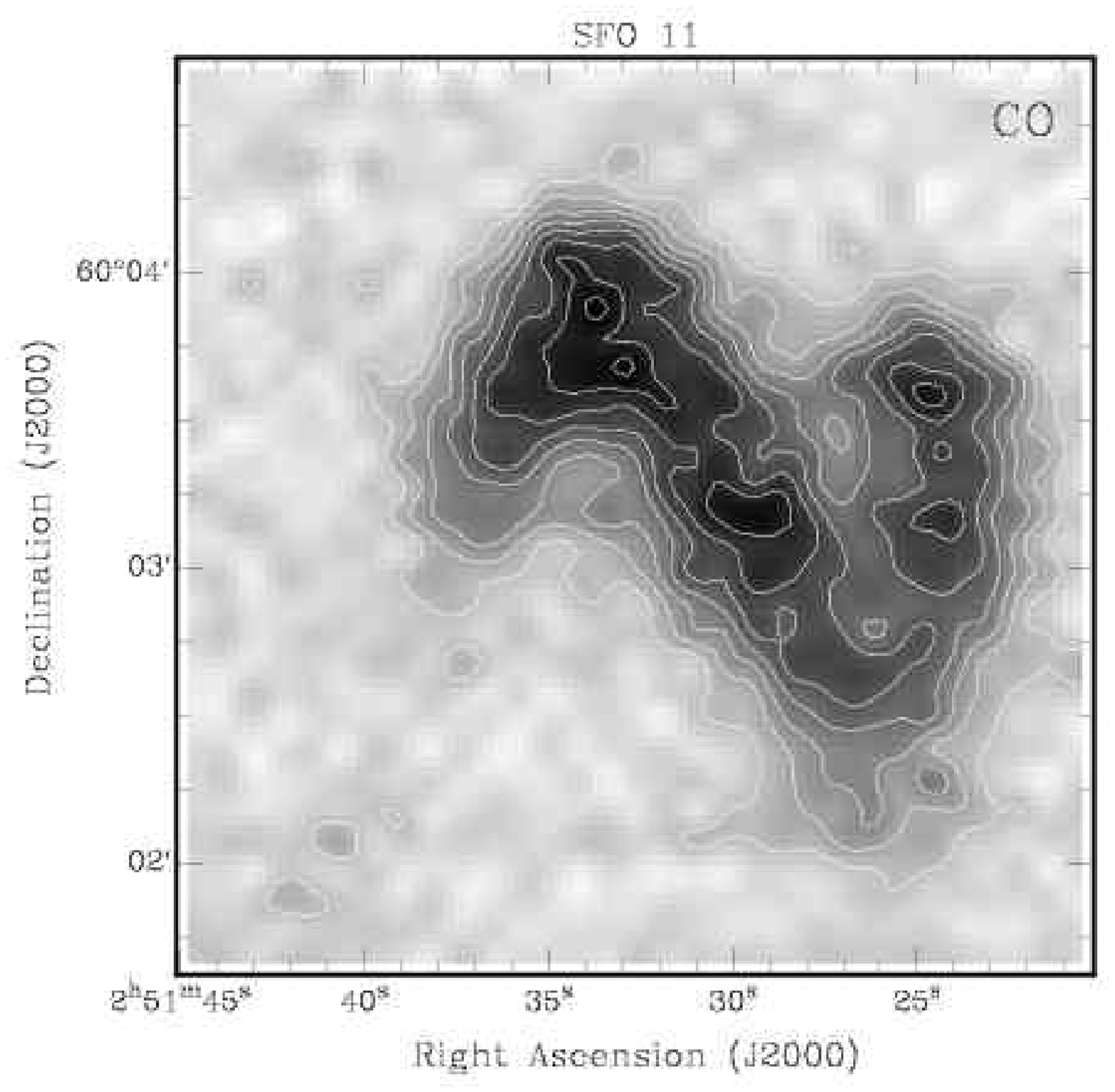}\hspace*{-0.35cm}\includegraphics*[scale=0.3,trim=0 0 0
48]{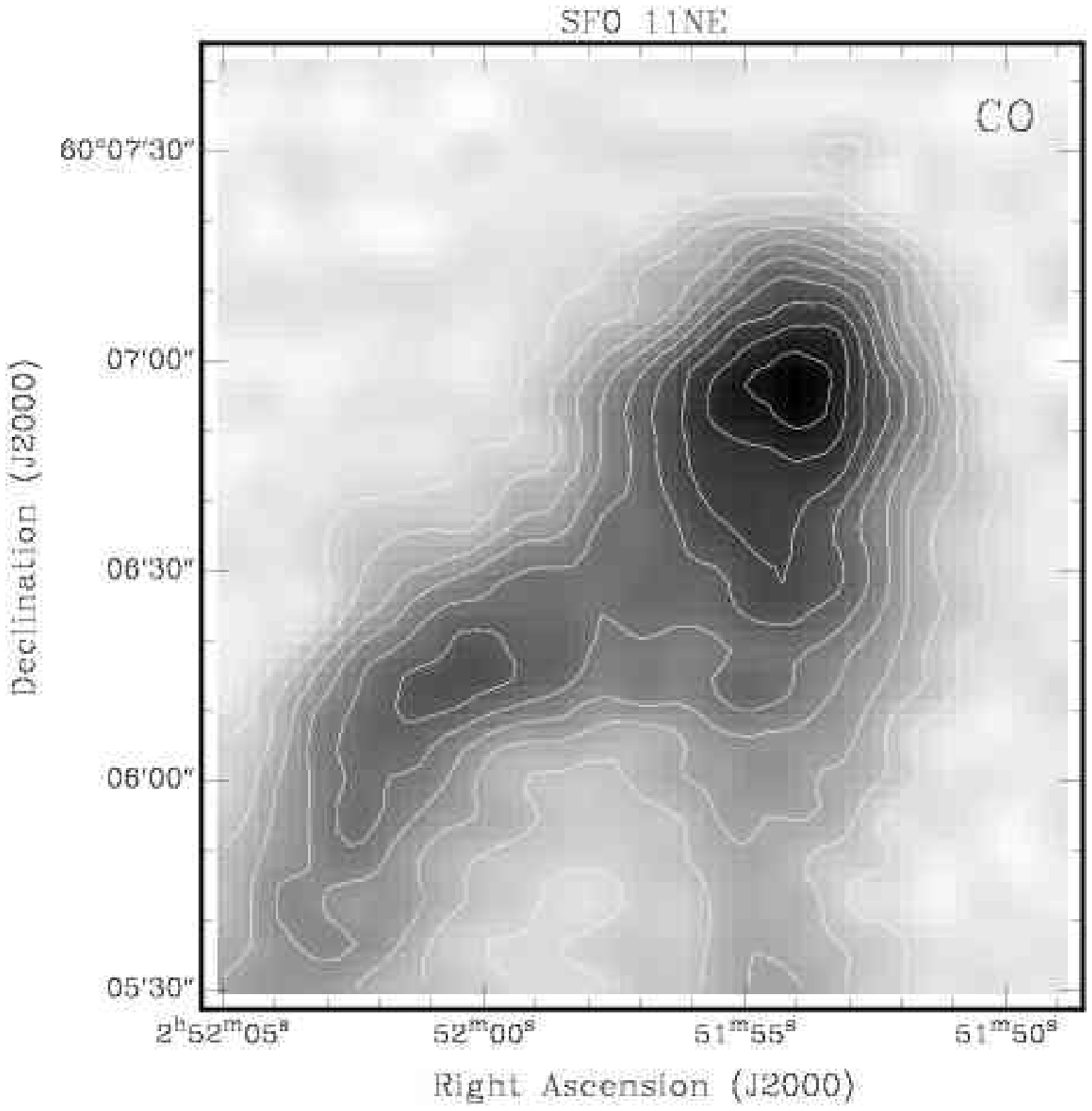}\hspace*{0.1cm}\includegraphics*[scale=0.3,trim=0 0 0
68]{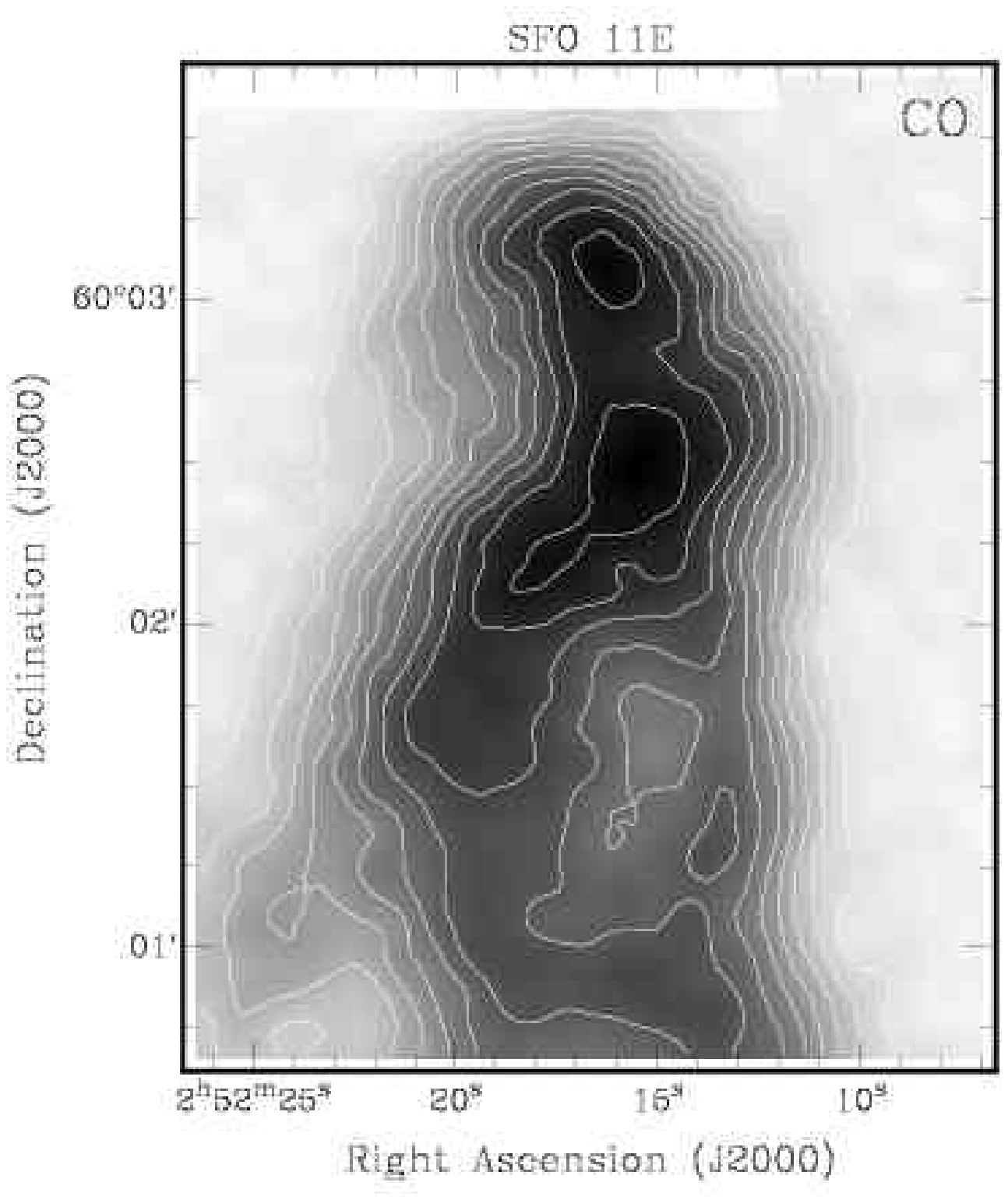}\vspace*{0.5cm}
\includegraphics*[scale=0.3,trim=0 0 0
65]{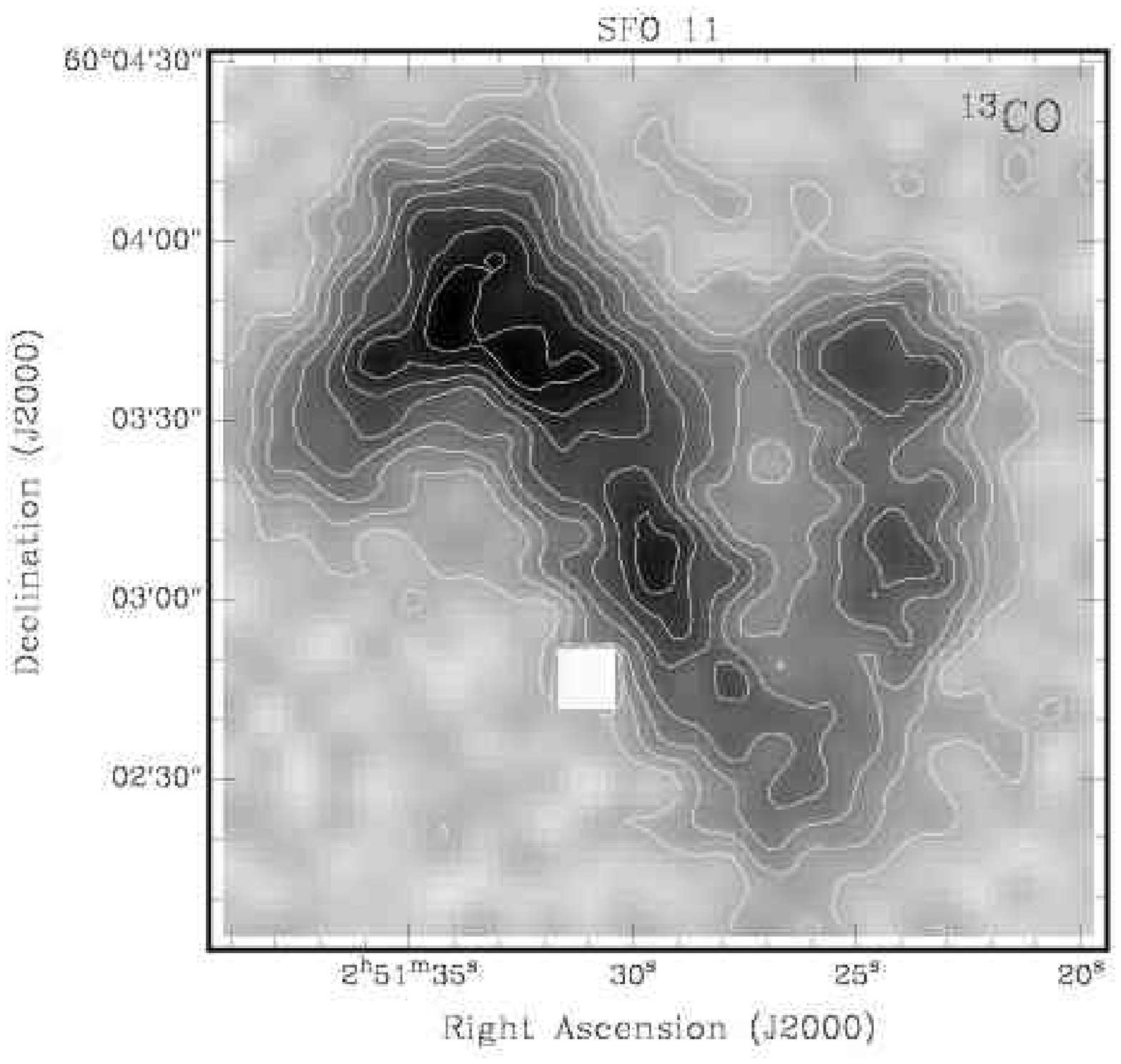}\hspace*{0.1cm}\includegraphics*[scale=0.3,trim=0 0 0
66]{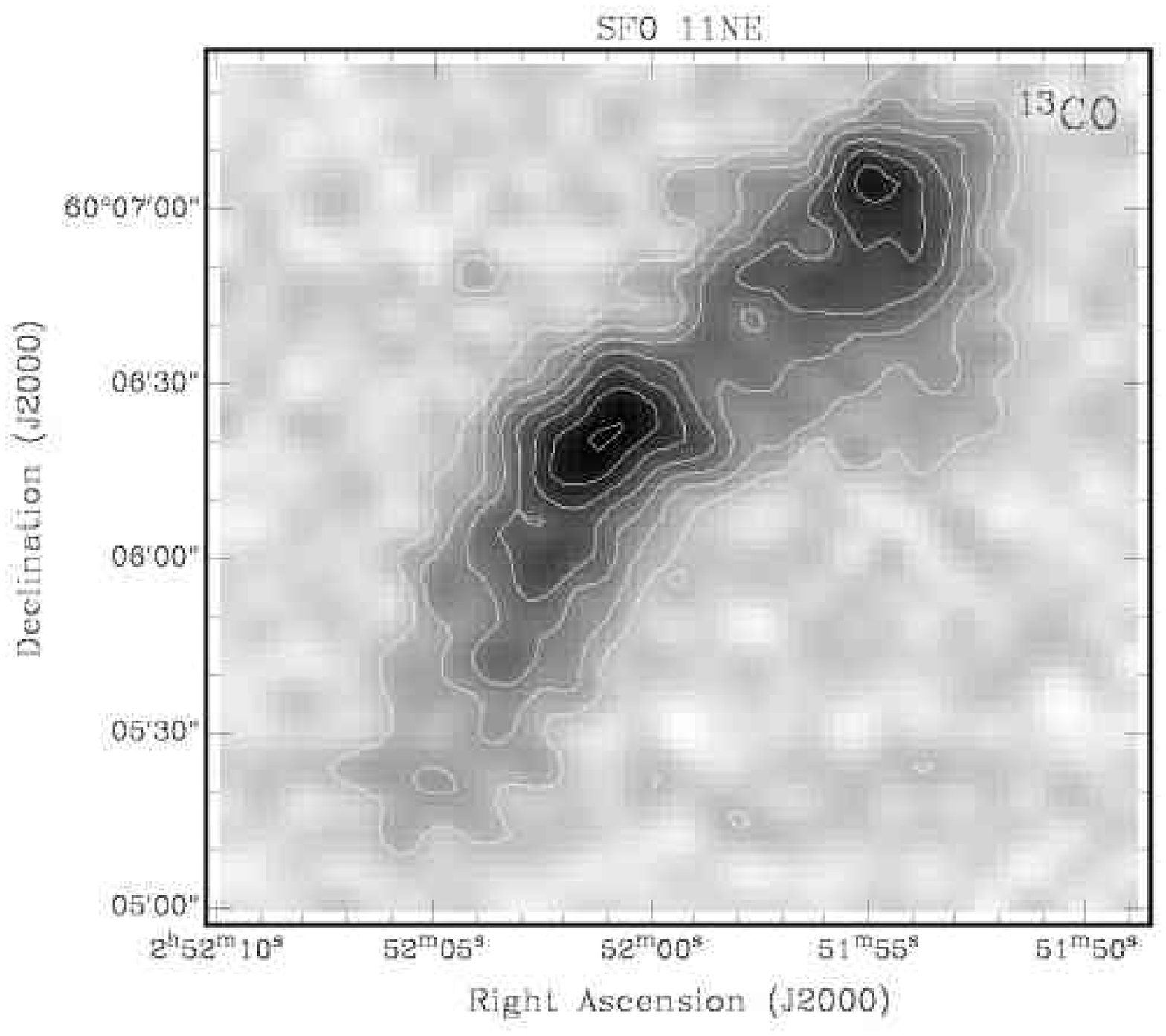}\hspace*{0.1cm}\includegraphics*[scale=0.3,trim=0 0 0
68]{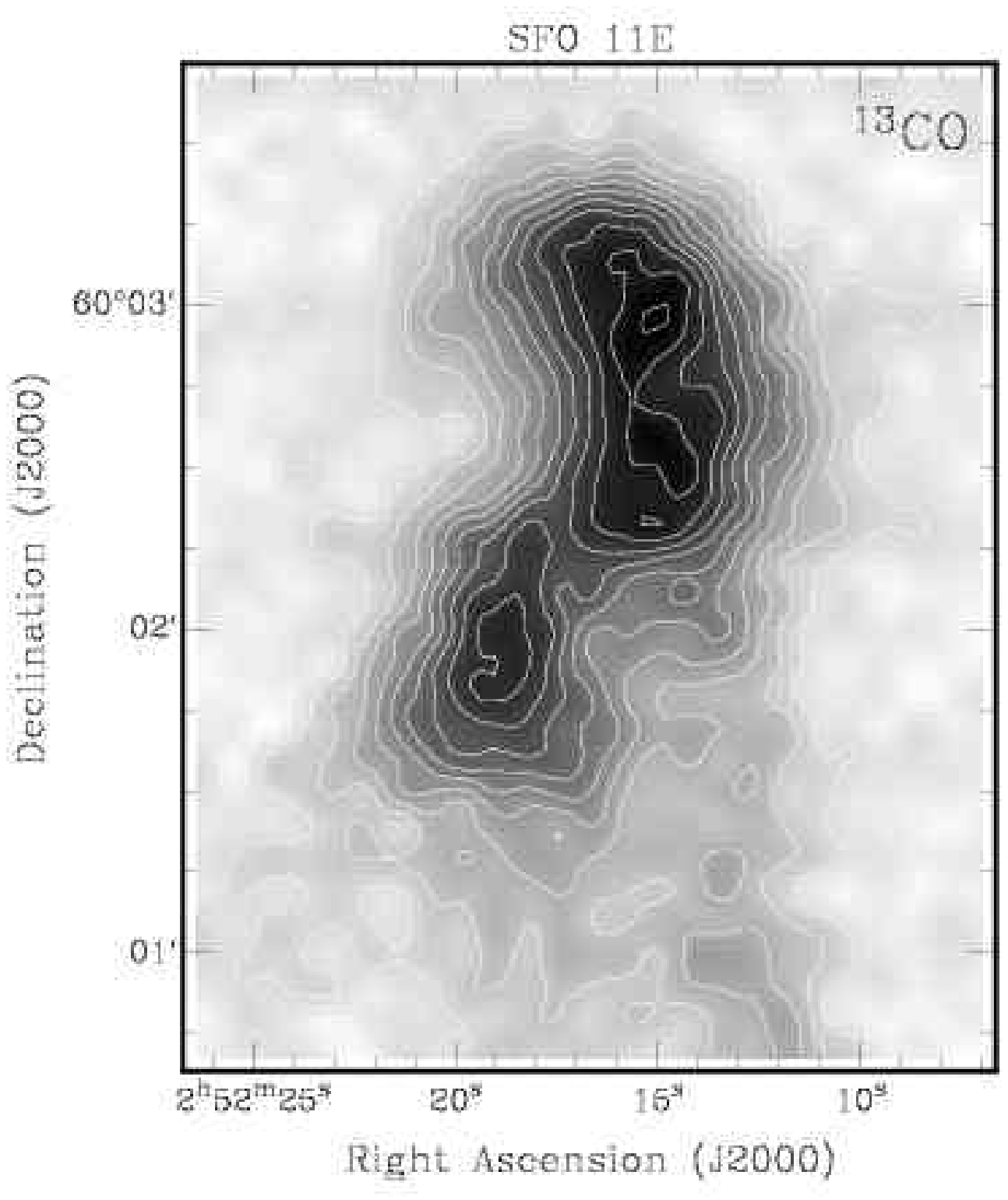}
      \caption{Velocity-integrated $^{12}$CO and $^{13}$CO maps of the three clouds SFO
      11, \object{SFO 11NE} and \object{SFO 11E} (from left to right).
      \emph{Top}: $^{12}$CO integrated intensity maps.
      \emph{Bottom}:$^{13}$CO integrated intensity maps. The maps were integrated over
      the following velocity ranges: $-42$ to $-38$ kms$^{-1}$ (SFO 11 and \object{SFO 11NE}) and
      $-38$ to $-32$ kms$^{-1}$  (SFO 11E).
     }
   \label{fig:intco}
   \end{figure*}

The pointing accuracy of the telescope was checked regularly and found to be better
than 3\arcsec. The data were calibrated to the antenna temperature scale $T_{\rm
A}^{*}$ using the standard chopper-wheel three-load technique of Kutner \& Ulich
(\cite{ku81}). Values on the $T_{\rm A}^{*}$ scale are corrected for the atmosphere,
resistive telescope losses and rearward spillowver and scattering. The data were then
corrected for forward spillover and scattering to the corrected receiver temperature 
scale $T_{\rm R}^{*}$, where $T_{\rm R}^{*} = T_{\rm A}^{*}/\eta_{\rm fss}$ and
$\eta_{\rm fss}$ is the forward spillover and scattering efficiency (0.8 for the JCMT at
230 GHz). All line temperatures quoted in this paper are on the  $T_{\rm R}^{*}$ scale
unless otherwise indicated. Absolute calibration was performed by regularly checking the
line temperatures of standard sources and comparing the observed values to standard
values. The observed line temperatures are accurate to within 10\%.

The data were reduced with the Starlink package SPECX (Prestage et al.~\cite{specx}). The distribution of
the CO emission
from each cloud was found to roughly follow that of the SCUBA dust continuum images,
allowing for the slight difference in beam sizes (21\arcsec\ for the 230 GHz CO
observations and 14\arcsec\ for the SCUBA images). The
peak line temperatures range from 15--23 K for $^{12}$CO and 9--12 K for $^{13}$CO. The
observed linewidths were extremely narrow, with typical FWHMs between 1--2 kms$^{-1}$.
The integrated
intensity maps are shown in Fig.~\ref{fig:intco} and the channel maps may be found in
Fig.~\ref{fig:channmaps}. 

\subsection{Narrowband H$\alpha$ imaging}

\begin{figure}

\includegraphics*[scale=0.4,trim=180 30 200 40]{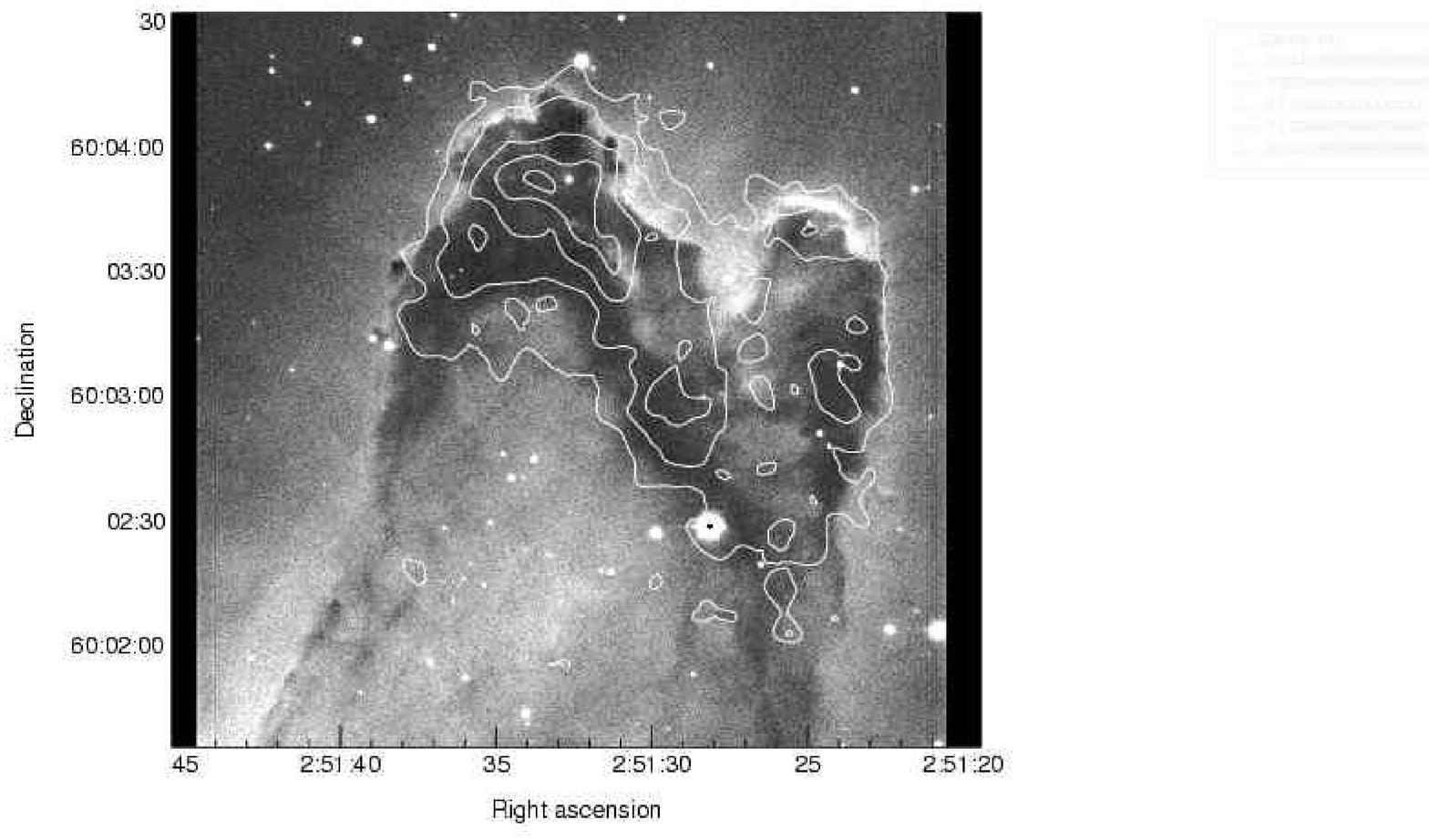}
\includegraphics*[scale=0.4,trim=180 30 200 40]{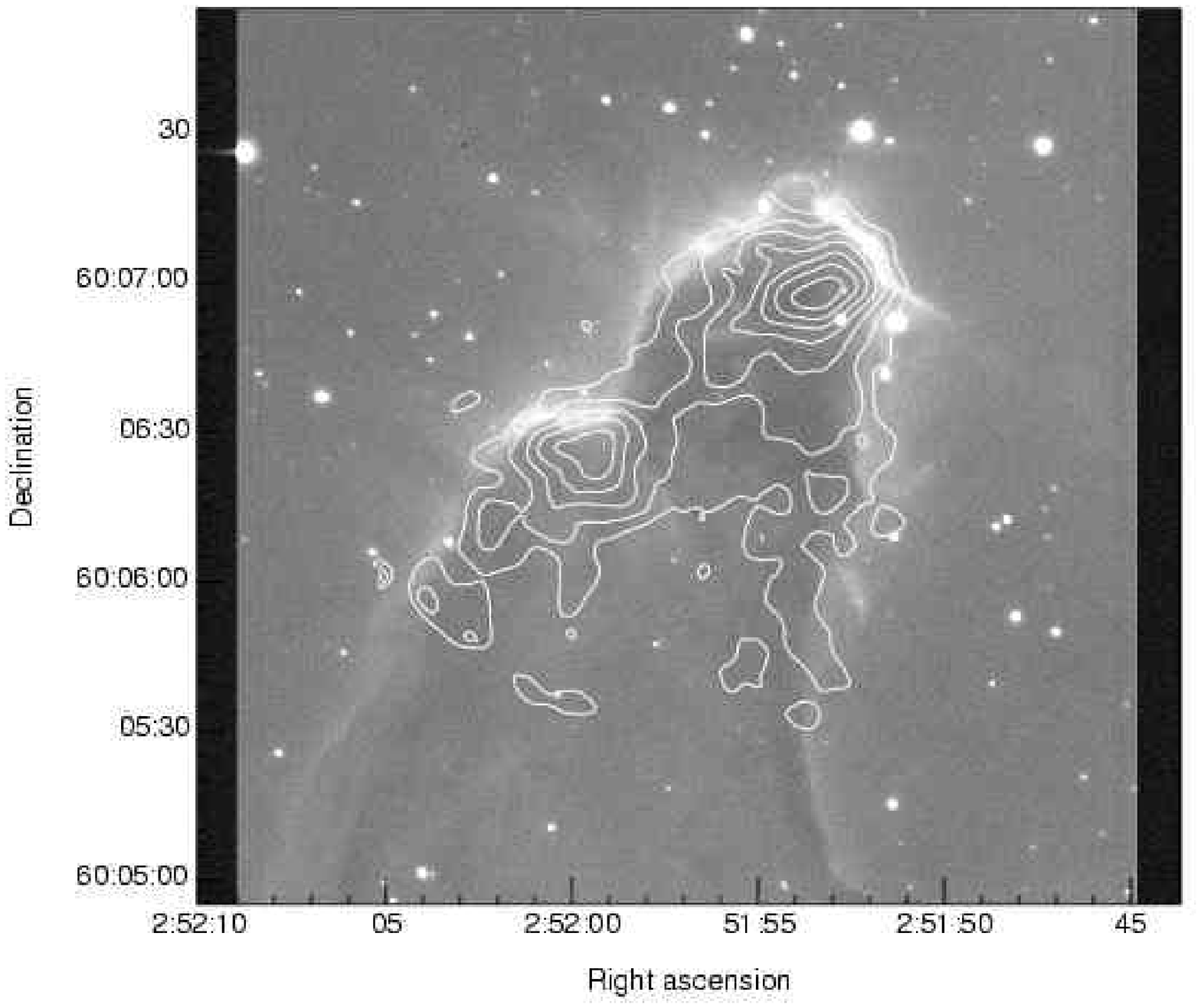}
\includegraphics*[scale=0.4,trim=180 30 200 40]{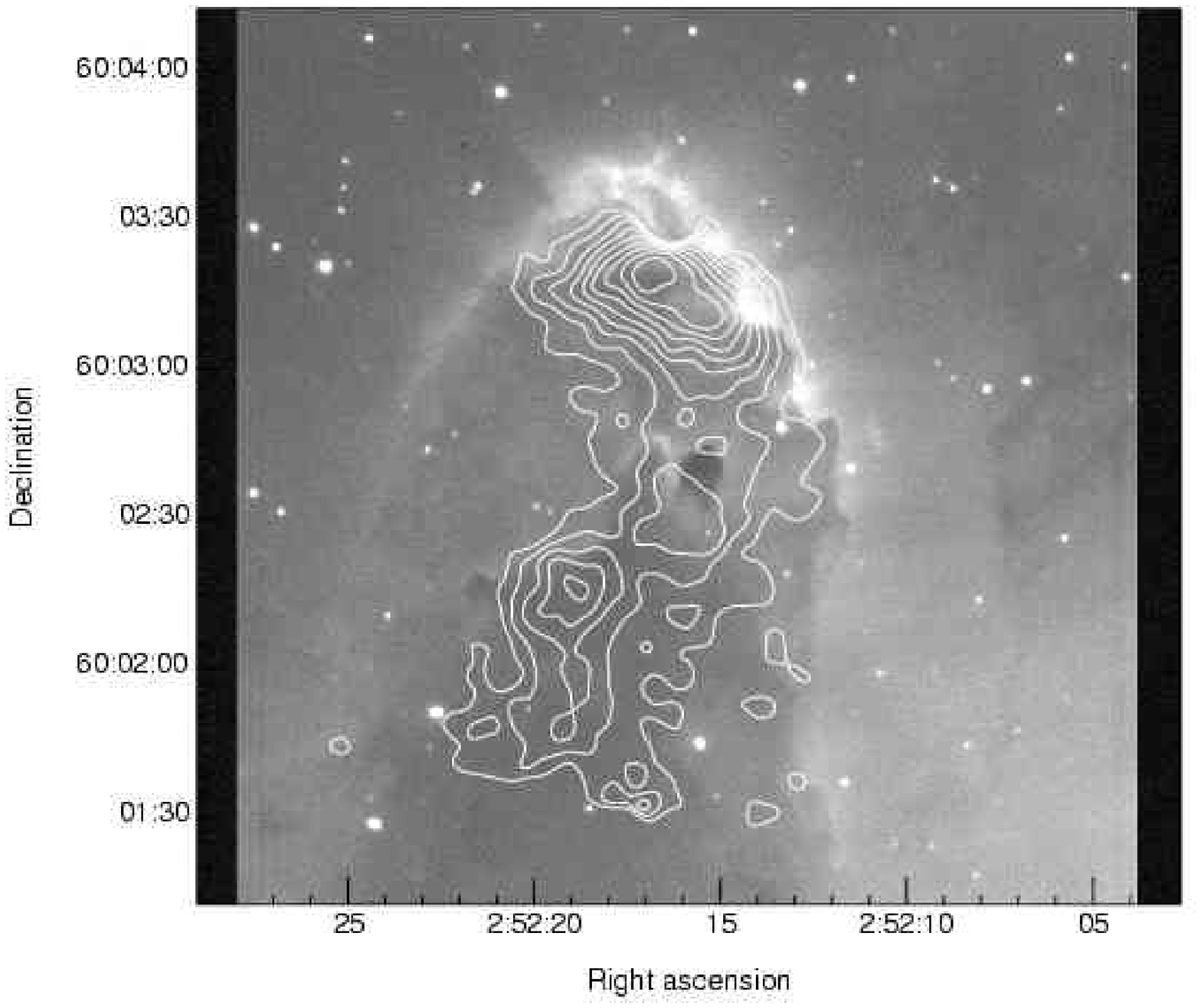}
\caption{Nordic Optical Telescope narrowband H$\alpha$ images of \object{SFO 11} (\emph{top}), SFO
11NE (\emph{middle}) and \object{SFO 11E} (\emph{bottom}) overlaid with SCUBA 850 $\mu$m contours. All images are displayed in J2000
coordinates. The greyscale range of \object{SFO 11} has been
reduced to emphasise the small dark knots seen near the bright rim of the cloud. The
greyscale ranges of the remaining two images have only been slightly compressed so that
the faint emission features next to the rims of the clouds (e.g.~the jet-like feature in the
image of \object{SFO 11NE}) are not swamped by the strong 
emission from the bright cloud rim.} 
\label{fig:halpha}
\end{figure}

We obtained narrowband H$\alpha$ images of each cloud at the 2.6 m Nordic Optical
Telescope (NOT) in order to trace the bright optical rim of the clouds at high resolution
and to search for the radial striations that are a clear signature of a photoevaporated
flow. The images were taken on January 1st 1996 using the Brocam1 camera and a Tektronix
1024$\times$1024 backside illuminated thinned CCD. The resulting images have a field of
view of 3$\times$3\arcmin\ and a pixel scale of 0\farcs18. The seeing during the
observations was typically 0\farcs75. The central wavelength and FWHM of the filter used
to isolate the narrowband H$\alpha$ emission were 656.4 nm and 3.3 nm respectively. The
total exposure time for each image was 600s.  The images were reduced, flatfielded and
calibrated in the standard manner using IRAF (Tody \cite{iraf}). The data were calibrated
against the white dwarf standard star G193-74 (Oke \cite{oke90}) observed at a similar
airmass. Astrometric calibration of the final flatfielded and processed images was carried
out for each image by measuring the positions of several known stars from the USNO
database and solving for the best fit using the astrometry routine in GAIA. The resulting
astrometry of each image is good to within a single 0\farcs18 pixel. The final processed
images with the SCUBA 850 $\mu$m emission overlaid as contours are shown in Fig.~\ref{fig:halpha}.

\subsection{Archival data}

Archival data were obtained to complement the JCMT SCUBA and CO observations. IRAS
HIRES images in all four wavebands (12, 25, 60 \& 100 $\mu$m) were obtained from the
NASA/IPAC Infrared Science Archive (\texttt{http://irsa.ipac.caltech.edu}) in
order to extend the FIR wavelength coverage of each source and enable the spectral
energy distribution (SED) to be measured. The angular resolution and signal to noise ratio of
HIRES data varies depending upon the number of deconvolution iterations used and the
position within the image. HIRES images are also subject to a number of processing
artifacts, most notable of which are negative ''bowls'' surrounding bright sources and
known as ringing. We obtained HIRES images using the default processing parameters (20
iterations) with typical angular resolutions at 60 and 100 $\mu$m of
90$\times$60\arcsec\ and 120$\times$100\arcsec, which is only sufficient to identify
the strongest cores in each SCUBA field  as point sources. The 100 $\mu$m images were found to be
 confused by a combination of ringing around a bright source found to the
north of \object{SFO 11} and strong emission from the rim of dust found to the south.
Consequently, the 
100 $\mu$m fluxes could not be accurately measured and 
the 100 $\mu$m data was discarded
from any further analysis.  At 60 $\mu$m the ringing and southern dust rim were 
less apparent and integrated fluxes could be measured. An offset of  5 Jy was applied to the
flux measurements of  \object{SFO 11NE SMM1} to take into account the flux depression caused by
the negative ringing bowl surrounding this source. Both effects were completely
negligible at 12 and 25$\mu$m.  The fluxes measured from the HIRES maps are contained in Table
\ref{tbl:hires}.

\begin{table}
\centering
\caption{Fluxes of the three cloud cores identified in the IRAS HIRES maps.}
\label{tbl:hires} 
\begin{tabular}{lccc}\hline\hline
Source ID & \multicolumn{3}{c}{IRAS HIRES flux (Jy)} \\
 & 12 $\mu$m & 25 $\mu$m & 60 $\mu$m \\\hline
SFO 11 SMM1 &  0.6$\pm$0.1 & 1.4$\pm$0.1 & 17.0$\pm$0.7 \\
SFO 11NE SMM1 & 0.3$\pm$0.1 & 1.0$\pm$0.1 & 10.0$\pm$0.4\\
SFO 11E SMM1 & 0.8$\pm$0.1 & 3.0$\pm$0.1 & 20.1$\pm$0.4 \\\hline
\end{tabular}
\end{table}

We also obtained 20 cm radio images of the three clouds from the NRAO VLA Sky Survey
(NVSS; Condon et al.~\cite{nvss}) postage stamp server at
(\texttt{http://www.cv.nrao.edu/nvss}), to derive the electron density and pressure in
the ionised gas surrounding the clouds.  The NVSS was a 20 cm sky survey complete North
of  $\delta$=$-40\degr$ carried out using the VLA in its D-configuration. The
resolution of NVSS is 45\arcsec\ and the limiting 1$\sigma$ noise of the survey is
$\sim$0.5 mJy. All three clouds are clearly detected in the NVSS data, as shown by the
contours in Fig.~\ref{fig:dssnvss}. The NVSS data are not sensitive to smooth
structures much larger than several arcminutes, which explains why the diffuse emission
from the HII region IC 1848 is not visible in Fig.~\ref{fig:dssnvss}. The peak fluxes
of the radio emission associated with the three clouds  \object{SFO 11}, \object{SFO 11NE} and \object{SFO 11E}
are 4.0, 4.9 and 11.6 mJy/beam respectively. Integrating the emission over each cloud
yields total fluxes of 7.7, 8.8 and 37.0 mJy respectively.

\begin{figure*}[ht!]

   \includegraphics*[scale=0.8,trim=150 0 200 0]{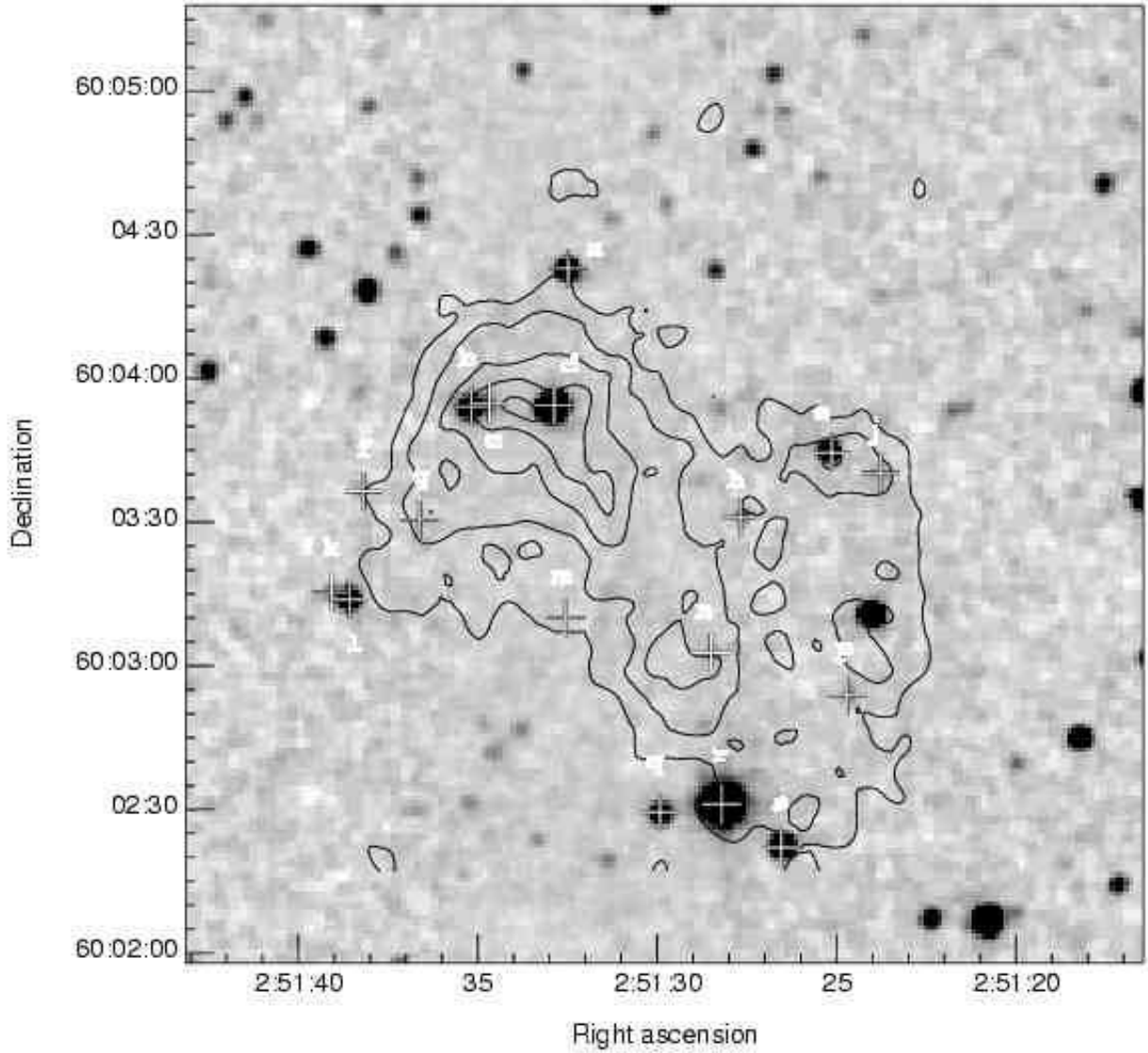}
      \caption{2MASS K-band image of the SFO 11 overlaid with SCUBA 850 $\mu$m
      contours. White crosses indicate the sources from the 2MASS Point Source Catalogue that
      are positionally associated with the clouds. Each source is identified by a letter,
      referred to in Table \ref{tbl:jhk}. 
       The SCUBA 850
      $\mu$m contours start at 45 mJy/beam and are spaced by 45 mJy/beam.}
         \label{fig:2mass1}
   \end{figure*}

\begin{figure*}[ht!]

   \includegraphics*[scale=0.8,trim=150 0 200 0]{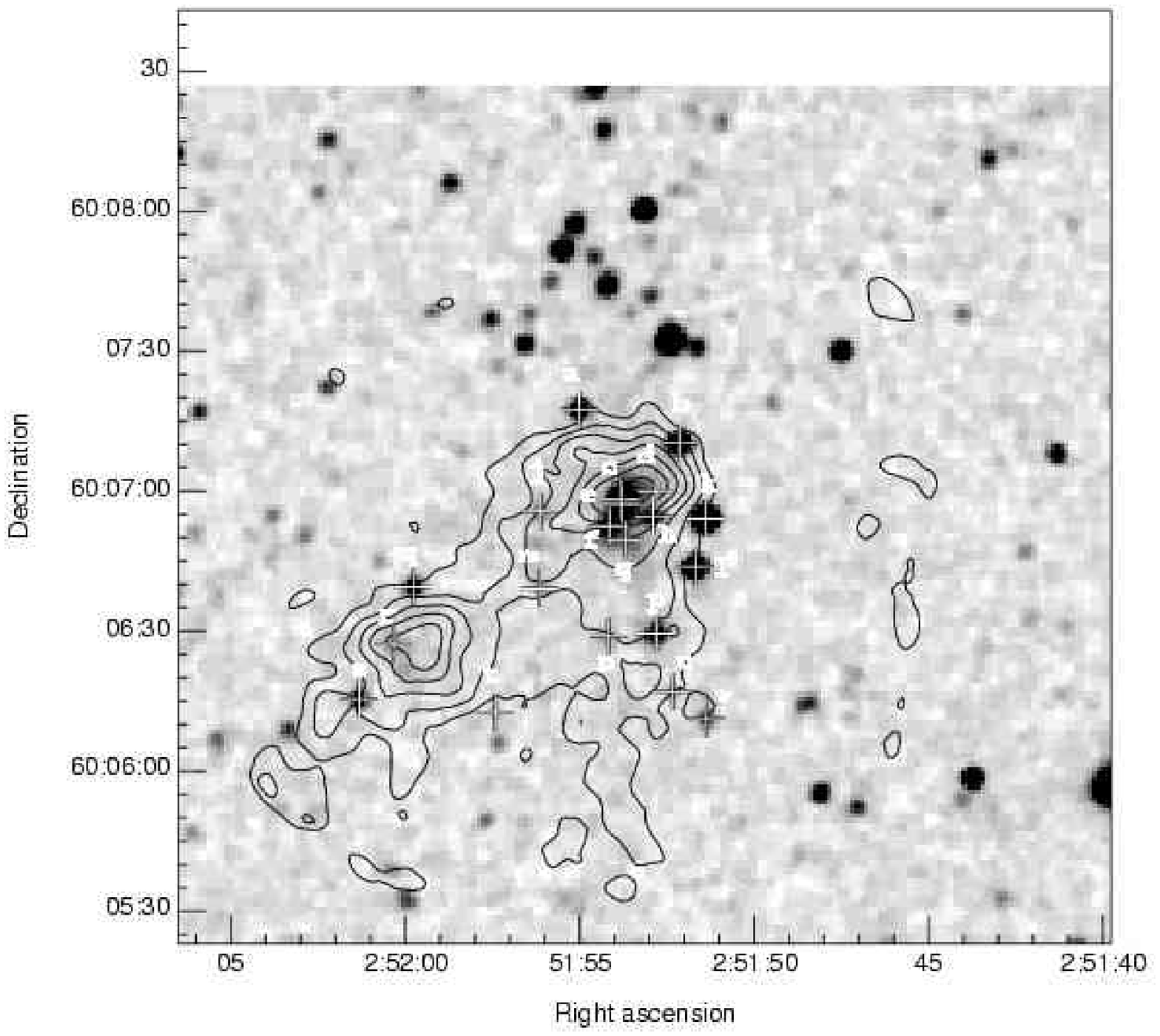}
      \caption{2MASS K-band images of SFO 11NE  overlaid with SCUBA 850 $\mu$m
      contours. White crosses indicate the sources from the 2MASS Point Source Catalogue that
      are positionally associated with the clouds. Each source is identified by a letter,
      referred to in Table \ref{tbl:jhk}.
       The SCUBA 850
      $\mu$m contours start at 45 mJy/beam and are spaced by 45 mJy/beam.}
         \label{fig:2mass2}
   \end{figure*}

\begin{figure*}[ht!]

   \includegraphics*[scale=0.8,trim=150 0 200 0]{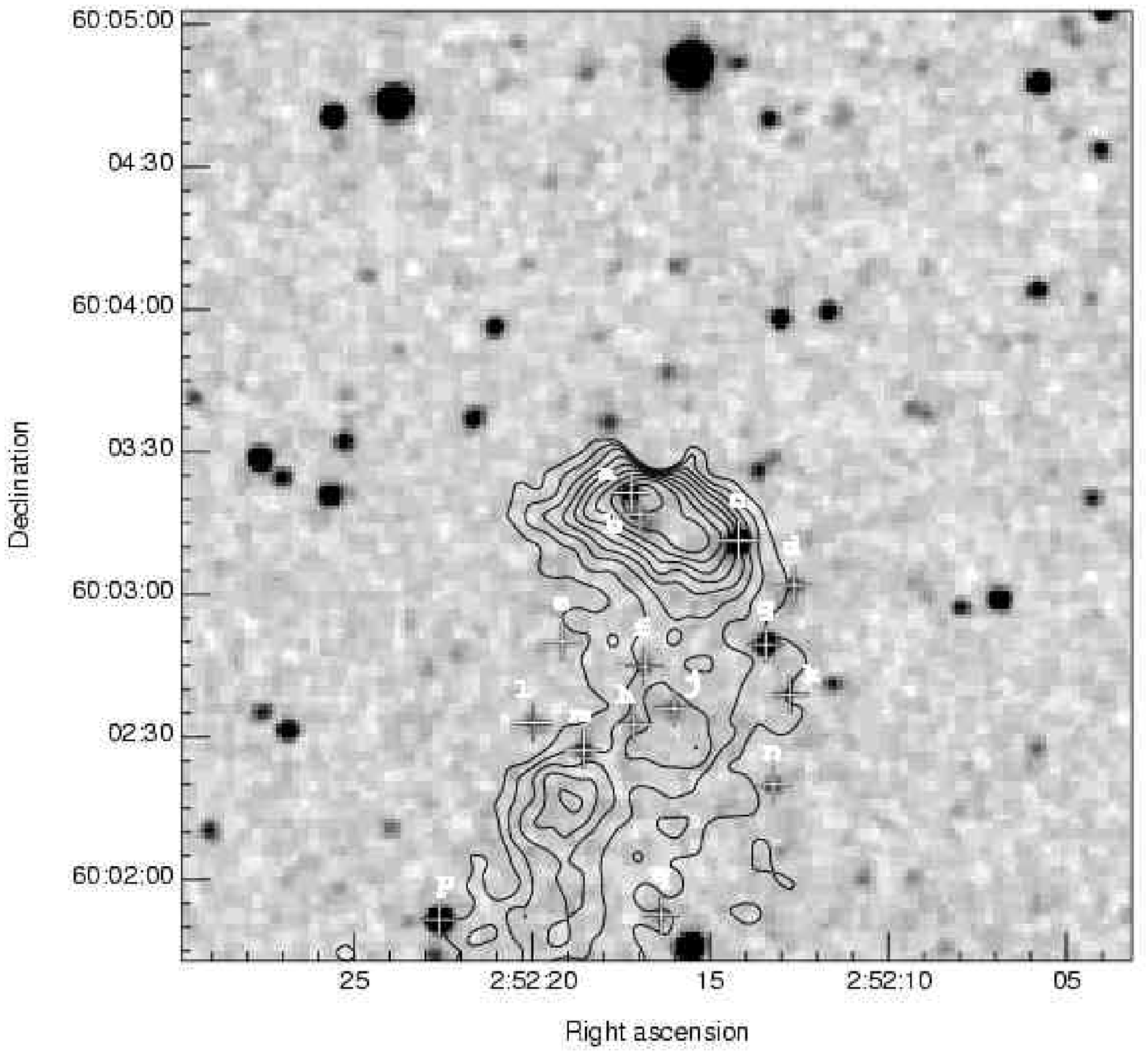}
      \caption{2MASS K-band images of SFO 11E overlaid with SCUBA 850 $\mu$m
      contours. White crosses indicate the sources from the 2MASS Point Source Catalogue that
      are positionally associated with the clouds. Each source is identified by a letter,
      referred to in Table \ref{tbl:jhk}. 
       The SCUBA 850
      $\mu$m contours start at 45 mJy/beam and are spaced by 45 mJy/beam.}
         \label{fig:2mass3}
   \end{figure*}

Near-infrared $J$, $H$ and $K_{s}$ 2MASS Quicklook images (Cutri et al.~\cite{2mass}) were obtained of
each cloud to search for protostars and  embedded young stellar objects (YSOs). The Quicklook images
and photometric  measurements of associated point sources were obtained from the  2MASS  Quicklook
image database and point source catalogue held at the NASA/IPAC Infrared Science Archive
(\texttt{http://irsa.ipac.caltech.edu}). The Quicklook images are compressed using a lossy compression
algorithm which does not conserve the low-level flux in the images. Accurate photometry from the
Quicklook images is not possible and so we obtained photometric measurements from the 2MASS Point
Source Catalogue. Each 2MASS image has a pixel scale of 1\arcsec\ and the limiting magnitudes of the
$J$, $H$ and $K_{s}$ images are 15.8, 15.1 and 14.3 respectively. The $K_{s}$ band images of the three
clouds, again overlaid by contours of SCUBA 850 $\mu$m emission,  are shown in
Figs.~\ref{fig:2mass1}--\ref{fig:2mass3}.

%  
%______________________________________________________________

\section{Analysis}
\label{sect:anal}

\subsection{Core positions, sizes and fluxes}

Individual sources were identified from visual inspection of the SCUBA maps 
as discrete objects (i.e.~bounded by unbroken contours) 
with peak fluxes greater than 3$\sigma$. The positions of
each sub-mm dust core were determined by fitting centroids to the 850 $\mu$m jiggle-maps
 as
these maps possess the highest signal to noise ratios. We designated each core as the
cloud name from Ogura, Sugitani \& Pickles (\cite{osp02}) followed by an SMM number to
indicate sub-mm detections (e.g.~ \object{SFO 11E SMM1}).  The positions of each core are listed
in Table \ref{tbl:fluxes}. In total 8 sub-mm cores were identified, with the brightest
cores generally located at the head of each bright-rimmed cloud.  The ``double-lobed''
core  \object{SFO 11E SMM1} is marginally resolved at 850 $\mu$m but it was not possible to fit a
centroid to each peak and so this source is treated hereafter as a single core. We
inspected the unsmoothed 450 $\mu$m map in order to try to separate these two
cores, however the signal to noise ratio was found to be too low to adequately
distinguish the cores. Higher
quality 450 $\mu$m images or mm-wave interferometry of  \object{SFO 11E SMM1} are needed  to
separate this core into its two components.

The size of each source was estimated by the FWHM of a Gaussian fitted to the azimuthal 850 $\mu$m flux
average (again, because of the higher signal to noise ratio). We took into account the beam size via
assuming a simple convolution of a Gaussian source with a Gaussian beam of 14\arcsec\  ($\Theta^{2}_{\rm
obs} = \Theta^{2}_{\rm beam} + \Theta^{2}_{\rm source}$). It was not possible to fit azimuthal averages to
the two cores  \object{SFO 11 SMM1} and  \object{SFO 11 SMM2} because of their  irregular shape. Therefore,
estimates for their sizes listed in Table 4 represent the average diameter as defined by their 50\% flux
contours. The angular diameter of each core was converted into an effective
physical diameter $D_{\rm eff}$ by assuming each core lies at the same distance as IC
1848, i.e. 1.9 kpc. The effective physical diameters for each core are given in Table
\ref{tbl:fluxes} and are found to be in the range 0.1--0.3 pc, which is typical for
cores found in star-forming regions (e.g.~Evans \cite{evans99}).

There is good agreement between the morphology of the SCUBA maps and the  $^{12}$CO
and $^{13}$CO maps. The difference in core positions measured from the SCUBA maps and CO
maps are well within the typical pointing errors ($\le$3\arcsec). The SCUBA maps
are of higher angular resolution than the CO maps (14\arcsec\ vs 21\arcsec) and the
typical pointing residuals were lower during the SCUBA observations. Quoted core
positions are those measured from the SCUBA jiggle-maps. The core sizes
determined via fits to the azimuthal averages of the $^{13}$CO emission are consistent
with those determined from the 850 $\mu$m jiggle-maps, whereas those determined from the
$^{12}$CO maps are in general $\sim$10\% larger than those measured from the 850 $\mu$m
images. It is likely that the $^{12}$CO, by virtue of its large optical depth, traces
the temperature of the opaque material lying close to the surface of the 
cloud rather than the column density.

We measured the peak and integrated continuum flux values for each core using the
Starlink package GAIA (Draper, Gray \& Berry \cite{gaia}).  Apertures were carefully
chosen by hand to avoid confusion from nearby sources. Background levels were estimated from
emission-free regions of each map and subtracted from the measured flux values.  The
jiggle-maps of \object{SFO 11E} were found to possess a significantly negative
background, which was most likely caused by  chopping onto extended cloud emission  from
the dust ridge found to the West of SFO 11E. Peak and integrated fluxes for the three
cores in this map were determined by adding a constant background level of 2.5 Jy/beam to
the 450 $\mu$m data and 46 mJy/beam to the 850 $\mu$m data. These background levels were
determined by fitting baselines to emission-free regions of the jiggle-maps. 

We estimate that the systematic errors in measuring the fluxes of all 8 cloud cores 
are no more than 30\% in the case of the 450 $\mu$m measurements and 10\% for the 850
$\mu$m measurements (including errors in the absolute flux calibration). The systematic
errors for the cores detected in the map of SFO 11E may be larger than these values, 
given the addition of a constant background level to the map, however we estimate that
the increased systematic error is no larger than 50\%. The cores  \object{SFO 11E SMM1}
and SMM2 could not be separated at 450 $\mu$m due to the lower signal-to-noise ratio in
this image. The integrated 450 $\mu$m flux for this core was determined by integrating
over an aperture of the same size and position as that used for the  850 $\mu$m
measurement. The peak and integrated fluxes are listed in Table \ref{tbl:fluxes}. 

\renewcommand{\thefootnote}{\alph{footnote}}
\begin{center}
\begin{table*}
\caption{Positions, peak and integrated sub-mm fluxes for the detected cores. The
effective diameters of the sub-mm cores are also given as $D_{\rm eff}$. Note that the
peak fluxes for both 450 and 850 $\mu$m observations are measured in Jy per 14\arcsec\
beam.}
\label{tbl:fluxes}
\begin{minipage}{\linewidth}
\begin{tabular}{lccccccl}\hline
 & & & \multicolumn{2}{c}{Peak Flux (Jy/beam)} & \multicolumn{2}{c}{Integrated Flux (Jy)} & $D_{\rm eff}$  \\
Clump ID & $\alpha_{\rm 2000}$ & $\delta_{\rm 2000}$ & 450 $\mu$m & 850 $\mu$m & 450 $\mu$m & 850 $\mu$m & (pc) \\ \hline \hline
 \object{SFO 11 SMM1} & 02 51 33.7 & +60 03 54 & 1.2$\pm$0.2 & 0.23$\pm$0.01 & 8.1$\pm$1.4 & 1.42$\pm$0.15 & 0.28\setcounter{footnote}{0}\footnotemark \\
 \object{SFO 11 SMM2} & 02 51 25.0 & +60 03 42 & 0.7$\pm$0.2 & 0.15$\pm$0.01 & 5.2$\pm$0.9 &  0.53$\pm$0.06 & 0.20\setcounter{footnote}{0}\footnotemark \\
 \object{SFO 11 SMM3} & 02 51 29.6 & +60 03 01 & 0.7$\pm$0.2& 0.13$\pm$0.01 & 2.5$\pm$0.6 & 0.27$\pm$0.05 & 0.20\footnotemark\\
& & & & & & &  \\
 \object{SFO 11NE SMM1} & 02 51 53.6 & +60 07 00 & 1.6$\pm$0.1& 0.34$\pm$0.02 & 4.0$\pm$0.5 & 0.92$\pm$0.10  & 0.24\setcounter{footnote}{1}\footnotemark \\
 \object{SFO 11NE SMM2} & 02 51 59.5 & +60 06 27 & 1.0$\pm$0.1 & 0.25$\pm$0.02 & 2.7$\pm$0.3 & 0.54$\pm$0.08 & 0.16\setcounter{footnote}{1}\footnotemark \\
& & & & & & &  \\
 \object{SFO 11E SMM1} & 02 52 16.8 & +60 03 19 & 5.8$\pm$0.3 & 0.48$\pm$0.02 & 15.5$\pm$1.4 & 1.29$\pm$0.10 & 0.27\setcounter{footnote}{1}\footnotemark \\
 \object{SFO 11E SMM2} & 02 52 15.9 & +60 02 29 & -- & 0.21$\pm$0.02 & 3.9$\pm$0.7\footnotemark & 0.34$\pm$0.25 & 0.12\setcounter{footnote}{1}\footnotemark \\
 \object{ \object{SFO 11E SMM3}} & 02 52 19.0 & +60 02 17 & 2.7$\pm$0.3 & 0.26$\pm$0.02 & 3.2$\pm$0.5 & 0.40$\pm$0.23 & 0.11\setcounter{footnote}{1}\footnotemark \\\hline
\end{tabular}
\footnotetext[1]{$D_{\rm eff}$ estimated from FWHM flux contours}
\footnotetext[2]{$D_{\rm eff}$ measured from Gaussian fit to azimuthal flux average}
\footnotetext[3]{450 $\mu$m flux measured by integrating over the same solid angle as for 850 $\mu$m}
\end{minipage}
\end{table*}
\end{center}
\renewcommand{\footnoterule}{\oldfootnoterule}

\subsection{Core velocities}

\begin{figure*}[ph!]
\centering
\hspace*{-10cm}{\large {\bf\textsf{SFO 11}}}

\vspace*{-0.65cm}\includegraphics*[scale=0.5,angle=-90,trim=35 0 0 0]{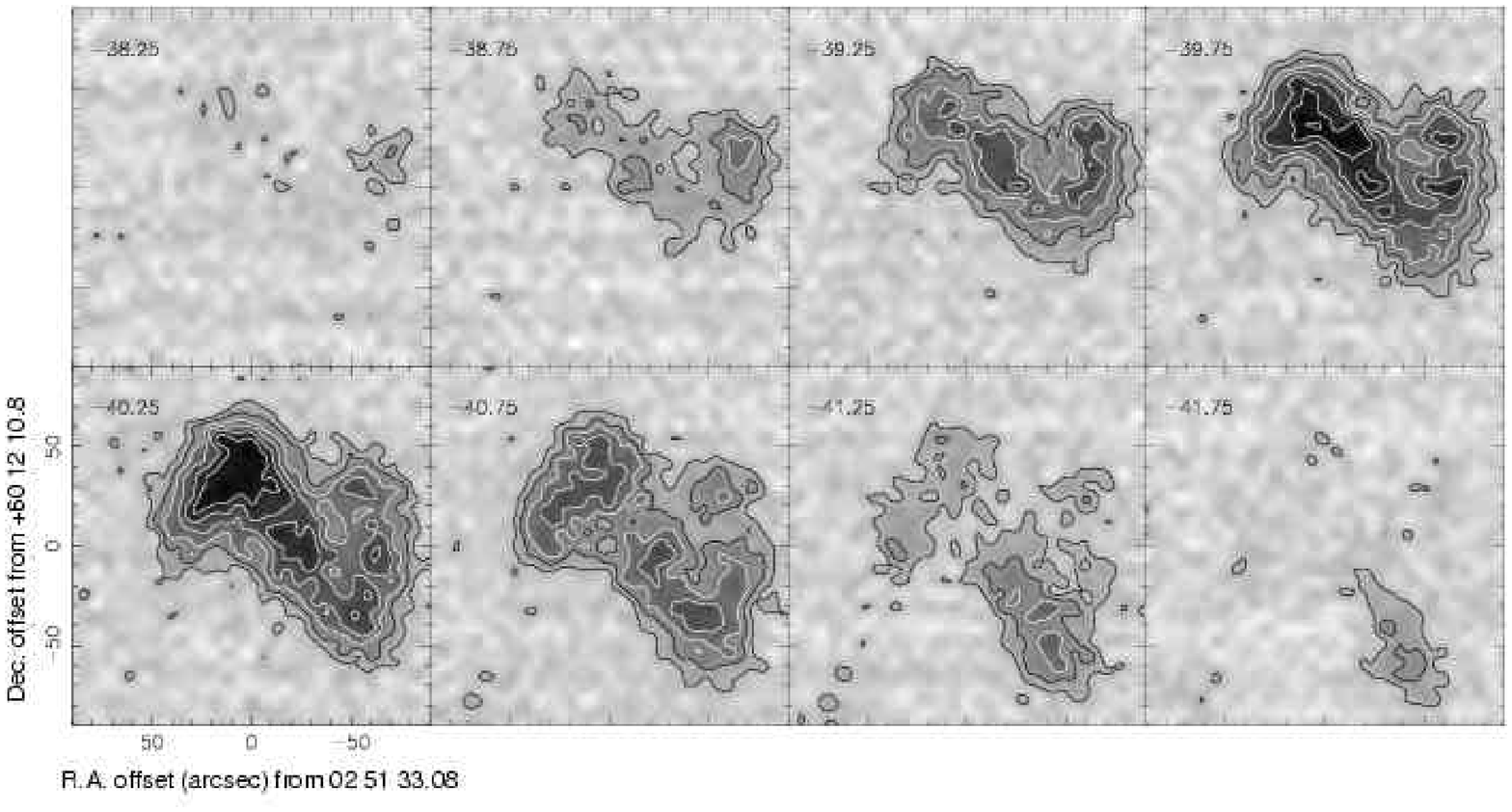}\vspace*{0.3cm}

\vspace*{0.2cm}\hspace*{-9.5cm}{\large {\bf\textsf{SFO 11NE}}}

\vspace*{-0.6cm}\includegraphics*[scale=0.5,angle=-90,trim=35 0 0 0]{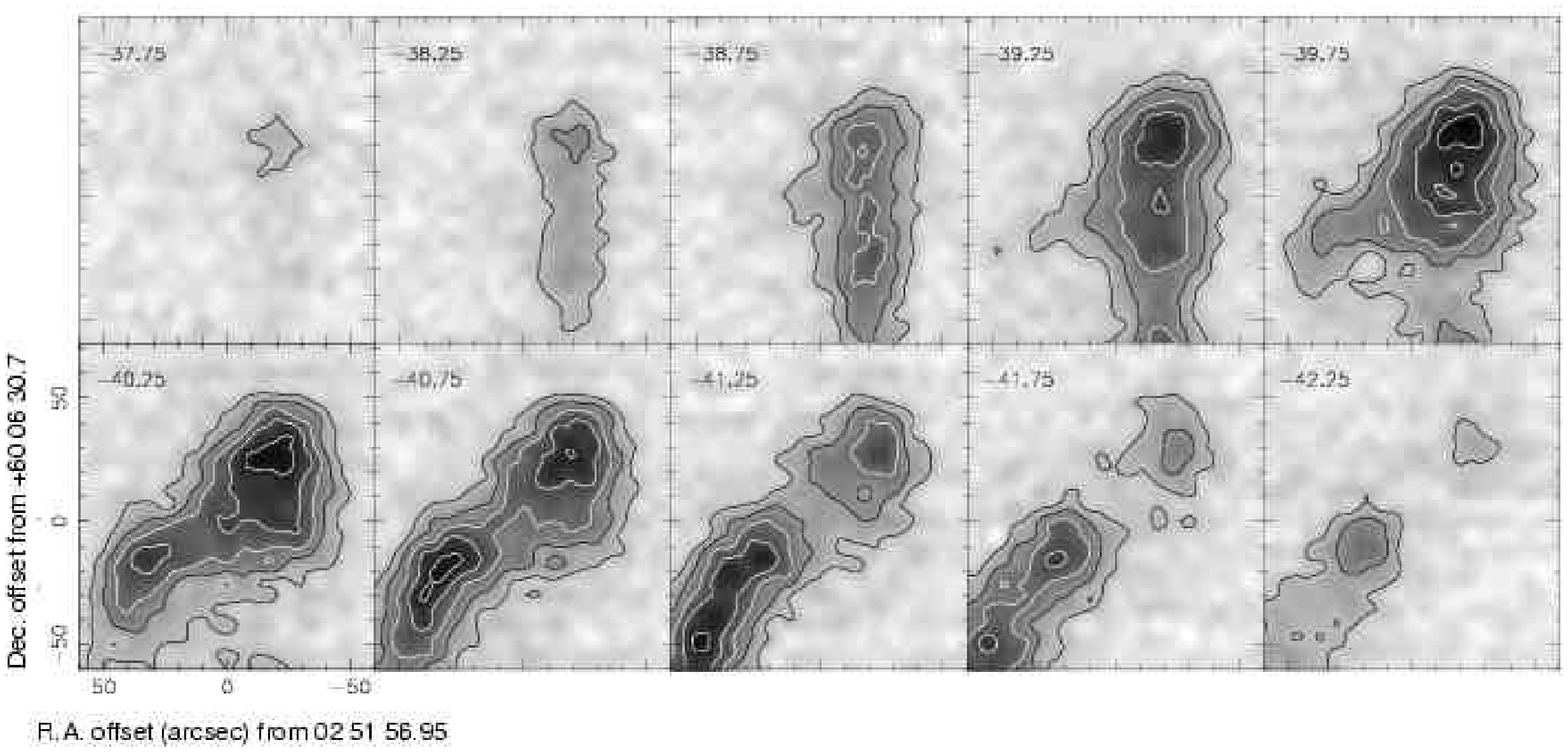}\vspace*{0.3cm}

\vspace*{0.2cm}\hspace*{-9.8cm}{\large {\bf\textsf{SFO 11E}}}

\vspace*{-0.6cm}\includegraphics*[scale=0.5,angle=-90,trim=35 0 0 0]{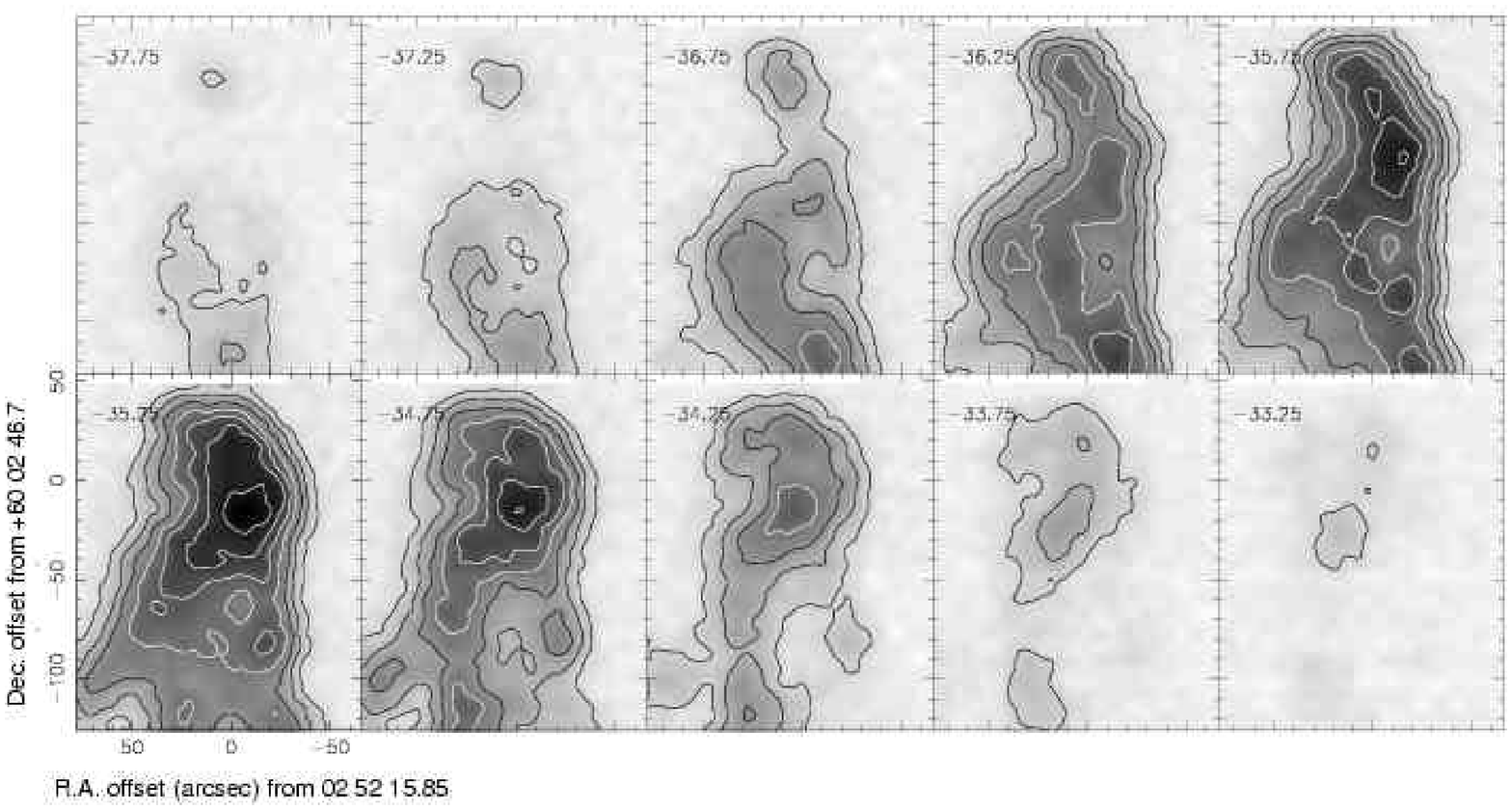}
\caption{Channel maps of the $^{12}$CO J=2--1 emission observed towards SFO11,
SFO 11NE and \object{SFO 11E}. The emission is integrated over 0.5 km s$^{-1}$
channels.}
\label{fig:channmaps}
\end{figure*}

The velocity distribution of the molecular gas in the clouds was examined by
constructing channel maps of the $^{12}$CO emission (Fig.~\ref{fig:channmaps}). The
emission was integrated over velocity bins of width 0.5 km$\,$s$^{-1}$ over  the entire
range of velocities observed in the CO spectra.   Channel-maps of the $^{13}$CO
emission were also inspected, however the reduced signal-to-noise ratio of the
$^{13}$CO maps meant that these data were  of little use for inspecting the velocity
distribution of the molecular gas.  The channel maps show that the bulk of the gas in
the clouds is at roughly the same velocity. The molecular gas of the clouds is clearly
revealed as possessing a clumpy structure in which the SCUBA sub-mm cores can be
identified with discrete features in the $^{12}$CO channel maps. It is thus likely that
the SCUBA sub-mm cores are true dense cores of dust and gas embedded within the
molecular clouds.

Whilst the bulk of the gas is roughly at the same velocity there are two
velocity-shifted features (with respect to the central cloud velocity) seen in the
channel maps of \object{SFO 11} and \object{SFO 11NE}. The most southerly clump of gas centred roughly at
the velocity of $-$40.75 km\,s$^{-1}$ in the channel map of \object{SFO 11} is displaced from the
molecular gas at the head of the cloud by $\sim$ 1 km\,s$^{-1}$ towards the blue end of
the spectrum. The North-South bar of CO emission centred at approximately -38.75 km\,s$^{-1}$ in the
channel map of \object{SFO 11NE} is similarly displaced towards the red end of the spectrum with
respect to the remainder of the cloud. It is likely that these velocity displacements
arise through the momentum transferred to the body of the clouds by the photoevaporated
surface layers. This phenomenon is well documented in  photoionisation models (Oort \&
Spitzer \cite{os55}; Bertoldi
\cite{bertoldi}; Lefloch \& Lazareff \cite{ll94}) and velocity displacements between
the head and tails of cometary clouds have been observed in several objects
(e.g.~Codella et al.~\cite{cbnst01}; White et al.~\cite{rosette}).  

The North-South bar seen in the channel map of \object{SFO 11NE} is also positionally coincident
with faint  H$\alpha$ emission at the western edge of the cloud and  weak 850 $\mu$m
emission stretching to the south of  \object{SFO 11NE SMM1} (see Fig.~\ref{fig:scumaps}). The
positional coincidence of these features suggests that as the gas at the western edge
of the cloud is pushed inwards by the photoevaporation of the surface layer its density
may be enhanced. Whilst this is in line with the predictions that the photoionisation
shock may form dense clumps or cores via implosion, this potential density enhancement
is not observed in either \object{SFO 11} or SFO11E. The rims of these clouds do not show any
evidence for enhanced CO or dust emission. In each case the sub-mm continuum and CO
emission decreases towards the rim with a sharp boundary at the bright optical rim of
the cloud. 

 Unlike other
CO observations of bright-rimmed clouds (e.g. White at al.~1997) there is no
evidence in the channel maps for bright molecular rims that are displaced in
velocity from the interior gas. The reasons behind this are probably the 
limited angular resolution of these
observations (21\arcsec) compared to the other observations and also the fact that
the CO J=2--1 lines probe lower temperature regions more likely to be located
deeper within the rims of the clouds that the higher J transitions observed by
White et al.~(\cite{rosette}).

\subsection{Evidence for molecular outflows?}

The channel maps and raw spectra were inspected for evidence of non-Gaussian line wings
that might indicate the presence of molecular outflows. Both  \object{SFO 11NE SMM1} and \object{SFO 11E}
SMM1 show some evidence for molecular outflows in the form of moderate velocity line
wings up to 4--5 kms$^{-1}$ from the line centre. The remaining cores in the sample do
not show any evidence for non-Gaussian line wings. Integrated intensity maps of both
the red and blue line wings were formed, but the angular resolution in the maps
(21\arcsec) is insufficient to separate any outflow lobes. 

It is possible that the wing emission results from photoevaporated molecular gas streaming
from the cloud surface. Prominent blue-shifted wings are predicted by radiative-driven
implosion models for the collapse phase of the cloud (Lefloch \& Lazareff \cite{ll94}).
The striations seen in the H$\alpha$ images show that there is a flow of photoionised 
(and presumably photoevaporated) gas from the surface of all three clouds.  If the line
wings arise from a photoevaporated gas flow then we might expect to see broad  line wings
toward all three clouds rather than just \object{SFO 11NE} and \object{SFO 11E}. 

\begin{figure}
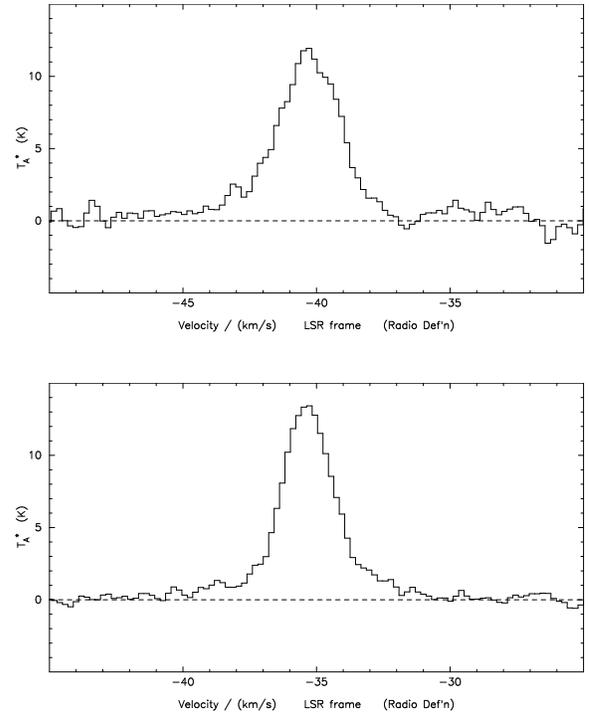

\centering
\vspace*{0.3cm}\includegraphics*[scale=0.3,angle=-90,trim=8 0 75 0]{h4521f9a.eps}\vspace*{0.5cm}

\includegraphics*[scale=0.3,angle=-90,trim=8 0 75 0]{h4521f9b.eps}
\caption{$^{12}$CO spectra towards positions in  \object{SFO 11NE SMM1} (\emph{top}) and
 \object{SFO 11E SMM1} (\emph{bottom}) illustrating the non-Gaussian line wings found in
these cores. These line wings are most likely to arise from molecular outflows,
however the angular resolution of the $^{12}$CO J=2--1 observations is not
sufficient to resolve the red and blue outflow lobes.}
\label{fig:oflowspec}
\end{figure}

\begin{figure}
\centering
\includegraphics*[scale=0.35,trim=110 0 150 0]{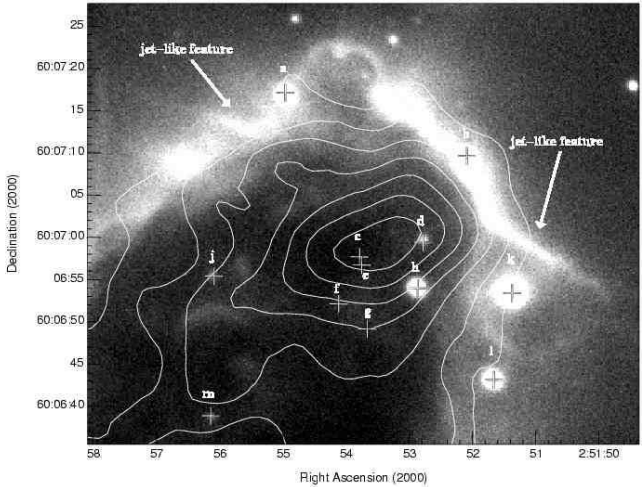}
\caption{A closeup of the H$\alpha$ image of the cloud \object{SFO 11NE}, showing the two jet-like
features located on either side of the cloud. The contours are of the SCUBA 850 $\mu$m
emission and the crosses indicate the positions of 2.2 $\mu$m 2MASS objects.}
\label{fig:halpha_zoom}
\end{figure}

The $H\alpha$ images offer support for the outflow hypothesis in the case of \object{SFO
11NE} SMM1. There are two jet-like features seen towards the head of \object{SFO 11NE} (see
Fig.~\ref{fig:halpha_zoom} for a closeup). The jet-like features are not normal to the
cloud surface or aligned with the direction of the UV illumination and so it is unlikely
that they result from photoevaporated flow from the cloud surface. The two features are
roughly aligned with each other which suggests that they may originate from a common perhaps
protostellar source. A similar feature was observed in the H$\alpha$ emission from the
bright-rimmed cloud TC2 in the Trifid nebula and attributed to the well-known photoionised
jet HH 399 (Lefloch et al.~\cite{lcrmch02}). It is  likely that we have observed a similar
phenomenon associated with \object{SFO 11NE} and that the jet-like features indicate the
location of an embedded protostar or YSO. The alignment of the two jet-like features does
not, however, correlate with the positions of either the sub-mm peak or objects
seen in the near-infrared 2.2 $\mu$m 2MASS image of \object{SFO 11NE}. This may indicate the
presence of an additional unseen  protostellar source within  \object{SFO 11NE SMM1},
located at the intersection of the two jet-like features.

In summary  we conclude that the line wings are more likely to result from protostellar
outflow than a photoevaporated molecular flow.  However, further investigation is required
to more conclusively support this hypothesis. Higher resolution, possibly interferometric, CO
observations are needed to resolve the individual outflow lobes and optical spectroscopy
of the two jet-like features associated with \object{SFO 11NE} is required to confirm that these
two features  are indeed  photoionised  jets.

\subsection{Temperature, mass and density}

The temperature, density and mass of each core
has been derived independently from both the SCUBA dust continuum images 
and the CO maps. In this section we present the results from the standard
analysis techniques used to derive these quantities. 

\subsubsection{Dust continuum}
\label{sect:scuba_anal}

In order to derive the physical properties of each core from their dust continuum emission
we fitted a single-temperature greybody model to their Spectral Energy Distributions (SEDs). For the
SMM1 cores we were able to construct SEDs from a combination of the  SCUBA 450 \& 850
$\mu$m and IRAS HIRES 60 $\mu$m integrated fluxes. We discarded the IRAS 12, 25 and 100
$\mu$m fluxes from the fitting procedure as the former two wavelengths are  not typically
well fitted by a single-temperature greybody and the latter wavelength was  hopelessly
confused due to the limited angular resolution. For the SMM2 and SMM3 cores we were
restricted to SEDs comprised of only the SCUBA 450 and 850 $\mu$m integrated fluxes.

In order to determine the temperature of the dust we initially fitted the SEDs of the
SMM1 cores with a model following a greybody function of the form: 

\begin{equation}
F_{\nu} = \Omega \, B_{\nu}(T_{\rm d}) (1 - e^{-\tau_{\nu}}),
\label{eqn:greybody}
\end{equation}

where $F_{\nu}$ is the flux emitted at frequency $\nu$, $\Omega$ is the solid angle
subtended by the core, $B_{\nu}(T_{\rm d})$ is the Planck function evaluated at frequency
$\nu$ and dust temperature $T_{\rm d}$, and $\tau_{\nu}$ is the optical depth at
frequency $\nu$ (e.g.~Dent, Matthews \& Ward-Thompson \cite{dmwt98}). A common practice
is to parameterise the optical depth $\tau_{\nu}$ in terms of the dust emissivity $\beta$
so that $\tau$ may be evaluated at arbitrary frequencies from a known reference frequency
$\nu_{\rm ref}$ and optical depth $\tau_{\rm ref}$, i.e.  $\tau_{\nu} = \tau_{\rm ref}
\left(\nu/\nu_{\rm ref}\right)^{\beta}$. The typical values for the dust emissivity
$\beta$ observed in molecular clouds range from 1--2 (e.g.~Hildebrand \cite{hildebrand83}).

The greybody analysis is an extreme simplification of the dust properties, assuming a
single effective temperature for the dust, a simple power-law extrapolation for the
frequency-dependent optical depth and a constant dust emissivity over a large wavelength
range. The greybody approach nevertheless has been used to  predict dust properties that
are often reasonably consistent with those derived from more complex radiative transfer
models (Dent, Matthews \& Ward-Thompson \cite{dmwt98}).   As the cores are only
moderately spatially resolved by our SCUBA observations and we possess only limited data
on the SEDs at 60, 450 and 850 $\mu$m it was felt that a more complex analysis was
unwarranted.

  As the SEDs of the SMM1 cores are only defined by three points we assumed
a fixed value of $\beta = 2$ to reduce the number of fitted parameters. 
For the remaining cores in the sample the dust temperature was determined by the
450 $\mu$m/850 $\mu$m flux ratio, again assuming $\beta = 2$.  The dust temperatures
evaluated by these methods are shown in Table \ref{tbl:dustprops}. The accuracy in the
dust temperature is roughly 1--2 K for the SMM1 cores where we could fit the greybody
model to three flux points. The uncertainties in the remainder of the sample are
considerably more inaccurate: two sub-mm flux points do not constrain high temperature
values particularly well due to a fifth power term (assuming $\beta=2$) 
in the flux ratio. The lower temperature bounds are better constrained and overall we
estimate that the temperature estimates for the SMM2 and SMM3 cores are good to within a
factor of 2.

\begin{table*}
\centering
\caption{Core properties determined from greybody fits to the SCUBA and IRAS
HIRES sub-mm and far-infrared fluxes.}
\label{tbl:dustprops}
\begin{tabular}{lccccc}\hline
Source ID & $T_{\rm d}$ & $M$ & $\log(n)$ & $A_{V}$& $A_{K}$\\
 & (K) & $M_\odot$ & cm$^{-3}$ &  &  \\ \hline \hline
 \object{SFO 11 SMM1} & 23  & 20.6  & 4.6 & 11.5 & 1.3 \\
 \object{SFO 11 SMM2} & 32   & 4.9     & 4.4    & 5.4     & 0.6    \\
 \object{SFO 11 SMM3} & 27   & 3.1     & 4.2    & 3.5     & 0.4    \\
 & & & & &   \\ 
 \object{SFO 11NE SMM1} & 23 & 13.4 & 4.6 & 10.2 & 1.1 \\ 
 \object{SFO 11NE SMM2} & 11   & 26     & 5.4    & 45.7    & 5.1    \\
& & & & &  \\
 \object{SFO 11E SMM1} & 23 & 18.7 & 4.6 & 11.2 & 1.3 \\
 \object{SFO 11E SMM2} & 76   &  1.1   & 4.4    & 3.4    & 0.4    \\
 \object{ \object{SFO 11E SMM3}}  & 19   & 7.7     & 5.3    & 25.8    & 2.9    \\ \hline
\end{tabular}
\end{table*}

Masses for the cores were determined by using the standard method of Hildebrand
(\cite{hildebrand83}) for an optically thin molecular cloud with a uniform temperature.
The total mass (dust plus gas mass) of the cloud $M$ is given by

\begin{equation}
M = \frac{d^{2}F_{\nu}C_{\nu}}{B_{\nu}(T_{\rm d})},
\label{eqn:dustmass}
\end{equation}
 
\noindent where $d$ is the distance to the cloud, $F_{\nu}$ is the observed flux density
 and $B_{\nu}(T_{\rm d})$ is the Planck function evaluated at frequency $\nu$ and
 dust temperature $T_{\rm d}$. The parameter $C_{\nu}$ is a mass conversion
 factor combining both the dust-to-gas ratio and the frequency-dependent dust
 opacity $\kappa_{\nu}$. As with the dust optical depth $\tau_{\nu}$, the mass
 conversion factor $C_{\nu}$ is assumed to follow a power-law over frequency,
 i.e.~$ C_{\nu} \propto \nu^{-\beta}$. 
 
Various values for $C_{\nu}$ are quoted in the
literature, ranging from 21.4 g\,cm$^{-2}$ (Kr\"ugel \& Siebenmorgen \cite{ks94}) to 
286 g\,cm$^{-2}$ (Draine \& Lee \cite{drainelee84}).  For comparison the canonical value
quoted in Hildebrand (\cite{hildebrand83}) is 116 g\,cm$^{-2}$.  All of these values have
been normalised to the appropriate value at 850 $\mu$m. The larger values of 
$C_{\nu}$ were calculated for the diffuse ISM and are not appropriate for
these investigations of denser molecular cores. The smaller values of $C_{\nu}$
are more appropriate for cold dense regions of the ISM. For this analysis we
have adopted a value for $C_{\nu}$ of 50 g\,cm$^{-2}$ at 850 $\mu$m, following the method of
Kerton et al.~(\cite{kerton}) and the approximate value for moderate density
regions ($n \sim 10^{5}$ cm$^{-3}$) quoted by Ossenkopf \& Henning
(\cite{oh94}). In any case the values for core mass given here may be linearly
rescaled by another mass conversion factor.

The core masses were evaluated from the integrated 850 $\mu$m fluxes quoted in Table
\ref{tbl:fluxes} and the dust temperatures given in Table \ref{tbl:dustprops}.
The solid angle subtended by each core was calculated from the effective core
diameter $D_{\rm eff}$ and the distance $d$ to each core was assumed to be that
of IC 1848, i.e.~1.9 kpc. The core masses are shown in Table
\ref{tbl:dustprops}. From the mass of each core and its effective diameter
$D_{\rm eff}$ we have also evaluated the H$_{2}$ number density $n$, also shown in
Table \ref{tbl:dustprops}. 

The uncertainties in the mass and density (neglecting the spread in quoted values
for C$_{\nu}) $ are dominated by temperature effects in the non-linear Planck function
in  Eq.~(\ref{eqn:dustmass}). A change of only a few degrees Kelvin in dust
temperature can alter the derived mass by a factor of 2 or more. For the SMM2 and SMM3
cores, where the temperature estimates may only be good to a factor of 2 the derived
mass has a  considerably wider range of up to a factor of 10, depending upon the
signal-to-noise ratio of the sub-mm fluxes. The worst offender is SFO 11E SMM2 where the
low signal-to-noise ratio means that our mass estimate is good to only within 1--11
M$_{\odot}$, whilst for SFO 11NE SMM2 the high signal-to-noise ratio allows us to constrain
the mass of the core to within 13--46 M$_{\odot}$. For comparison the uncertainty in mass due
to the estimated error in the integrated flux is typically 10--20\%,  although for the
fainter cores SFO 11 E SMM2 \& SMM3 this is raised considerably.

It is also possible to derive the visual and infrared extinctions toward each
core from their sub-mm fluxes, as the sub-mm flux provides a direct estimate of
the column density of dust toward each core. The relationship between the
observed sub-mm flux density and visual extinction is given by Mitchell et
al.~(\cite{mjmft01}) as

\begin{equation}
F_{\nu} = \Omega\, B_{\nu}(T_{\rm d}) \,\kappa_{\nu}\, m_{\rm H} \,
\frac{N_{\rm H}}{E\left(B-V\right)}\,\frac{1}{R}\,A_{V}.
\label{eqn:extn}
\end{equation}

The total
opacity of both gas and dust is represented by
$\kappa_{\nu}$, which is functionally equivalent to the reciprocal of the
mass conversion factor $C_{\nu}$. The mass of interstellar material is
written as $m_{\rm H}$ which is in units of hydrogen nuclei, $N_{\rm H}/E\left(B-V\right)$ is the
conversion factor between column density of hydrogen nuclei and the selective absorption
and $R = A_{v}/E(B-V)$ is the ratio between visual extinction and and the selective
absorption. Following Mitchell et al.~(\cite{mjmft01}) and references contained therein
we assume values of $N_{\rm H}/E\left(B-V\right) = 5.8 \times 10^{21}$
cm$^{-2}$\,mag$^{-1}$ and $R = 3.1$. Also of interest are the infrared extinctions
towards each core, as the recent full data release of the 2MASS all sky survey makes it
possible to search each core for embedded protostars or YSOs. The relationship between
sub-mm flux and visual extinction $A_{v}$ contained in equation \ref{eqn:extn} may be
converted to K-band extinction $A_{K}$ by multiplying by the ratio between visual and
infrared extinctions. We assume this ratio to be $A_{v}/A_{K} = 8.9$, following Rieke \&
Lebofsky (\cite{rl85}). Source-averaged visual and K-band extinctions calculated using
Eq.~(\ref{eqn:extn}) and the integrated 850 $\mu$m flux of each core are shown in Table
\ref{tbl:dustprops}.

The typical K-band extinction is around one or two magnitudes, with  \object{SFO 11NE SMM2} possessing
the largest value of $A_{K} = 5.2$ mag. Is 2MASS likely to be a reasonably complete
survey of these cores? We may examine this possibility by using the Zinnecker et
al.~(\cite{zinn93}) relation between stellar mass and absolute $K$ magnitude. Using this
relation a 1 $M_{\odot}$ star 3 $\times 10^{5}$ years old has an apparent K-band magnitude of
$\sim$13 at 1.9 kpc. The limiting K-band magnitude of 2MASS is approximately 14.3 (Cutri et
al.~\cite{2mass}) and so for most of the cores in the sample 2MASS is complete down to
the order of a solar mass or so. Of course the extinctions calculated here assume that
the embedded protostars or YSOs are obscured by a column density throughout the full
depth of the core ($D_{\rm eff}$). If this is not the case then 2MASS may be
sensitive to significantly lower mass protostars or YSOs. We will return to the 
2MASS data in more detail in  Sect.~\ref{sect:2mass}.

\subsubsection{CO measurements}

\begin{table*}[ht]
\centering
\caption{Source-averaged line parameters central velocity $V_{\rm LSR}$, FWHM and
$T_{R}^{*}$ determined by fitting Gaussians to the $^{12}$CO and $^{13}$CO J=2--1 source-averaged
spectra.}
\label{tbl:linepars}
\begin{tabular}{lcccccc}\hline
 & \multicolumn{3}{c}{$^{12}$CO J=2--1} & \multicolumn{3}{c}{$^{13}$CO J=2--1} \\
Source ID & $V_{\rm LSR}$  & Peak $T_{\rm R}^{*}$  & FWHM &$V_{\rm LSR}$  & Peak $T_{\rm R}^{*}$ & 
 FWHM \\ 
 & (kms$^{-1}$) & (K) & (kms$^{-1}$) & (kms$^{-1}$) & (K) & (kms$^{-1}$) \\\hline \hline
 \object{SFO 11 SMM1} & -40.0 & 16.5 & 1.6 & -40.0 & 5.9 & 1.2 \\
 \object{SFO 11 SMM2} & -40.0 & 14.0 & 1.8 & -39.9 & 4.6 & 1.4 \\
 \object{SFO 11 SMM3} & -39.8 & 9.5 & 2.0 & -39.7 & 3.3 & 1.6 \\
 & & & & & &  \\ 
 \object{SFO 11NE SMM1} & -40.1 & 9.5 & 2.5 & -40.0 & 5.3 & 1.8 \\
 \object{SFO 11NE SMM2} & -40.9 & 10.3 & 2.5 & -41.0 & 3.9 & 2.1 \\
& & & & & & \\
 \object{SFO 11E SMM1} & -35.3 & 10.1 & 2.2 & -35.2 & 4.9 & 1.7 \\
 \object{SFO 11E SMM2} & -35.3 & 17.5 & 2.2 & -35.3 & 5.6 & 1.7 \\
 \object{ \object{SFO 11E SMM3}} & -35.8 & 12.4 & 2.7 & -35.6 & 5.0 & 2.0 \\ \hline
\end{tabular}
\end{table*}

In order to determine the temperature and density of the molecular gas from the CO maps
it is first necessary to estimate the optical depth of the $^{12}$CO and $^{13}$CO
transitions. The optical depth may be derived from the ratio of the brightness
temperatures of two isotopomeric lines, assuming a fixed isotopic ratio. The
$^{12}$C/$^{13}$C ratio was taken to be the standard interstellar value of 60 (Frerking,
Langer \& Wilson \cite{flw82}). The relation between brightness
temperature and optical depth follows the form:

\begin{equation}
\frac{T_{12}}{T_{13}} = \frac{1 - e^{-\tau_{12}}}{1 - e^{-\tau_{13}}}
\label{eqn:optdepth}
\end{equation}

\noindent where $T_{12}$, $T_{13}$, $\tau_{12}$ and $\tau_{13}$ are the brightness temperatures
(or corrected receiver or antenna temperatures) and peak optical depths of the $^{12}$CO
and $^{13}$CO lines respectively. $\tau_{12}$ and $\tau_{13}$ are related by $\tau_{12} =
X \tau_{13}$ where $X$ is the  $^{12}$C/$^{13}$C ratio.

A source-averaged spectrum was constructed for each core by averaging the spectra across
each core with the package \emph{kview} (Gooch \cite{kview}). The peak line temperature
$T_{R}^{*}$, FWHM and central velocity of each source-averaged line were measured by
fitting Gaussian line profiles to the data. The source-averaged line parameters are
shown in Table \ref{tbl:linepars} and optical depths for the $^{12}$CO and $^{13}$CO
J=2--1 lines are shown in Table \ref{tbl:linedata}. The line ratios reveal that the
$^{12}$CO line is optically thick towards all of the cores, with typical optical depths
between 25 and 50. $^{13}$CO, on the other hand, is only moderately optically thick with
typical values of 0.4--0.8.

The excitation temperature of the gas was determined from the peak brightness
temperature ($T_{B})$ of the optically thick $^{12}$CO line. Here, we assume 
that the gas in the cores  is in Local Thermodynamic Equilibrium (LTE) and so
can be described by a single excitation temperature $T_{\rm ex}$. The
excitation temperatures were calculated using the Rayleigh-Jeans relation
between brightness and excitation temperature for optically thick lines,
neglecting any background contribution. It was also assumed that the brightness
temperature $T_{B} \simeq T_{\rm R}^{*}$. The
excitation temperature $T_{\rm ex}$ may also be assumed to be roughly equal to
the kinetic temperature $T_{\rm kin}$.

\begin{table}[h]
\begin{minipage}{\linewidth}
\caption{Excitation temperatures ($T_{\rm ex}$), optical depths, source-averaged 
column and number
densities derived from the $^{12}$CO and $^{13}$CO line data. The column and number
densities are of H$_{2}$ and are calculated assuming a $^{12}$CO/$^{13}$CO ratio of 60
and a $^{12}$CO/H$_{2}$ abundance ratio of 10$^{-4}$.}
\label{tbl:linedata}
\begin{tabular}{lccccc}\hline

Source ID &  $T_{\rm ex}$ & \multicolumn{2}{c}{Optical depth} & $\log(N)$ &
$\log(n)$ \\
 & (K) & $^{12}$CO & $^{13}$CO & (cm$^{-2}$) & (cm$^{-3}$) \\ \hline\hline
 \object{SFO 11 SMM1} &  22 & 26.4 & 0.4 & 21.4 & 3.4 \\
 \object{SFO 11 SMM2} &  19 & 24.1 & 0.4 & 21.3 & 3.5 \\
 \object{SFO 11 SMM3} &  14 & 25.1 & 0.4 & 21.1 & 3.4 \\
 & & & & &  \\ 
 \object{SFO 11NE SMM1} & 14 & 48.2 & 0.8 & 21.5 & 3.6 \\
 \object{SFO 11NE SMM2} & 15 & 28.5 & 0.5 & 21.4 & 3.7 \\
& & & & & \\
 \object{SFO 11E SMM1} & 15 & 39.4 & 0.7 & 21.4 & 3.5 \\
 \object{SFO 11E SMM2} & 23 & 23.2 & 0.4 & 21.5 & 3.9 \\
 \object{ \object{SFO 11E SMM3}} & 17 & 31.1 & 0.5 & 21.5 & 3.9 \\ \hline
\end{tabular}
\end{minipage}
\end{table}

The column density of the molecular gas was determined using standard LTE analysis of
the $^{13}$CO J=2--1 observations. To relate the observed integrated line intensity to the
optical depth and column density we used the equation given by White \& Sandell
(\cite{ws95}), where the column density $N$ may be written

\begin{equation}
N = 3.34 \times 10^{14} \int T_{R}\,dV \frac{e^{Jh\nu/2kT_{\rm ex}}}{\nu \mu^{2} \left( 1- e^{-h\nu/kT_{\rm
ex}} \right)} \frac{\tau}{1-e^{-\tau}},
\label{eqn:coldensity}
\end{equation}

\noindent where $N$ is in cm$^{-2}$, $\int T_{R}\,dV$ is the integrated intensity of the line
measured in K\,kms$^{-1}$, $\nu$ is the frequency of the line in GHz, $\mu$ is the dipole moment
of the molecule in Debyes, $J$ is the lower rotational quantum number (1 for the J=2--1
transition), $T_{\rm ex}$ is the excitation temperature in K, and $\tau$ is the peak optical
depth of the line as evaluated from Eq.~(\ref{eqn:optdepth}). The dipole moment of the CO molecule
$\mu$ was taken to be 0.11 D and a value of 220.3987 GHz was used for the  rest frequency of the
$^{13}$CO J=2--1 transition (from the JPL Molecular Spectroscopy database available at
\texttt{http://spec.jpl.nasa.gov}). Integrated line intensities were determined from Gaussian
fits to the source-averaged $^{13}$CO spectra using the relation $\int T_{R}\,dV = 1.06
\,T_{R}^{*}\, \Delta v$, where $\Delta v$ is the FWHM of the line. The column densities so
derived are thus source-averaged rather than beam-averaged. The $^{13}$CO column densities
calculated using Eq.~(\ref{eqn:coldensity}) were then converted into a $^{12}$CO column density by
multiplying by the standard $^{12}$C/$^{13}$C interstellar ratio of 60 and then into an H$_{2}$
column density by assuming the typical $^{12}$CO/H$_{2}$ abundance ratio of 10$^{-4}$
(e.g.~Bachiller \& Cernicharo \cite{bc86}). The resulting H$_{2}$ column densities are given in
Table \ref{tbl:linedata}.

The source-averaged column densities were scaled to a source-averaged number density of
H$_{2}$ molecules by assuming that the cores are spherical with a depth of $D_{\rm
eff}$. The source averaged number densities are shown in Table \ref{tbl:linedata}. 
These $^{13}$CO-derived values for the H$_{2}$ number density are roughly an order of 
magnitude smaller than those obtained from the sub-mm continuum emission in
Sect.~\ref{sect:scuba_anal}.  At first sight this may be taken for evidence that the CO
abundances in the cores may be depleted by freeze-out onto grain mantles,  however the
uncertainties in the dust-derived densities, which may be up to an order of magnitude,
preclude a definitive statement. A more likely cause of the low $^{13}$CO abundance is 
that the $^{13}$CO at the molecular boundary of the cloud  is selectively
photodissociated by the incident FUV field. The FUV photodissociation rate of $^{13}$CO
may be up to an order of magnitude higher than that of $^{12}$CO (van Dishoeck \& Black
\cite{vdhb88}).  Observations of less abundant CO isotopomers and more detailed
modelling of the dust emission and photochemistry of the molecular gas are required to
address these issues further.

\subsection{Infrared objects associated with the clouds}
\label{sect:2mass}

The 2MASS Point Source Catalogue (Cutri et al.~\cite{2mass}) was used to locate infrared
sources associated with each cloud. Our analysis of the SCUBA 850 $\mu$m fluxes in
Sect.~\ref{sect:scuba_anal} indicates that the 2MASS catalogue is complete for embedded
objects of down to roughly a solar mass. A total of 52 objects were found to be located
within the optical boundaries of the clouds and are labelled in Figs.~\ref{fig:2mass1}--\ref{fig:2mass3}.

\begin{figure}
\includegraphics*[angle=-90,scale=0.65]{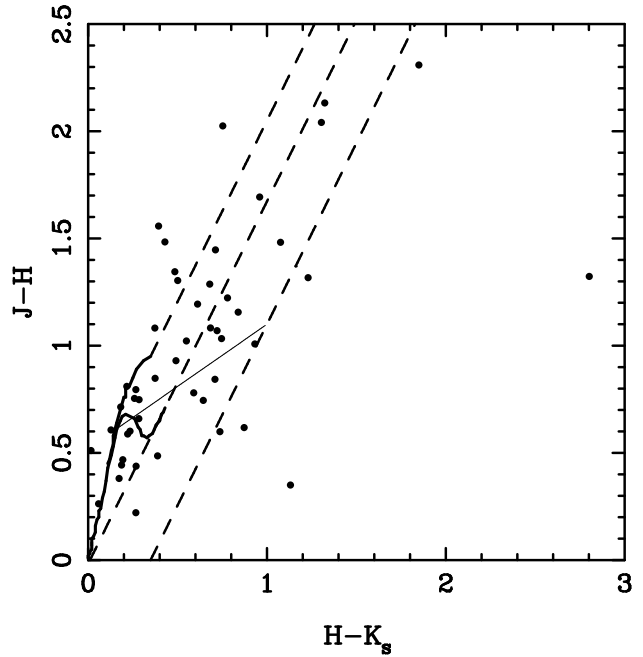}
\caption{$J$--$H$ versus $H$--$K_{s}$ diagram of the 2MASS sources associated with the three bright-rimmed
clouds. The thick solid lines represent the unreddened locuses of main-sequence and giant stars
from Koornneef (\cite{koornneef83}). The thin solid line indicates the Classical T-Tauri
locus of Meyer, Calvet \& Hillenbrand (\cite{mch97}). Reddening tracks are shown by dashed
lines.}
\label{fig:jhk}
\end{figure}

We classified each infrared object using the $J$--$H$ vs $H$--$K_{s}$ colour diagram
method described in Lada \& Adams (\cite{la92}). The $J$, $H$ and $K_{s}$ magnitudes of
each object were taken from the 2MASS Point Source Catalogue and their $J$--$H$ \&
$H$--$K_{s}$ colours are plotted in Fig.~\ref{fig:jhk}. Reddening tracks for giant and
main-sequence stars were determined from the photometric data of Koornneef
(\cite{koornneef83}). A similar track was determined for Classical T-Tauri stars using
the published locus from  Meyer, Calvet \& Hillenbrand (\cite{mch97}). 

As an object is subjected toward higher extinctions, either by increasing interstellar or
circumstellar dust, it moves up toward the top of Fig.~\ref{fig:jhk} parallel to the
reddening tracks. Objects found between the left-hand and middle reddening tracks are thus
likely to be reddened giant or main sequence stars (Lada \& Adams \cite{la92}). Classical T
Tauri stars (also known as Class II objects) are more reddened than main sequence stars, due
to  the presence of excess near-infrared emission from their circumstellar disks. Those
objects found between the middle and right-hand tracks are candidate Classical T Tauri stars
(Meyer, Calvet \& Hillenbrand (\cite{mch97}). Class I protostars are surrounded by extended
envelopes as well as circumstellar disks and are hence more reddened than the Classical T
Tauri stars. Objects lying to the right of the rightmost dashed line in Fig.~\ref{fig:jhk})
are thus class I protostellar candidates.

\begin{table*}
\centering
\caption{2MASS sources identified as either candidate T-Tauri stars (TTS) or class I protostellar
candidates
(YSO) in the
$JHK_{s}$ diagram of Fig.~\ref{fig:jhk}. Sources with no quoted errors for the $J-H$ or $H-K$ colours have
an unreliable (null) photometric error listed in the 2MASS catalogue and the colours of these objects should
be regarded as highly uncertain. Where the errors in the $J-H$ and $H-K$ colours allow multiple
classifications as T-Tauri stars, protostars or reddened main sequence stars (RMS), all possible
classifications are noted. The Figure Label refers to the 
alphabetic label of each source in Figs.~\ref{fig:2mass1}--\ref{fig:2mass3} and the 2MASS PSC ID indicates the formatted J2000 coordinates of each
source. T-Tauri candidates marked with an asterisk ($^{*}$) were identified as TTS in the
H$\alpha$ grism study of Ogura, Sugitani \& Pickles (\cite{osp02})}   
\label{tbl:jhk}
\begin{tabular}{llcccl}\hline
\multicolumn{2}{l}{Figure Label} & 2MASS PSC ID & $J-H$ & $H-K_{s}$ & Type \\\hline\hline
SFO 11 & d & 02513283+6003542  & 1.48$\pm$0.04  & 1.08$\pm$0.04 &  TTS$^{*}$ \\
 & f & 02513815+6003364 & 0.44$\pm$0.25 & 0.19 & RMS/TTS\\
 & j & 02512374+6003399  & 0.35$\pm$0.26  & 1.13 &  RMS/TTS/YSO \\
 & n & 02512848+6003024  & 0.62$\pm$0.20  & 0.87 &  RMS/TTS/YSO \\
SFO 11NE & b  & 02515212+6007102  & 1.16$\pm$0.04  & 0.84$\pm$0.04 &  TTS \\
& c & 02515380+6006581  & 2.31  & 1.85 &  RMS/TTS/YSO \\
& e & 02515377+6006571  & 1.32  & 2.80 &  RMS/TTS/YSO \\
& f & 02515413+6006524  & 2.04$\pm$0.18  & 1.30$\pm$0.12 &  RMS/TTS \\
& h & 02515288+6006545  & 0.22$\pm$0.09  & 0.27$\pm$0.13 &  RMS/TTS \\
& j & 02515610+6006555  & 1.194$\pm$0.18 & 0.61$\pm$0.16 & RMS/TTS \\
& m & 02515614+6006390  & 1.08$\pm$0.18  & 0.68$\pm$0.16 &  RMS/TTS \\
& p & 02515279+6006294  & 2.13$\pm$0.12  & 1.32$\pm$0.05 &  RMS/TTS \\
& q & 02515975+6006394  & 1.07$\pm$0.06  & 0.72$\pm$0.06 &  RMS/TTS$^{*}$ \\
& s & 02520131+6006154  & 1.22$\pm$0.08  & 0.78$\pm$0.07 &  RMS/TTS \\
& t & 02515741+6006125  & 0.78$\pm$0.26  & 0.59$\pm$0.26 &  RMS/TTS \\
& v & 02515136+6006112  & 0.38$\pm$0.06 & 0.17$\pm$0.08 & RMS/TTS \\
SFO 11E & c & 02521422+6003114  & 1.03$\pm$0.05  & 0.75$\pm$0.05 &  TTS \\
& e & 02521916+6002501  & 1.082$\pm$0.23 & 0.37 & RMS/TTS/YSO\\
& f & 02521685+6002449  & 1.32$\pm$0.24  & 1.23$\pm$0.16 &  TTS/YSO \\
& h & 02521715+6002328  & 0.75$\pm$0.18  & 0.64$\pm$0.20 &  RMS/TTS \\
& j & 02521602+6002361  & 0.93$\pm$0.14 & 0.49$\pm$0.16 & RMS/TTS\\
& k & 02521275+6002390  & 0.60$\pm$0.24 & 0.74$\pm$0.26 &  RMS/TTS/YSO \\
& l & 02521997+6002331  & 0.49$\pm$0.13  & 0.39$\pm$0.15 &  RMS/TTS \\
& m & 02521856+6002274  & 1.01$\pm$0.14  & 0.93$\pm$0.12 &  TTS/YSO \\ 
& p & 02525660+6001515  & 0.44$\pm$0.04 & 0.27$\pm$0.04 & RMS/TTS \\
& q & 02521639+6001526  & 0.84$\pm$0.18  & 0.71$\pm$0.19 &  RMS/TTS \\ \hline

\end{tabular}
\end{table*}

The $J$--$H$ vs $H$--$K_{s}$ demarcation between reddened main-sequence stars, classical T Tauri
stars and class I protostars is not as clear cut as Fig.~\ref{fig:jhk} suggests. Lada \&
Adams (\cite{la92})  note that although main sequence stars and YSOs occupy different
regions of the colour-colour diagram, the individual YSO types (such as weak-line T Tauris,
classical T Tauris, class I protostars and Herbig AeBe stars) may overlap somewhat.
Nevertheless the $J$--$H$ vs $H$--$K_{s}$ colour-colour diagram is a useful technique to 
identify those cores currently forming stars, whether their evolutionary state is at the
Class I stage or advanced toward the T Tauri stage. In Table \ref{tbl:jhk} we list the infrared
objects associated with each cloud that satisfy the Lada \& Adams (\cite{la92}) and Meyer,
Calvet \& Hillenbrand (\cite{mch97}) criteria for class I protostars or T Tauri stars. Where the errors in
the $J$--$H$ and $H$--$K_{s}$ colours preclude a unique classification we have listed the alternatives.

{Table \ref{tbl:jhk} reveals that the 2MASS photometry of the objects associated with the clouds is highly
uncertain and with the existing data it is not possible to make a definitive statement about the young
stellar or protostellar content of each cloud. Many of the objects associated with the clouds may be
equally classified as either reddened main sequence stars or T Tauri stars. Nevertheless we note that  each
cloud possesses at least one object whose colours are not consistent with those of a reddened main sequence
star (source d associated with SFO 11; sources b and c associated with SFO 11NE; sources c, f and m
associated with SFO 11E).  Following the classification system of  Lada \& Adams (\cite{la92}) and Meyer,
Calvet \& Hillenbrand (\cite{mch97}) the infrared colours of these objects are consistent with either T
Tauri stars or class I protostellar objects. Two of these candidates are also confirmed as a weak-line T Tauri
(02513283+6003542) and classical T Tauri (02515975+6006394) via their H$\alpha$ emission lines (Ogura,
Sugitani \& Pickles \cite{osp02}). 

The 2MASS data lends support to the hypothesis that each cloud is a site of recent or ongoing star
formation, as at least one young stellar object or protostar is identified with each cloud. If even a
fraction of the T Tauri candidates are indeed bona fide T Tauri stars then each cloud may be home to a
cluster of young stellar objects and/or protostars. More accurate photometry at $J$, $H$, $K$ and perhaps
in the $L$ band (Meyer, Calvet \& Hillenbrand \cite{mch97}) is required to investigate the nature of the
infrared objects associated with these clouds. 

The bright-rimmed cloud core  \object{SFO 11NE SMM1} is host to the greatest number of embedded YSO and
protostar candidates. These infrared sources are distributed along the long axis of the cloud as seen in the
SCUBA maps (see Fig.~\ref{fig:2mass2}) in a manner suggestive of the small scale sequential star formation
scenario of Sugitani, Tamura \& Ogura (\cite{sto95}), i.e.~with the stellar clusters elongated toward the
exciting star of the bright-rimmed cloud.  However, without a more accurate classification of the protostar
and T Tauri candidates it is not possible to determine whether there is a smooth progression of 
earlier evolutionary
type deeper into the cloud  as would be expected in the small scale sequential star formation scenario
(Sugitani, Tamura \& Ogura \cite{sto95}). More accurate photometry and higher resolution observations are
required to investigate the potential of SFO 11NE SMM1 as a candidate for the
small scale sequential star formation process proposed by  Sugitani, Tamura \&
Ogura (\cite{sto95}).

\subsection{The pressure balance between ionised and molecular gas}
\label{sect:pressure}

Radiative driven implosion models of bright-rimmed cometary clouds (Bertoldi \cite{bertoldi};
Lefloch \& Lazareff \cite{ll94}) predict that the evolution of the cloud is largely controlled
by the pressure balance between the cloud interior and exterior.  As the clouds are exposed to
the UV flux from a nearby OB star (or stars) their surfaces become ionised. A photoionised
sheath of gas, known as an ionised boundary layer or IBL, develops around the cloud and a
significant fraction of the impinging UV photons may be trapped in this layer. Photoionised and
photoevaporated gas  also flows normally from the cloud surface following the decreasing density
found  at an increasing distance  from the cloud surface. The clearly visible striations seen
perpendicular to the cloud rims in the narrowband H$\alpha$ images (Fig.~\ref{fig:halpha}) show
that this outwards flow of photoionised gas is occurring in the three clouds in this study. 

As the IBL and the photoionised flow develop, a photoionisation shock is driven into the
molecular gas of the clouds. Depending upon the balance between the interior molecular
pressure of the cloud and the exterior pressure of the photoionised sheath or IBL this
shock may either stall at the surface or propagate through the molecular gas, followed
by a D-critical ionisation front. If the cloud is underpressured with respect to the IBL
or in pressure equilibrium with the IBL, the photoionisation shock and following
D-critical ionisation front progress through the cloud, leading to its complete
ionisation and dispersal within a few Myr. On the other hand if the cloud is
overpressured with respect to the IBL then the shock stalls at the cloud surface until
the growing density (and hence pressure) of the IBL reach  equilibrium with that
of the cloud. When pressure equilibrium is reached the ionisation front becomes
D-critical and the shock and ionisation front will continue their propagation into the
cloud. The important result from the RDI models is that the evolution of the
bright-rimmed clouds depends mainly upon the    \emph{duration} of their UV illumination.

Establishing the presence of a shock propagating through the molecular gas is extremely
important from the point of view of investigating the cloud evolution and whether any star
formation in the clouds is likely to have been triggered by  photoionisation shocks. The
presence of a photoionisation shock can be inferred by the pressure balance between the
cloud interior and exterior. If the cloud is underpressured (or at the same pressure) with
respect to the surrounding medium then it is highly likely that a photoionisation shock and
D-critical ionisation front are being driven into the cloud. Conversely if the cloud is
overpressured with respect to the surrounding medium then the cloud must be in the
compression phase and the shock is stalled at the cloud surface. 

The radio emission mapped in the NVSS data probes the conditions in the ionised boundary
layer of the clouds, whilst the JCMT $^{13}$CO maps reveal the conditions within the
molecular interior of the clouds. In this section we use the NVSS and JCMT observations to
determine the pressure in the ionised and molecular gas so that we may investigate the
pressure balance of the clouds and establish whether photoionisation shocks are currently
propagating through the clouds.
    
\subsubsection{Ionising flux, electron density and ionised gas pressure}
\label{sect:ionising}

As well as the pressure in the ionised gas the free-free radio flux from the ionised boundary layer allows the
impinging flux of ionising photons and the electron density of the layer to be determined. Because of the low
resolution of the NVSS data (45\arcsec\ FWHM beam) it is important to stress that the quantities derived for the
clouds are global averages and do not represent local point-to-point values (e.g.~for the individual cores
within each SCUBA jiggle-map).    The 20 cm radio emission appears to be elongated along the cometary axis
of each cloud (somewhat marginally in the case of SFO 11NE), but is mostly centred upon the head of each
bright-rimmed cloud. The morphology of the 20cm  emission associated with  \object{SFO 11E} suggests that this
cloud lies in the same line-of-sight as the bright southern ionisation-bounded ridge of  IC 1848. The radio
emission associated with  \object{SFO 11E} is much stronger than the other two clouds and it is likely that the
emission is enhanced by the line-of-sight effects of sampling through a much deeper column of ionised gas along
the southern ridge.

To evaluate the strength of the ionising flux impinging upon the clouds and also
the electron
density and pressure in the photoionised boundary layer we use the general equations from
Lefloch et al.~(\cite{llc97}).  Rearranging their Eq.~(1), the ionising photon flux
 $\Phi$ arriving at the cloud rim  may be written in units of cm$^{-2}\,$s$^{-1}$ as:
 
\begin{equation}
\label{eqn:phi}
\Phi = 1.24 \times 10^{10} \,S_{\nu}\, T_{\rm e}^{\,0.35}\, \nu^{0.1} \,\theta^{-2}, 
\end{equation}

where $S_{\nu}$ is the integrated radio flux in mJy, $T_{\rm e}$ is the effective
electron temperature of the ionised gas in K, $\nu$ is the frequency of the free-free
emission in GHz and $\theta$ is the angular diameter over which the emission is
integrated in arcseconds.

The electron density ($n_{\rm e}$) of the ionised boundary layer surrounding the cloud
may also be derived from the integrated radio flux $S_{\nu}$ by subsituting for the
ionising photon flux in Eq.~(6) of Lefloch et
al.~\cite{llc97}. The electron density in cm$^{-3}$ is given by:

\begin{equation}
\label{eqn:ne}
n_{\rm e} = 122.41 \,\,\sqrt{\frac{S_{\nu}\, T_{\rm e}^{0.35}\, \nu^{0.1}\, \theta^{-2}}{\eta R}},
\end{equation}

where those quantities common to both Eqs.~(\ref{eqn:phi}) and (\ref{eqn:ne}) are in
the same units, $R$ is the radius of the cloud in pc and  $\eta$ is the effective
thickness of the ionised boundary layer as a fraction of the cloud radius (typically $\eta \sim 0.2$, Bertoldi
\cite{bertoldi}). From the electron density we may evaluate the pressure in the ionised
boundary layer via $P_{i} = 2\rho_{i} c_{i}^{2}$, where $\rho_{i}$ is the density in the
boundary layer and $c_{i}$ is the sound speed of the ionised gas (typically $\sim$ 11.4
kms$^{-1}$). 

The measured integrated 20 cm fluxes for each of the three clouds \object{SFO 11}, \object{SFO 11NE} and
SFO 11E are given in Sect.~\ref{sect:obs} as 7.7, 8.8 and 37.0 mJy respectively. Values for
the ionising flux $\Phi$, electron density and ionised gas pressure $P_{i}/k$ were
calculated using the above equations and assuming  an effective electron temperature
$T_{\rm e} = 10^{4}$ K and  an ionised boundary layer thickness $\eta = 0.2$. The
results of these calculations are shown in Table \ref{tbl:pressure}.
Given that SFO11E lies along the
southern ionisation boundary of IC1848 it is also likely that both the integrated 20 cm
flux and  radio-derived ionising flux are overestimated for this cloud. If we assume that
SFO 11E is at roughly the same distance from  \object{HD17505} as the other two clouds, then
the typical ionising flux illuminating \object{SFO 11E} should be similar to that of the two other
clouds ($\sim$3$\times 10^{9}$ cm$^{-2}$ s$^{-1}$). In this case the measured 20 cm flux is
overestimated by roughly a factor of two and the electron density and pressure should be
reduced by a factor of $\sqrt{2}$.

IC1848 is  excited by the young open cluster OCl 364, whose dominant member is the O6V star \object{HD17505}.
 Inspection of the orientation of the three bright-rimmed clouds with respect to the stars in the open cluster
shows that it is likely  that  \object{HD17505} is the star primarily responsible for ionising the cloud
surfaces: the cometary axes of SFO 11 and SFO 11NE point directly toward  \object{HD17505}. The axis of
\object{SFO 11E} does not point directly toward \object{HD17505}, although the brightest rim of the cloud is
found on the side facing  \object{HD17505} (see Fig.~\ref{fig:halpha}). It is likely that another nearby OB star
is also exciting SFO 11E, but as the H$\alpha$ and 20 cm emission are located on the  \object{HD17505}-facing
side of the cloud  \object{HD17505} is almost certainly the predominant exciting star of this cloud.  The nearby
O9V star HD17520 is also a potential exciting star, although following Panagia (\cite{panagia}),  
\object{HD17505} is expected to emit roughly eight times the ionising photon flux of HD17520 and is the dominant
member of the pair. The ionising photon flux predicted by Panagia (\cite{panagia}) for an 06V star is
$1.7\times10^{49}$ photons s$^{-1}$, scaling this to the projected distance of  \object{HD17505} from the three
clouds (which is $\sim$11 pc) and assuming that there is negligible absorption of the UV radiation from the
intervening HII region we predict  that the ionising photon flux impinging upon the clouds is 1.2$\times10^{9}$
cm$^{-2}$ s$^{-1}$. 

This value is reasonably  consistent with that estimated from  the 20 cm free-free flux
measured from \object{SFO 11E} (0.8 $\times 10^{9}$ cm$^{-2}$ s$^{-1}$), but a factor of 3--5 larger
than the value for $\Phi$ estimated from the 20 cm flux associated with the clouds \object{SFO 11}
and \object{SFO 11NE}.  Projection effects are the  most likely cause for the over-prediction of the
ionising flux from the spectral type of the illuminating star. Even a modest inclination of
the star-cloud vector to the line-of-sight results in an increase of the ``true'' distance
between star and cloud and hence a significant decrease in the predicted UV flux impinging
upon the cloud.  The measured and predicted ionising fluxes may be used to estimate the
true distance and inclination angle of the star-cloud vector to the line of sight. Again
assuming an average illuminating UV flux of 3$\times 10^{9}$ cm$^{-2}$ s$^{-1}$ at the
cloud surfaces we estimate that the true (or at least an upper limit) distance  of the clouds from the ionising 06V star
 \object{HD17505} is $\sim$22 pc and the inclination of the star-cloud vector to the line of sight is
30$\degr$.

\subsubsection{Molecular gas pressure}

The pressure of the molecular gas ($P_{m}$) is comprised of contributions from both
turbulent and thermal components. For cold gas, such as that of the cores in our
sample, there is a negligible thermal contribution to either the pressure or observed
line-width. The molecular pressure may thus be written as the product of the square of the
turbulent velocity dispersion ($\sigma^{2}$) and the density of the molecular gas
($\rho_{m}$); i.e. $P_{m} \simeq \sigma^{2} \rho_{m}$. The turbulent velocity dispersion may be
written in terms of the observed line width $\Delta v$ as $\sigma^{2} = \left< \Delta v
\right> ^{2} /
(8 \ln 2)$.  In order to avoid optical depth selection effects and sample the gas
throughout the cloud the line width should be determined from an optically-thin line. The
line widths used here are those measured from the source-averaged spectra of the (at most)
moderately optically thick $^{13}$CO line.

The H$_{2}$ number densities of the clouds as derived from the $^{13}$CO observations
are all typically a few times 10$^{3}$ cm$^{-3}$, however those derived from greybody
fits to the dust emission are roughly an order of magnitude higher. In order to estimate
the \emph{maximum} pressure within the clouds we have used the dust-derived density
rather than the $^{13}$CO-derived value which may be affected via depletion (either onto
dust grain ice mantles or via selective photodissociation) or optical depth effects. As
the radio emission is concentrated around the cores at the head of each cloud (the SMM1
cores) we have used the density values derived for each of these cores. We caution
however that the density of the gas (and hence the molecular pressure) is dependent upon
the validity of the assumed spherical core geometry, the value of the dust mass
coefficient $C_{\nu}$ and the dust temperature $T_{\rm d}$ derived from the greybody
fit to the SED. The molecular pressure should only be considered to be accurate to
within a factor of 3 at best. 

\begin{table*}
\caption{Values for the ionising flux impinging upon the clouds, electron density and
ionised and molecular gas pressures. The ionised gas pressure is derived from the NVSS
20 cm data and the molecular pressure from the $^{13}$CO linewidth.}
\label{tbl:pressure}
\begin{tabular}{lcccc}\hline
Cloud & Ionising flux & Electron density  & Ionised pressure
& Molecular pressure \\
 & $\Phi$ (cm$^{-2}$s$^{-1}$) & $n_{\rm e}$ (cm$^{3}$) & $P_{i}/k$ (cm$^{-3}\,$K) &
 $P_{m}/k$ (cm$^{-3}\,$K) \\\hline\hline
SFO 11 & $2.2\times10^{8}$ & 266 & $8.2\times10^{6}$ & $2.5\times10^{6}$\\
SFO 11NE & $3.5\times10^{8}$ & 354 & $1.1\times10^{7}$ & $5.7\times10^{6}$\\
SFO 11E & $8.5\times10^{8}$ & 478 & $1.5\times10^{7}$ & $5.1\times10^{6}$\\\hline
\end{tabular}
\end{table*}

 The molecular pressures for each cloud are all within a few times 10$^{6}$ cm$^{-3}\,$K  and are
shown in Table \ref{tbl:pressure}. Comparing the values of $P_{m}$ and $P_{i}$ for each cloud, and
taking the error in the molecular pressure into account, reveals that the pressures are
approximately equal. In this scenario the ionised and molecular gas are in pressure balance and the
conditions are consistent for the propagation of photoionisation-induced shocks into the clouds.
However, from the currently available  data we cannot rule out the possibility that the clouds may be
either underpressured or marginally overpressured with respect to their IBLs. In the former case
photoionisation-induced shocks may propagate into the clouds, whilst in the latter the shocks are
stalled at the ionised boundary layer (Lefloch \& Lazareff \cite{ll94}). 

It is unfortunate that the results of this analysis are not more concrete. It is possible that
photoionisation-induced shocks are propagating into the clouds, given the similar values of $P_{m}$
and $P_{i}$ for each cloud. However, if the molecular pressure $P_{m}$ is greater than the estimates
in Table \ref{tbl:pressure} the data do not preclude the likelihood that the shocks are stalled in
the ionised boundary layer. We will discuss both of these possibilities further in 
Sect.~\ref{sect:cloudprops}.

%______________________________________________________________

\section{Discussion}
\label{sect:discuss}

In this section we draw together the results from the previous analyses to investigate
the nature of the clouds, their potential for star formation and speculate upon their
evolution and 
eventual fate.

\subsection{Cloud morphology and physical properties}
\label{sect:cloudprops}

The general morphology of the clouds indicates that the predominant UV flux illuminating
the clouds originates from the single 06V star  \object{HD17505}. All three clouds possess a similar
cometary morphology with their long axes pointing in the general direction of  \object{HD17505}.
There is some ambiguity regarding the cloud \object{SFO 11E}, whose axis does not point directly
toward  \object{HD17505}, although we note that the brightest H$\alpha$ and 20 cm emission
originates from the  \object{HD17505}-facing side of the cloud and so  \object{HD17505} is likely to be the
predominant exciting star.
Assuming that there is little absorption of the UV flux in the intervening HII region the
strength of the 20 cm free-free emission associated with the clouds suggests that their
true distance from  \object{HD17505} is around 22 pc, roughly double the projected distance. Although
the three clouds lie close together on the sky, their arrangement is probably a
chance superposition.  For \object{SFO 11E} the Digitised Sky Survey and H$\alpha$ images show that 
bright optical emission from the ionised gas is located on the facing side of the cloud,
whereas for \object{SFO 11NE} and  \object{SFO 11} the emission is shielded by the dark molecular gas.  SFO
11E must be located behind the illuminating star with the long axis of the cloud pointing
toward the observer for its ionised surface to be visible in the optical. The remaining two
clouds are probably located in front of the ionising star with their long axes pointing
away from the observer and the dark molecular gas obscuring the bright optical emission
from the face exposed to the ionising star. 

This scenario is consistent with the observed difference in the $V_{LSR}$ of the $^{12}$CO
and $^{13}$CO emission from the three clouds. \object{SFO 11E} is redshifted  by $\sim$5 km\,s$^{-1}$
with respect to the other two clouds. A well-known phenomenon in cometary clouds is their
velocity displacement from their ionising star, as the star photoevaporates material from
the cloud  surface the so-called rocket effect (Oort \& Spitzer \cite{os55}) accelerates
the cloud away from the star by up to $\sim$10 km\,s$^{-1}$ (Bertoldi \cite{bertoldi}). If
SFO 11E is located behind the ionising star, whereas \object{SFO 11} and \object{SFO 11NE} are located in
front of the ionising star the two groups of clouds will be accelerated away from each
other, leading to the observed difference in $V_{LSR}$ between the clouds.

Comparing the overall morphology of the clouds as seen in the H$\alpha$ images, SCUBA
and CO maps with that predicted by the RDI models of Lefloch \& Lazareff (\cite{ll94})
suggests that the clouds have been exposed to the UV flux from  \object{HD17505} for
between 1--2$\times 10^{5}$ years. The ``inverted V'' appearance of \object{SFO 11}
closely resembles the RDI model snapshot at $t=0.183$ Myr (Fig.~4c of Lefloch \&
Lazareff \cite{ll94}). \object{SFO 11NE} and SFO 11E display a more rounded
``pillar''  morphology than the sharply swept-back appearance of SFO 11 which appears
slightly detached from the H$\alpha$ emission to the south. The rounded pillar
morphology of SFO 11NE and SFO 11E is similar to the Lefloch \& Lazareff RDI model
snapshot at 0.036 Myr since initial ionisation. This may suggest that either SFO 11NE
and SFO 11E have been exposed to the UV flux for a shorter period of time, or that their
internal pressures were initially higher than that of \object{SFO 11} and the ionisation
front was stalled at the cloud surface until the internal and external pressures reached
equilibrium.  

This phase in the models corresponds to the early collapse phase of the clouds, prior to
the maximum compression of the cloud and the subsequent cometary stage (Lefloch \&
Lazareff \cite{ll94}).   The cores found at the head of SFO 11NE and \object{SFO 11E}
are not elongated along the axis of UV illumination, contrary to the predictions of the
RDI models, where the dense cores are formed via radiative-driven  compression of the
molecular gas. These cores may not have been formed by the RDI process and could thus be
pre-existing structures within the clouds. However, we note that the cores are only just
resolved in the 14\arcsec\ SCUBA maps and higher resolution mapping to more accurately
determine their degree and axis of elongation may be required to address this point.

Internally, the three clouds possess a clumpy  structure comprised of two or three
dense molecular cores. The cores located at the ``head'' of each cloud (i.e. those
closest to the ionising star) are in general more massive and slightly denser than the
cores found deeper inside the clouds, even taking into account the rather large
uncertainty in the mass of the SMM2 and SMM3 cores. The larger concentration  of material towards the ``heads'' or ``tips'' of the clouds
is similar to that seen in other bright-rimmed clouds and globules, e.g.~the Eagle
Nebula (White et al;.~\cite{whiteeagle}; Fukuda, Hanawa \& Sugitani \cite{fhs02}),  the
Rosette Nebula (White et al.~\cite{rosette}) and RNO 6 (Bachiller, Fuente \& Kumar
\cite{bfk02}).

The cores found  at the head of each cloud are in approximate pressure balance with the  exterior
ionised gas in the ionised boundary layer. The error in the determination of the molecular pressure
(which arises mostly from the uncertainty in the H$_{2}$ density of the molecular gas) means that we
cannot say for definite whether the clouds are over- or under-pressured with respect to their
surroundings. Thus from the pressure evidence alone it is difficult to ascertain if
photoionisation-induced shocks are currently propagating into the clouds. However, the close
correspondence of the cloud morphologies to the collapse phase in the Lefloch \& Lazareff 
(\cite{ll94}) RDI model lends weight to the supposition that ionisation shocks are propagating into
the clouds and causing their collapse. On balance we conclude that 
photoionisation-induced shocks are likely propagating into the clouds. We base the
following discussion of the cloud properties and evolution (see Sect.~\ref{sect:evolution})  upon
this hypothesis.

 In order to estimate the likely extent to which the shocks have propagated into the clouds we may derive an
estimate of the shock velocity from the pre- and post-shock pressures of the neutral gas (White et
al.~\cite{whiteeagle}). We follow Eqs.~(22) and (23) from  White et al.~(\cite{whiteeagle}) and
their assumption that  the ratio of the post- and pre-shock densities ranges from $2\,$--$\,\infty$,
which leads to a maximum error in the estimated shock velocity of a factor $\sqrt{2}$. Using these
assumptions we derive typical shock velocities in the range 1.2--1.6 kms$^{-1}$ for neutral gas of
density $4 \times 10^{4}$ H$_{2}$ molecules cm$^{-3}$.  The shock velocity is much greater than the
sound speed of the molecular gas, which is typically $\sim$0.3 km\,s$^{-1}$ for molecular hydrogen
at 20 K, implying that the photoionisation-induced shocks propagating into the clouds are
supersonic. The shock crossing time for the cores found at the head of the clouds is $\sim$ $2
\times 10^{5}$ years, based upon a shock velocity of 1.4 km\,s$^{-1}$ and a typical core diameter of
0.25 pc.

It it also possible to independently estimate the duration over which the clouds have been exposed
to the UV flux from a simple treatment of the expansion time of the HII region IC 1848. In the
following, we assume that the expansion of IC 1848 was powered primarily by the 06V star
\object{HD17505} and that the clouds lie at a distance of 22 pc from the star, consistent with the
ionising flux measured from the 20 cm free-free continuum and the spectral type of the ionising
star. As this distance was evaluated using the assumption that the absorption of the UV radiation by
the intervening material within the HII region is negligible, it should be regarded strictly as an
upper limit to the actual star-cloud distance. We note that the distance between \object{SFO 11E}
and  \object{HD17505} is  not certain as the 20 cm flux associated with this cloud may be
over-estimated by association with the bright rim of the ionisation boundary to the south. The
initial expansion of the HII region is rapid out to the radius of the Str\"omgren sphere, which for
an O6V star is typically 1--4 pc, assuming the density of the surrounding material is between
10$^{2}$ and 10$^{3}$ cm$^{-3}$. Following this rapid expansion the ionisation front moves outward
much more slowly, at around the sound speed of the ionised gas which is typically $\sim$11.4
km\,s$^{-1}$.  

If the clouds are indeed located around 22 pc from their ionising star and the ionising
front expanded at the sound speed of 11.4 km\,s$^{-1}$  after the initial Str\"omgren
expansion it will have taken $\sim$1.5 Myr for the front to reach the clouds. This is
comparable to the estimated lifetime of IC 1848 (Vall\'ee, Hughes \& Viner \cite{vhv79}) and indicates that the clouds have only
just become exposed to the ionising UV flux of HD 17505. It must be noted that our estimate
of the expansion timescale for the ionisation front to reach the clouds is a rather
simplistic estimate; assuming a constant expansion velocity of the front at the sound speed,
an initial density of 10$^{2}$ cm$^{-3}$ for the material surrounding the O-star and a
star-cloud distance based upon negligible absorption of the UV flux emitted from the O-star..
However the presence of the bright ridge of nebular emission to the south of the clouds
suggests 
that the ionisation front has only just reached the base of the clouds and
supports the argument that the clouds have only just become exposed to the UV flux. We can
thus derive an alternative UV illumination timescale for the clouds from the time taken for
the ionisation front to travel the length of the clouds to the bright southern ridge. 
Assuming that the clouds are not
significantly foreshortened by projection effects, the time taken for an ionisation front
moving at the sound speed to traverse the distance between the head of \object{SFO 11NE} (the
furthest cloud from the ridge)  and the bright nebular ridge is $\sim$300,000 years. 

This is comparable to the UV illumination timescale predicted by the cloud morphology from
the RDI model of Lefloch \& Lazareff (\cite{ll94}) and reinforces the conclusion that the
overall evolution of the clouds will be governed by their radiative-driven implosion. Both
timescales are also comparable to the shock crossing time of the cores ($\sim 2 \times
10^{5}$ years). This implies that there
has been sufficient time since the clouds were illuminated for the shocks to propagate deep
within the clouds and to perhaps substantially affect their interior evolution and
star-forming history.

\subsection{Star formation in the clouds?}
\label{sect:star_formn}

From the SCUBA maps we have identified 8 dust cores that have similar characteristics
(i.e.~density, temperature and spatial diameter)  to protostellar cores observed in other
molecular clouds (e.g.~Evans 1997).  Three of the cores (\object{SFO 11NE SMM1},  \object{SFO 11NE SMM2} and
 \object{SFO 11E SMM1}) appear centrally-condensed, also suggesting their protostellar nature. The
cores found at the heads of these bright rimmed clouds are at the high mass end of the
scale compared to the cores observed in other molecular clouds and Bok globules, which
typically mass $\sim 10$ M$_{\odot}$. This is consistent with the overall tendency toward
higher masses for other known bright-rimmed cloud cores (e.g.~Sugitani et
al.~\cite{smnto00}; Lefloch et al.~\cite{lcrmch02}; White et al.~\cite{whiteeagle};
Lefloch, Lazareff \& Castets \cite{llc97}).

There is some evidence for molecular outflow within the two cores  \object{SFO 11NE SMM1} and \object{SFO 11E}
SMM1 in the form of moderate velocity line wings in the $^{12}$CO spectra. However, outflow
lobes are not seen in either the $^{12}$CO channel maps  (Fig.~\ref{fig:channmaps}) or in
integrated intensity maps of the line wings, which may be due to the limited 21\arcsec\
resolution of the CO maps. The line wings could instead be explained by photoevaporated CO
flowing away from the surface of the clouds, rather than protostellar molecular
outflows.

There are are two jet-like features observed in the H$\alpha$ image of SO 11NE (see
Fig.~\ref{fig:halpha_zoom}), which support the outflow hypothesis within the SMM1 core. The
jet-like features resemble the well-known photoionised jet HH 399 associated with the
bright-rimmed cloud TC2 in the Trifid Nebula (Lefloch et al.~\cite{lcrmch02}). The jet-like
features associated with \object{SFO 11NE} may thus arise from a similar situation as that found
in TC2. The alignment of the two jet-like features, while roughly corresponding with each other
and indicating that they may both arise from the same source, does not correspond with that of
either the 850 $\mu$m continuum peak of  \object{SFO 11NE SMM1} or any 2MASS Point Source
Catalogue objects . This could indicate the location of another protostellar object within
\object{SFO 11NE} SMM1, which is perhaps either a less evolved class 0 object or more heavily
obscured within the core. Further investigation of these jet-like features,  higher resolution
CO observations and deeper infrared images are a priority to confirm the presence of outflows
and identify their driving sources. 

The physical conditions within the cores are consistent with those of an early  phase of
protostellar evolution. Integrating over the greybody fits to the observed SED of the
three SMM1 cores  (Sect.~\ref{sect:scuba_anal}) yields the bolometric luminosity, $L_{\rm
bol}$, of each core (330, 170 and 430 L$_{\odot}$ for  \object{SFO 11 SMM1},  \object{SFO 11NE SMM1} and SFO
11E SMM1 respectively).  The bolometric luminosities are higher than typical Class 0 and
I objects, which are typically a  few tens of L$_{\odot}$ (Andr\'e, Ward-Thompson \&
Barsony \cite{awb93}; Chandler \& Richer \cite{cr00}), possibly indicating that the
cores are forming multiple stars or intermediate-mass stars (Sugitani et
al.~\cite{smnto00}). 

The 2MASS results provide some support for this hypothesis, by showing that
there are multiple YSOs and protostallar candidates embedded within the cores. In SFO11
the core SMM1 is associated with a weak-line T Tauri star and the two cores SMM2 and
SMM3 are both associated with candidate class I protostars. A small cluster of YSOs and
class I protostar candiates is embedded in the core \object{SFO 11NE SMM1}, with some
evidence for a possible class 0 object at the intersection of the two jet-like features.
\object{SFO 11NE SMM2} does not show any evidence for ongoing star formation, apart from
a T Tauri candidate located to the south west (source s in Fig.~\ref{fig:2mass2}). A T
Tauri candidate is found at the edge of  \object{SFO 11E SMM1}, but there is no evidence
for embedded star formation in this core beyond the moderate velocity line wings seen in
the CO observations,  as previously discussed. The cores  \object{SFO 11E SMM2}
and SMM3 are not associated with embedded protostars or YSOs, although two candidate
class I  protostars  (sources f and k in Fig.~\ref{fig:2mass2}) are located to the
north in the ridge of sub-mm emission stretching from  \object{SFO 11E SMM1} to 
\object{SFO 11E SMM2}.

In summary, there is reasonable evidence for either ongoing or recent star formation within all
three clouds. Two of the cloud cores (\object{SFO 11NE SMM1} and \object{SFO 11E} SM1) are associated with
line wings that are suggestive of molecular outflow. All of the sub-mm cores detected by
SCUBA possess similar physical characteristics to protostellar cores found in other
molecular clouds (e.g.~Evans \cite{evans99}) and this implies that the three cores in our
study that do not show any signs of current star formation (\object{SFO 11NE SMM2},  \object{SFO 11E SMM2} and
 \object{ \object{SFO 11E SMM3}}) may well be good prospects for future star formation. In order to investigate
their star-forming nature and to unambiguously classify the protostellar population of
the remaining cores more  accurate infrared photometry and sub-mm observations are
required to constrain their SEDs, temperatures and luminosities.

\subsection{Could the star formation have been induced by  \object{HD17505}?}

 In the previous two subsections we dwelt upon the physical properties of the clouds
and their star-forming nature. Here, we consider the effect that the UV illumination has
had upon the evolution of the clouds and their eventual fate. In the following
discussion we assume that the molecular gas of the clouds is in pressure balance with
the exterior ionised gas and that photoionisation-induced shocks are propagating into
the clouds. In Sect.~\ref{sect:cloudprops} we showed that this possibility is likely,
based upon the similarity of the cloud morphologies to those predicted by RDI models
(Bertoldi \cite{bertoldi}; Lefloch \& Lazareff \cite{ll94}) and the approximate pressure
equilibrium between the ionised exterior and molecular interior of the clouds. We are
confident  that the past and future evolution of the clouds may be intepreted in the
context of these RDI models. In this section and the next we will extend the model
predictions  to examine two important areas: could the star formation seen in the clouds
have been triggered by the UV illumination and what is the eventual fate of the clouds?

The SMM1 cores found at the head of each cloud are the logical candidates to examine for
signs that the star  formation within them was triggered by their UV illumination, as
the remaining cores are mostly shielded from the UV illumination by the SMM1 cores and,
apart from  \object{SFO 11 SMM3},
there is no evidence that they are forming stars.  It is difficult to assess whether the
SMM1 cores were actually formed by the RDI process; the morphology of  \object{SFO 11 SMM1}
suggests that it may have formed via RDI as the inverted-V morphology of the cloud is
extremely close to that predicted by Lefloch \& Lazareff (\cite{ll94}). However, the
remaining cores do not show signs of elongation along the axis of the UV illumination
(as predicted by the RDI models) and their origin remains unclear. It is impossible to
determine if the cores were formed by radiative-driven collapse or are simply
pre-existing structures found in the parent molecular cloud of IC 1848. No matter how
the SMM1 cores were originally formed it is clear that they are all host to either
current or recent star formation. 

One piece of circumstantial evidence supporting the hypothesis that the star formation in
the SMM1 cores was induced by the UV illumination from the nearby O star is that the
estimated age of the protostars and YSOs is similar to the timescale over which the clouds
have been illuminated. From a simple consideration of the expansion velocity of the
ionisation front we estimate that the three bright-rimmed clouds have been exposed to the
UV radiation for around 3$\times$10$^{5}$ years.  \object{SFO 11NE SMM1} is associated with Class I
protostars and T Tauri stars (class II YSOs), which typically have a characteristic age of
1--2$\times$10$^{5}$ years (Andr\'e, Ward-Thompson \& Barsony \cite{awb00}) and a few
times 10$^{5}$ to a few times 10$^{6}$ years respectively (Andr\'e \& Montmerle
\cite{am94}).   \object{SFO 11 SMM1} and  \object{SFO 11E SMM1} are both associated with  T Tauris and
possibly less evolved class 0 protostars. The evolutionary stage of the star formation
within the cores is thus consistent with that expected for star formation induced by the
UV illumination.  This similarity, however,  does not offer conclusive proof that the star
formation within the cores was induced, but merely that the induced scenario is plausible
with the apparent UV illumination timescale of the clouds.

We may also investigate the stability of the cores against gravitational collapse in
order to ascertain whether the UV illumination caused them to become unstable.
The RDI models  of Lefloch \& Lazareff (\cite{ll94}), Bertoldi \& McKee (\cite{bk90}) and
Bertoldi (\cite{bertoldi}) do not specifically address the gravitational stability of the
clouds with respect to eventual star formation, except to derive stability criteria for
equilibrium clouds (Bertoldi \& McKee \cite{bk90}). As a simple approach we 
use the virial
theorem to explore the
stability of a molecular cloud against the rise in external pressure caused by
photoionisation of the cloud surface.  For an
unmagnetised spherical isothermal cloud with uniform density and in virial equilibrium the
virial theorem reduces to (e.g. Hartmann \cite{hartmann00});

\begin{equation}
\label{eqn:virialeqn}
4 \pi R_{c}^{3} P_{\rm ext} = 3 c_{s}^{2} M_{c} - \frac{3}{5} \frac{GM_{c}}{R_{c}}
\end{equation}

where the radius and mass of the cloud are represented by $R_{c}$ and $M_{c}$
respectively, the sound speed of the gas is $c_{s}$, $P_{\rm ext}$ is the pressure of the
external medium surrounding the cloud and $G$ is the gravitational constant. By
considering the derivative of  the external pressure with respect to cloud radius it is
easy to show that there is a critical external pressure for the cloud above which the
cloud is not stable to collapse. This critical pressure may be written in terms of the
FWHM linewidth $\Delta v$ (where the turbulent sound speed $c_{s}^{2} = \left<\Delta
v\right>^{2}/8 \ln 2$) and the mass $M_{c}$ of the cloud as

\begin{equation}
P_{\rm crit} \simeq 3.33\times10^{-3}\,\frac{\left<\Delta v\right>^{8}}{G^{3} M_{c}^{2}}.
\end{equation}

However, if we know the external pressure $P_{\rm ext}$ acting upon the cloud, we may
write Eq.~(\ref{eqn:virialeqn}) in the more familiar terms of a ``virial mass'',
i.e. the mass at which the critical pressure for cloud collapse is equal to the external
cloud pressure. The mass determined in this manner is comparable to the more traditional
virial mass of the cloud as derived from the FWHM linewidth and cloud radius, but takes
into account the pressure of the medium surrounding the cloud. We refer to this
pressure-sensitive mass as the ``pressurised virial mass'' to distinguish it from the
more traditional definition of the virial mass. An increase in the cloud mass
consequently reduces the critical external pressure required for the cloud to collapse;
if the external pressure is fixed and the cloud mass is such that the critical pressure
is lower than the external pressure the cloud will collapse. Thus, clouds with true
masses exceeding their pressurised virial mass are unstable against gravitational
collapse. The pressurised virial mass $M_{\rm pv}$ is written as:

\begin{equation}
M_{\rm pv} \simeq 5.8\times10^{-2}\,\frac{\left<\Delta v\right>^{4}}{G^{\,3/2}\,P_{\rm
ext}^{\,1/2}}. 
\label{eqn:pvm}
\end{equation}

As the pressurised virial mass depends upon the fourth power of the FWHM linewidth $\Delta
v$, its accuracy is highly dependent upon the accuracy of the linewidth. From a
consideration of the errors in gaussian fits to the $^{13}$CO lines we estimate that the
resulting pressurised virial masses derived from Eq.~(\ref{eqn:pvm}) are accurate to
within a factor of 2. The pressurised virial mass may be compared to the standard virial
 mass $M_{\rm vir}$, which
is given as: 

\begin{equation}
M_{\rm vir} \simeq 210\, R_{c}\, \left<\Delta v\right>^{2},
\end{equation}

where $M_{\rm vir}$ is in M$_{\odot}$, $R_{c}$ is the cloud radius in pc and $\Delta v$
is the FWHM linewidth in km\,s$^{-1}$ (e.g.~Evans \cite{evans99}).  In order to explore
the evolution of the cores before and after their UV illumination we consider the
difference between their standard and pressurised virial masses. We equate the standard
virial mass with the pre-illumination conditions of the cores and the pressurised virial
mass with the current core characteristics. Implicit assumptions in these scenarios are 
that the cores were subject to a negligible external pressure prior to the illumination
and that their physical properties  (such as mass, linewidth and radius) were not
significantly different prior to the illumination to their current values.  We use the
standard ``unpressurised'' virial mass to illustrate the pre-illumination case, rather
than estimate the initial external pressure acting upon the clouds, as the initial
conditions exterior to the cores are difficult to establish.

Whilst this approach is simplistic, assuming that the pressure of the ionised boundary
layer applies to the entire surface of the core instead of the face illuminated by the
UV flux and that the cores are pre-existing entities, it nevertheless allows us to
explore whether the rise in external pressure is significant in terms of the eventual
stability of the cores.  Values for both the pressurised and standard virial masses of
the SMM1 cores are found in Table \ref{tbl:pvm}. The FWHM linewidths used to calculate
the virial masses were the source-averaged $^{13}$CO linewidth, the core radii as
measured from the 850 $\mu$m continuum maps and the external pressures as derived from
the NVSS 20 cm fluxes. For ease of comparison the core masses  determined from the 850
$\mu$m continuum emission in Sect.~\ref{sect:anal} have also been included in Table
\ref{tbl:pvm}.      

\begin{table*}
\caption{Values of the pressurised virial mass $M_{pv}$, 
the standard unpressurised virial mass M$_{\rm
vir}$ and the H$_{2}$ mass derived from the sub-mm continuum measurements $M_{\rm submm}$ 
for each of the SMM1 cores.}
\label{tbl:pvm}
\begin{tabular}{lccc}\hline
Core & $M_{pv}$ & $M_{\rm vir}$ & M$_{\rm submm}$ \\
 & (M$_{\odot}$)  & (M$_{\odot}$)  & (M$_{\odot}$) \\
\hline\hline
 \object{SFO 11 SMM1} & 10.4 & 42.3 & 20.6 \\
 \object{SFO 11NE SMM1} & 45.4 & 81.6 & 13.4 \\
 \object{SFO 11E SMM1} & 31.2 & 85.0 & 18.7 \\ \hline
\end{tabular}
\end{table*}

The standard virial masses of the cores are much larger than the masses derived from the
sub-mm continuum. This implies that prior to the UV illumination of the clouds the cores
were likely to be stable against collapse. The pressurised virial masses of the cores are
lower than their standard virial masses and this indicates the destabilising effect of the
high external pressures upon the gravitational stability of the cores.  \object{SFO 11NE SMM1} and
 \object{SFO 11E SMM1} are the least affected by the rise in external pressure; their pressurised
virial masses are a larger fraction of the standard virial mass and a factor of 2--3
greater than the mass derived from their sub-mm continuum emission.  \object{SFO 11 SMM1}, on the
other hand, has a pressurised virial mass comparable to its sub-mm mass and half that of
the standard virial mass.  However, within the uncertainties, it is possible that \object{SFO
11E SMM1} is also unstable against collapse. \object{SFO 11 SMM1} and \object{SFO
11E SMM1} are thus the only SMM1 cores whose
collapse may have been induced by the rise in external pressure.

From these results the immediate conclusions that may be drawn are that all of the cores
were stable against collapse prior to their UV illumination and that  \object{SFO 11 SMM1} and \object{SFO 11E SMM1}
are
currently unstable due to the rise in its external pressure. However these conclusions
must be interpreted with caution. It is obvious from the 2MASS K-band images  that the
cores are fragmented and that collapse of the individual fragments has begun to form
multiple protostars within the clouds. The picture of core stability drawn from the
virial masses of the cores applies to the stability of the overall core, not to
individual fragments within each core. The assumption is also that the cores pre-date
the UV illumination, this may not be true for  \object{SFO 11 SMM1} as this core displays the
hallmarks of RDI formation. What we may conclude from the core virial masses is that the
rise in external pressure has not played an active role in the overall evolution of \object{SFO
11NE SMM1}, but may have contributed to the gravitational instability of \object{SFO
11 SMM1} and \object{SFO 11E SMM1}. It is impossible to determine whether the star formation seen within the cores
was induced by the UV illumination without a more detailed theoretical treatment. We have
nevertheless shown that is it possible for the rise in external pressure to have caused
the collapse of  \object{SFO 11 SMM1} and \object{SFO 11E SMM1}.

\subsection{The evolution of the clouds and their eventual fate}
\label{sect:evolution}

The long-term evolution of the clouds is expected to broadly follow that described in the RDI models
of Bertoldi (\cite{bertoldi}) and Lefloch \& Lazareff (\cite{ll94}). The conditions within
the ionised boundary layer and the dense molecular cores at the head of the clouds are
consistent with those required for a D-critical ionisation front to propagate into the
interiors of the clouds, slowly ionising and dispersing the molecular gas. Lefloch \&
Lazareff (\cite{ll94}) estimate the mass loss rate of such clouds as

\begin{equation}
\dot{M} = 4.4 \times 10^{-3} \,\, \Phi^{1/2} \,R_{c}^{\,3/2} \,\,\,\,{\rm M_{\odot}} \,{\rm
Myr^{-1}},
\end{equation}

where the ionising photon flux $\Phi$ is measured in units of cm$^{-2}$\,s$^{-1}$ and the
cloud radius $R_{c}$ is in pc. For the SMM1 cores at the heads of the clouds in our study
(SFO 11, \object{SFO 11NE} and \object{SFO 11E}) we
estimate their current mass loss rates as 3.5, 3.5 and 6.6 M$_{\odot}$\,Myr$^{-1}$
respectively. If the ionising flux impinging upon \object{SFO 11E} is reduced to that of the remaining
two clouds to account for the possibility that the 20 cm flux associated with this cloud is
overestimated (see Section \ref{sect:ionising}) then the mass loss rate reduces to $\sim$3.5
M$_{\odot}$\,Myr$^{-1}$. Lower limits to the lifetime of the cores may be determined from
$M/\dot{M}$, suggesting that the lifetime of the cores is at least 4--6 Myr.

This lifetime is based upon the steady propagation of the ionisation front into the
core, there is some evidence from the H$\alpha$ images that this may not be the case for
SFO 11 or \object{SFO 11E}. The rims of both of these clouds are irregular, indicating that the
ionisation front may have developed Rayleigh-Taylor instabilities (Lefloch \& Lazareff
\cite{ll94}). \object{SFO 11} displays a lumpy, irregular structure with many tiny globules that
appear to have broken free from the main body of the cloud. \object{SFO 11E} possesses a
``corrugated'' rim on its NW face with bright pockets of H$\alpha$ emission located in
the troughs of the corrugations.  Lefloch \& Lazareff (\cite{ll94}) consider the effect
of instabilities in the  ionisation front and show that clouds with unstable ionisation
fronts experience an enhanced mass loss rate after the initial collapse phase. 
The morphologies of the clouds suggest that they are in the early collapse phase and so
the mass loss from \object{SFO 11} and \object{SFO 11E} may increase rapidly in their near future. 

Nevertheless, it is expected that the SMM1 cores at the head of the clouds will survive
for at least another 1--2 Myr and if instabilities do not develop in  \object{SFO 11NE SMM1} its
expected lifetime is at least 4 Myr. The current star formation occurring within the
cores is at the early class I/II phase. The accretion stage in star formation lasts for
typically a few 10$^{5}$ years (Andr\'e, Ward-Thompson \& Barsony \cite{awb00}) and so any
existing or imminent star formation within the cores should be relatively undisturbed by
the mass loss. The clouds are primarily illuminated by the nearby 06V star  \object{HD17505}, which
has an estimated main-sequence lifetime of 3.8 Myr, following McKee, Van Buren \& Lazareff
(\cite{mkvbb84}) and Panagia (\cite{panagia}). Taking the age of IC 1848 to be $\sim$ 1.5 Myr
(Vall\'ee, Hughes \& Viner \cite{vhv79}),  \object{HD17505} is roughly half-way through its
alloted lifespan. The lifetime of the cores is thus sufficiently long that they may even
perhaps outlast the UV illumination from their ionising O star.

%______________________________________________________________

\section{Summary and conclusions}
\label{sect:conc}

We have carried out an in-depth study of three bright-rimmed clouds (SFO 11, \object{SFO 11NE}
and \object{SFO 11E}) associated with the HII region IC 1848, in order to search the clouds for
evidence of star formation and attempt to place the star formation in context with the
likely past and future evolution of the clouds. We observed the three clouds using SCUBA
on the JCMT to map their continuum emission at 450 and 850 $\mu$m; the JCMT to map the
J=2--1 lines of $^{12}$CO and $^{13}$CO; and the Nordic Optical Telescope to image the
H$\alpha$ emission from the cloud rims using a narrowband filter. We have also obtained
archival data of the 20 cm radio emission from the NVSS; far and mid-infrared IRAS HIRES
maps; and 2.2 $\mu$m 2MASS images of the clouds. We draw the following conclusions from
our observations:

\begin{enumerate}

\item The overall morphology of the clouds as seen in the H$\alpha$ images,  SCUBA and
CO maps is reasonably consistent with the predictions of the radiative-driven implosion
(RDI) model of Lefloch \& Lazareff (\cite{ll94}). The cloud morphology resembles  the
early collapse phase of the Lefloch \& Lazareff (\cite{ll94}) model, some 1--2 $\times$
10$^{5}$ years after the UV illumination began.  However, the cores found at the head of
the clouds \object{SFO 11NE} and \object{SFO 11E} are not elongated in the direction of the UV
illumination and this indicates that these cores may not have formed by the RDI process
and could thus pre-date the UV illumination of the clouds. The H$\alpha$ images show the
typical radial striations expected for a photoevaporated flow of gas from the cloud
surface. 

\item There is reasonable evidence for ongoing star formation within all three clouds at the
class I or class II evolutionary stage. The SCUBA maps reveal 8 dust cores within the
optical boundaries of the clouds which possess the same physical properties as
protostellar cores. The dust cores found toward the head of the bright-rimmed clouds are
typically larger and more massive than those found further away from the bright rims.
2MASS near-infrared images reveal several  candidate class I protostars and T Tauri
stars associated with the clouds, although the majority of these objects are also consistent with reddened
main sequence stars. More accurate infrared photometry is required to resolve ambiguities in
classification. Molecular outflows may be associated with the two cores
 \object{SFO 11NE SMM1} and  \object{SFO 11E SMM1}, although higher spatial resultion is required to resolve
the individual outflow lobes. A possible photoionised jet is seen in the H$\alpha$ image
of \object{SFO 11NE} and optical spectroscopy to confirm the nature of this feature is a priority.

\item The cometary morphology of the clouds suggests that their primary ionising star is
\object{HD17505}. The orientation of \object{SFO 11E} is not aligned with the direction
toward  \object{HD17505} and this may indicate that \object{SFO 11E} is also affected by
another nearby OB star. The H$\alpha$ emission from the bright rim of \object{SFO 11E} is
brighter on the  \object{HD17505}-facing side of the cloud, suggesting that HD 17505 is the
dominant exciting star. The 20 cm free-free emission from the clouds is consistent with a
true distance from HD 17505 of approximately 22 pc. The star-cloud distance as seen in
projection is approximately 11 pc.

\item  All three clouds are in approximate pressure equilibrium between their ionised exterior and
molecular interior, which is consistent with the RDI model prediction that D-critical ionisation
fronts and photoionisation-induced shocks are currently propagating into the clouds. From the
current data we cannot rule out the possibility that the clouds are over-pressured with respect to
their ionised boundary layers  and in this scenario the photoionisation-induced shocks are stalled
at the ionised boundary layer (Bertoldi \cite{bertoldi}; Lefloch \& Lazareff \cite{ll94}).
Nevertheless the close correspondence between the morphologies of the clouds and the RDI models
suggests that the clouds are currently being shocked and on balance we conclude that it is likely
that photoionisation-induced shocks are propagating into the clouds. We estimate induced shock
velocities of $\sim$ 1.4 km\,s$^{-1}$, which have a crossing time of $\sim 2 \times 10^{5}$  years
for the diameters of the cores found at the head of the clouds. The crossing times are similar to
predictions of the cloud lifetimes determined from their morphology and indicates that there has
been sufficient time since the clouds were illuminated for the shocks to propagate deep within the
clouds.

\item It is difficult to ascertain whether or not the current star formation observed
within the clouds was induced by the photoionisation process. The duration of the cloud
UV illumination, their shock crossing times and the estimated age of the protostars/YSOs
that have formed within the clouds are all similar (typically a few times 10$^{5}$
years). We have shown via a simple pressure-based argument that the UV illumination may
have rendered  \object{SFO 11 SMM1} and \object{SFO 11E SMM1} unstable against collapse 
but was not likely to have
significantly affected  \object{SFO 11NE SMM1}. A more detailed theoretical
treatment of the cloud collapse, extending the existing RDI models in this direction, is
required to make more concrete predictions in this area.

\item Based upon the current mass loss from the SMM1 cores, predicted by the Lefloch \&
Lazareff (\cite{ll94}) RDI model, we estimate their lifetimes as between 4 and 6 Myr.
Potential  ionisation front instabilities seen in the H$\alpha$ images of \object{SFO 11} and
SFO 11E may substantially reduce this lifetime. The ongoing star formation within the clouds
should proceed relatively undisturbed by the dispersal of the clouds. In the absence of
instabilities developing  within the ionisation front, which significantly accelerate the mass
loss rate (Lefloch \& Lazareff \cite{ll94}), the dense core at the head of SFO 11NE may even
outlive its ionising O star. 

\end{enumerate}

\begin{acknowledgements}  

The authors would like to thank an anonymous referee for a most thorough reading of this paper and several
useful suggestions, particularly regarding the 2MASS photometry; James Urquhart for his critical insight 
and useful discussions; and the JCMT support staff for a pleasant and productive observing run. The
Digitized Sky Survey was produced at the Space Telescope Science Institute under U.S. Government grant NAG
W-2166. The images of these surveys are based on photographic data obtained using the Oschin Schmidt
Telescope on Palomar Mountain and the UK Schmidt Telescope. The plates were processed into the present
compressed digital form with the permission of these institutions. Quicklook 2.2 $\mu$m 2MASS images were
obtained as part of the Two Micron All Sky Survey (2MASS), a joint project of the University of
Massachusetts and the Infrared Processing and Analysis Center/California Institute of Technology, funded by
the National Aeronautics and Space Administration and the National Science Foundation. IRAS HIRES images
were obtained from the NASA/IPAC Infrared Science Archive, which is operated by the Jet Propulsion
Laboratory, California Institute of Technology, under contract with the National Aeronautics and Space
Administration. This research has made use of the SIMBAD astronomical database service operated at CCDS,
Strasbourg, France  and the NASA Astrophysics Data System Bibliographic Services.

\end{acknowledgements}

\end{document}